\newcommand\apjcls{1}
\newcommand\aastexcls{2}
\newcommand\othercls{3}

\newcommand\papercls{\aastexcls}
\documentclass[tighten, times, preprint2]{aastex6}  

\if\papercls \apjcls
\usepackage{apjfonts}
\else\if\papercls \othercls
\usepackage{epsfig}
\fi\fi
\usepackage{ifthen}
\usepackage{natbib}
\usepackage{amssymb, amsmath}
\usepackage{appendix}
\usepackage{etoolbox}
\usepackage[T1]{fontenc}
\usepackage{paralist}

\if\papercls \apjcls
\newcommand\aas{\ref@jnl{AAS Meeting Abstracts}}
\newcommand\dps{\ref@jnl{AAS/DPS Meeting Abstracts}}
\newcommand\maps{\ref@jnl{MAPS}}
\else\if\papercls \othercls
\usepackage{astjnlabbrev-jh}
\fi\fi

\if\papercls \aastexcls
\hypersetup{citecolor=blue, 
            linkcolor=blue, 
            menucolor=blue, 
            urlcolor=blue}  
\else
\usepackage[
bookmarks=true,           
bookmarksnumbered=true,   
colorlinks=true,          
citecolor=blue,           
linkcolor=blue,           
menucolor=blue,           
urlcolor=blue,            
linkbordercolor={0 0 1},  
pdfborder={0 0 1},
frenchlinks=true]{hyperref}
\fi

\if\papercls \othercls
\newcommand{\eprint}[1]{\href{http://arxiv.org/abs/#1}{#1}}
\else
\renewcommand{\eprint}[1]{\href{http://arxiv.org/abs/#1}{#1}}
\fi

\providecommand{\adsurl}[1]{\href{#1}{ADS}}

\makeatletter
\patchcmd{\NAT@citex}
  {\@citea\NAT@hyper@{%
     \NAT@nmfmt{\NAT@nm}%
     \hyper@natlinkbreak{\NAT@aysep\NAT@spacechar}{\@citeb\@extra@b@citeb}%
     \NAT@date}}
  {\@citea\NAT@nmfmt{\NAT@nm}%
   \NAT@aysep\NAT@spacechar\NAT@hyper@{\NAT@date}}{}{}

\patchcmd{\NAT@citex}
  {\@citea\NAT@hyper@{%
     \NAT@nmfmt{\NAT@nm}%
     \hyper@natlinkbreak{\NAT@spacechar\NAT@@open\if*#1*\else#1\NAT@spacechar\fi}%
       {\@citeb\@extra@b@citeb}%
     \NAT@date}}
  {\@citea\NAT@nmfmt{\NAT@nm}%
   \NAT@spacechar\NAT@@open\if*#1*\else#1\NAT@spacechar\fi\NAT@hyper@{\NAT@date}}
  {}{}
\makeatother

\makeatletter
\DeclareRobustCommand{\lowcase}[1]{\@lowcase#1\@nil}
\def\@lowcase#1\@nil{\if\relax#1\relax\else\MakeLowercase{#1}\fi}
\makeatother

\DeclareSymbolFont{UPM}{U}{eur}{m}{n}
\DeclareMathSymbol{\umu}{0}{UPM}{"16}
\let\oldumu=\umu
\renewcommand\umu{\ifmmode\oldumu\else\math{\oldumu}\fi}

\if\papercls \othercls

\else

\fi

\let\oldsim=\sim
\renewcommand\sim{\ifmmode\oldsim\else\math{\oldsim}\fi}
\let\oldpm=\pm
\renewcommand\pm{\ifmmode\oldpm\else\math{\oldpm}\fi}
\newcommand\by{\ifmmode\times\else\math{\times}\fi}

\newcommand\tablebox[1]{\begin{tabular}[t]{@{}l@{}}#1\end{tabular}}
\newbox{\wdbox}
\renewcommand\c{\setbox\wdbox=\hbox{,}\hspace{\wd\wdbox}}
\renewcommand\i{\setbox\wdbox=\hbox{i}\hspace{\wd\wdbox}}

\newcount\timect
\newcount\hourct
\newcount\minct
\newcommand\now{\timect=\time \divide\timect by 60
         \hourct=\timect \multiply\hourct by 60
         \minct=\time \advance\minct by -\hourct
         \number\timect:\ifnum \minct < 10 0\fi\number\minct}

\catcode`@=11

\newcommand\comment[1]{}

\newcommand\commenton{\catcode`\%=14}

\renewcommand\math[1]{$#1$}
\newcommand\mathshifton{\catcode`\$=3}

\let\atab=&
\newcommand\atabon{\catcode`\&=4}

\let\oldmsp=\sp
\let\oldmsb=\sb
\def\sp#1{\ifmmode
           \oldmsp{#1}%
         \else\strut\raise.85ex\hbox{\scriptsize #1}\fi}
\def\sb#1{\ifmmode
           \oldmsb{#1}%
         \else\strut\raise-.54ex\hbox{\scriptsize #1}\fi}
\newbox\@sp
\newbox\@sb
\def\sbp#1#2{\ifmmode%
           \oldmsb{#1}\oldmsp{#2}%
         \else
           \setbox\@sb=\hbox{\sb{#1}}%
           \setbox\@sp=\hbox{\sp{#2}}%
           \rlap{\copy\@sb}\copy\@sp
           \ifdim \wd\@sb >\wd\@sp
             \hskip -\wd\@sp \hskip \wd\@sb
           \fi
        \fi}
\def\msp#1{\ifmmode
           \oldmsp{#1}
         \else \math{\oldmsp{#1}}\fi}
\def\msb#1{\ifmmode
           \oldmsb{#1}
         \else \math{\oldmsb{#1}}\fi}

\def\supon{\catcode`\^=7}

\def\subon{\catcode`\_=8}

\def\supsubon{\supon \subon}

\newcommand\actcharon{\catcode`\~=13}

\newcommand\paramon{\catcode`\#=6}

\comment{And now to turn us totally on and off...}

\newcommand\reservedcharson{ \commenton  \mathshifton  \atabon  \supsubon 
                             \actcharon  \paramon}

\catcode`@=12
\reservedcharson

\if\papercls \apjcls

\else

\fi

\if\papercls \othercls
\else
  \newcommand\inpress{n}
  \if\inpress y
    \received{\today}
    \revised{}
    \accepted{}
    \if\papercls \apjcls
    \slugcomment{}
    \fi
  \else
  \slugcomment{\tablebox{In preparation for {\em ApJ}. DRAFT of {\today}.}}
  \fi
\fi

\newcommand\chisq{\ifmmode{\chi\sp{2}}\else\math{\chi\sp{2}}\fi}
\newcommand\redchisq{\ifmmode{ \chi\sp{2}\sb{\rm red}}
                    \else\math{\chi\sp{2}\sb{\rm red}}\fi}
\newcommand\Teq{\ifmmode{T\sb{\rm eq}}\else$T$\sb{eq}\fi}
\newcommand\mjup{\ifmmode{M\sb{\rm Jup}}\else$M$\sb{Jup}\fi}
\newcommand\rjup{\ifmmode{R\sb{\rm Jup}}\else$R$\sb{Jup}\fi}
\newcommand\msun{\ifmmode{M\sb{\odot}}\else$M\sb{\odot}$\fi}
\newcommand\rsun{\ifmmode{R\sb{\odot}}\else$R\sb{\odot}$\fi}
\newcommand\mearth{\ifmmode{M\sb{\oplus}}\else$M\sb{\oplus}$\fi}
\newcommand\rearth{\ifmmode{R\sb{\oplus}}\else$R\sb{\oplus}$\fi}

\shorttitle{Revealing reionization with the IGM thermal history}
\shortauthors{Boera {\em et al.}}

\usepackage{subfigure}
\usepackage{amssymb}
\renewcommand{\vec}[1]{\mathbf{#1}}
\begin{document}

\title{Revealing reionization with the thermal history of the intergalactic medium: new constraints from the Lyman-$\alpha$ flux power spectrum}

\author{Elisa~Boera\altaffilmark{1},
George~D.~Becker\altaffilmark{1},
James~S.~Bolton\altaffilmark{2}
and
Fahad~Nasir\altaffilmark{1}
}

\affil{\sp{1}  Department of Physics and Astronomy, University of California, Riverside, CA 92521, USA\\
       \sp{2} School of Physics and Astronomy, University of Nottingham, University Park, Nottingham, NG7 2RD, UK}

\email{elisa.boera@gmail.com}

\begin{abstract}
We present a new investigation of the thermal history of the intergalactic medium (IGM) during and after reionization using the Lyman-$\alpha$ forest flux power spectrum at $4.0\lesssim z\lesssim5.2$. Using a sample of 15 high--resolution spectra, we measure the flux power down to the smallest scales ever probed at these redshifts ($-1\lesssim \log(k/$km$^{-1}$s)$\lesssim -0.7$). These scales are highly sensitive to both the instantaneous temperature of the IGM and the total energy injected per unit mass during and after reionization. We measure temperatures at the mean density of $T_{0}\sim7000$-8000 K, consistent with no significant temperature evolution for redshifts $4.2\lesssim z\lesssim5.0$. 
We also present the first observational constraints on the integrated IGM thermal history, finding that the total energy input per unit mass increases from $u_{0}\sim4.6$ ${\rm eV}$ $m_{\rm p}^{-1}$ to 7.3 eV $m_{\rm p}^{-1}$ from $z\sim 6$ to 4.2 assuming a $\Lambda$-CDM cosmology.
We show how these results can be used simultaneously to obtain information on the timing and the sources of the reionization process. Our first proof of concept using simplistic models of instantaneous reionization produces results comparable to and consistent with the recent Planck constraints, favoring models with $z_{\rm rei}\sim 8.5^{+1.1}_{-0.8}$.  
\end{abstract}
\keywords{cosmology: observations ---
          cosmology: early universe   ---
          (galaxies) quasars: absorption lines --- 
          methods: observational ---
          methods: statistical}

\section{INTRODUCTION}
\label{introduction}

The epoch of hydrogen reionization, represents one the most dramatic phases of evolution of the Universe. During this period, the UV radiation from the first luminous sources reionized the neutral hydrogen (and He{\sc \,i}) atoms in the diffuse intergalactic medium (IGM), driving the transition from a neutral to a highly ionized Universe. Understanding sources and timing of this transformation can reveal crucial information on the properties of the first objects and the environment in which they were formed.
When and how reionization happened therefore remains a primary subject of interest in extragalactic astrophysics \citep[for a review, see][]{Becker15Rev}.  

The most direct probes of the highly ionized IGM have been obtained from observations of intergalactic Lyman--$\alpha$ (Ly$\alpha$) absorption along the lines of sight to high--redshift quasars.
Measurements of Ly$\alpha$ transmission along some lines of sight suggest that reionization was largely complete by $z\sim6$ \citep[e.g.,][]{McGreer15}. On the other hand, large fluctuations in IGM opacity remain at $z \lesssim 6$, suggesting that lingering evidence of reionization may remain in the IGM to somewhat lower redshifts \citep[]{Fan, Becker15,Bosman18,Eilers18}. 
While current constraints from cosmic microwave background (CMB) observations are consistent with a rapid reionization at redshift $z_{\rm rei}\simeq7.7\pm 0.7$
 \citep{Planck18}, measurements of the fraction of neutral hydrogen at high redshift have also been obtained from the presence of Ly$\alpha$ damping wings \citep{Mortlock11,Simcoe12,Greig17,Davies18} and from the weakening of Ly$\alpha$ emission lines in $z\sim 6-8$ galaxies \citep[e.g.,][]{Caruana14, Schmidt16, Sadoun17,Mason18}. 
The available data seem to generally support a late reionization scenario (with the bulk of reionization happening at $z\sim 6-8$) but are still consistent with a relatively broad range of reionization histories. 

The sources responsible for reionization also remain uncertain. Star--forming galaxies have commonly been considered the most likely candidate \citep[e.g.,][]{FinkelsteinRev16, Bowens16, Bowens15}. Scenarios in which active galactic nuclei (AGN) make a substantial contribution, however, continue to be considered \citep[e.g.,][]{Giallongo15, Madau15, Parsa17,DAloisio17}. 

Further insight may be gained from using the IGM thermal history to constrain the reionization process. The temperature of the IGM should increase significantly via photo--ionization heating during hydrogen reionization and, because its cooling time is long, the low density gas retains some useful memory of when and how it was reionized \citep{MiraldaRees94,Abel99, Sanderbeck16}. At the mean density of the IGM the characteristic signature of reionization is expected to be an increase in temperature of tens of thousands of Kelvin as an ionization front sweeps through \citep[e.g.,][]{Daloisio18}, followed by cooling (over $\Delta z\sim1-2$) towards a thermal asymptote set primarily by the balance between photo--heating by the UV background (UVB) and adiabatic cooling due to the expansion of the Universe (e.g., \citealp{McQuinn09}).  
The interplay among these effects is expected to lead to a power--law temperature--density (T--$\rho$) relation for the low density gas ($\Delta=\rho/\bar{\rho}\lesssim 10$) of the form
\begin{equation}\label{eq:TD}
T(\Delta)=T_{0}\Delta^{\gamma -1} ,
\end{equation}
where $T_{0}$ is the temperature at the mean density and ($\gamma$-1) is the slope of the relation \citep{HuiGnedin1997,Puchwein14,McQuinn16}.

Following reionization, the increase in gas pressure due to the boost in temperature smooths out of the gas on small scales \citep[e.g.,][]{Gnedin98,Rorai13,Kulkarni15}. The degree of ``Jeans smoothing'' in the IGM prior to a given redshift is sensitive to timing and the total heat injection during and after reionization. Measurements of both the gas temperature evolution and the Jeans smoothing at redshifts approaching reionization ($z\gtrsim4$) can therefore constrain the timing of this process and potentially provide information on the nature of the ionizing sources.

In the last two decades the Ly$\alpha$ forest in quasar spectra has been the main laboratory for the study of the thermal state of the IGM. In combination with cosmological hydrodynamical simulations, previous efforts have used a variety of statistical approaches to measure the IGM temperature--density relation at $1.5\lesssim z\lesssim5.4$ from the shapes of the Ly$\alpha$ absorption lines \citep[e.g.,][]{Schaye00,Ricotti00,McDonald01,Theuns02,Bolton08,Lidz10,Becker11,Rudie12,Bolton13,Boera14,Boera16,Hiss17,Rorai17,Rorai18}. However, the widths of these features are sensitive to both the instantaneous temperature of the gas (thermal broadening) and Jeans smoothing (which increases the Hubble broadening) due to the heat injection at higher redshifts. In previous works the impact of pressure smoothing has generally either not been included or has been considered a source of systematic error. For example, \citealt{Viel13} and \citealt{Irsic17} account for this effect by adding the redshift of reionization as a nuisance parameter for their warm dark matter constraints.

On the other hand, the first direct measurement of the characteristic filtering scale over which the gas is pressure smoothed ($\lambda_{\rm P}$) has recently been obtained from the analysis of the Ly$\alpha$ absorption correlations using close quasar pairs at $z\sim2-3$ \citep{Rorai17S}. This method largely disentangles the impacts of thermal broadening and pressure smoothing; however, the lack of known quasar pairs at higher redshifts prevents it from being used at redshifts closer to hydrogen reionization.

An alternative means of simultaneously constrain temperature and Jeans smoothing is presented by the Ly$\alpha$ flux power spectrum (\citealt{Puchwein14}; \citealt{Nasir16}, hereafter N16; \citealt{Walter18}).
 N16 demonstrated using hydrodynamical simulations that the Ly$\alpha$ flux power spectrum exhibits different scale dependences for the temperature and Jeans smoothing. In particular, probing small scales (wavenumber log($k$/km$^{-1}$s)$\gtrsim-1$) increases the sensitivity to different reionization scenarios (see also \citealt{Onorbe17} for an independent analysis). 
Although the one--dimensional flux power spectrum statistic has been already explored in several works \citep[e.g.,][]{Kim04,McDonald06,Viel13,Delabouille15b,Irsic17,Yeche17}, the lack of high resolution, high signal--to--noise (S/N) Ly$\alpha$ forest spectra has so far prevented these small scales from being measured at redshifts approaching reionization (but see \citealt{Walther18} for an analysis at $z<4$). 

In this paper we present a a new measurement of the Ly$\alpha$ flux power spectrum at $z\sim4-5.2$ obtained from a sample of high resolution, high S/N spectra. We extend the measurement to previously unexplored small scales (log($k$/km$^{-1}$s)$=-0.7$). By comparing the data to predictions from a suite of hydrodynamical simulations we investigate the IGM temperature evolution and, simultaneously, its integrated thermal history. We then demonstrate how the combined constraints offer new insights on the timing and sources of the hydrogen reionization process. 

For this work we have adopted the parametrization of the Jeans smoothing effect described in N16. We characterize the integrated thermal history of the IGM using the cumulative energy per unit mass, $u_{0}$, injected into the gas at the mean cosmic density during and after the reionization process. As we demonstrate, this quantity can be directly used to constrain reionization models. 

This paper is organized as follows. In Section 2 we introduce the observational sample of high-resolution spectra. An overview of the simulations used to test and interpret the measurements is presented in Section 3. In Section 4 we introduce the power spectrum method, discussing the effect of the most relevant thermal parameters. In Section 5 we present the observational power spectrum results and discuss the strategies applied to take into account and reduce systematic uncertainties. The calibration and analysis of the synthetic power spectrum models are described in Section 6. The Markov Chain Monte Carlo (MCMC) analysis, comparing models with the observational measurements is described in Section 7, where we also present our main results for the IGM temperature at the mean density and the integrated thermal history. As an example of how our thermal constraints can be used to test reionization histories, we apply our results to instantaneous reionization models in Section 8. We summarize our findings and conclude in Section 9. Tests for various systematic effects are described in the appendices.

\section{OBSERVATIONAL SPECTRA}
\label{sec:observations}
We obtained high--resolution spectra of a sample of 15 quasars spanning emission redshifts $4.8\lesssim z_{\rm em} \lesssim 5.4$. The quasars and their basic properties are listed in Table \ref{table:QSO}. The spectra for eleven of the objects were obtained with the Keck High Resolution Echelle Spectrometer (HIRES; \citealt{Vogt94}) while the remaining four were taken with the Ultraviolet and Visual Echelle Spectrograph (UVES; \citealt{Dekker2000}) on the Very Large Telescope (VLT). 

The spectra were reduced using a custom set of IDL routines that include optimal sky subtraction \citep{Kelson03} and extraction techniques \citep{Horne86}. 
For each object a single one--dimensional spectrum was extracted simultaneously from all exposures after individually applying telluric absorption corrections and relative flux calibration to the two--dimensional frames. Telluric corrections were modeled based on the ESO SKYCALC Cerro Paranal Advanced Sky Model \citep{Noll12,Jones13}. For the UVES data we found that flux calibration derived from standard stars yielded sufficiently accurate agreement between overlapping orders.  For HIRES, however, this approach produced well-known moderate ($\sim$10\%) inter-order flux discrepancies. For all except one of our HIRES quasars, therefore, we used lower--resolution spectra from Keck/ESI, VLT/X-Shooter, or Gemini/GMOS to derive a custom response function for each exposure. The remaining object, J2111$-$0156, was calibrated using a response function from a standard star. We verified that our final flux power spectra remained essentially unchanged if standard star flux calibration was used for every object. We therefore do not expect this aspect of the reduction to significantly impact our results.

The HIRES objects were observed using a 0$\farcs$86 slit, giving a nominal resolution FWHM of $\sim$6 km s$^{-1}$. The UVES spectra were taken with a 1$\farcs$0 slit, giving a nominal resolution of $\sim$7 km s$^{-1}$. The telluric models for the UVES data, however, indicated somewhat higher resolution consistent with a typical seeing of 0$\farcs$8. Consequently, we adopt a resolution of 6 km s$^{-1}$ for the full data set, which is sufficient to resolve small--scale features in the Ly$\alpha$ forest. We therefore expect that even the smallest scale of the flux power spectrum measured in this work (log($k$/km$^{-1}$s$)=-0.7$, or $\Delta v \sim 30$ km s$^{-1}$) will not be strongly affected by the finite spectroscopic resolution (but see Section \ref{sec:resolutionCorr}). For all the quasars, the echelle orders were redispersed onto a common wavelength scale with a dispersion of 2.5 km s$^{-1}$ per pixel.

According to the analysis presented in N16 using mock observations with a redshift path $\Delta z=4$, a continuum--to--noise ratio (C/N) of $\sim$15 per 3 km s$^{-1}$ pixel is necessary to break the degeneracy between thermal broadening and pressure smoothing and measure the thermal parameters with a statistical uncertainty of $\sim$20$\%$. Conservatively, we have chosen our sample imposing this minimum threshold inside the Ly$\alpha $ forest region. 
 
We have fitted the continuum in our spectra using spline fits guided by power-law extrapolations of the continuum redwards of the Ly$\alpha$ emission line. Given the high levels of absorption at $z\gtrsim4$ the continuum measurements are necessarily characterized by large uncertainties ($\sim$10--20$\%$). We therefore use these estimations only to derive a rough estimate of the C/N level. In measuring the power spectrum, as described in Section \ref{sec:Rolling}, we adopt an approach that does not require a priori knowledge of the continuum.  
The redshift coverage and the median C/N for the Ly$\alpha$ forest region of our sample is reported in Table \ref{table:QSO}. The majority of the spectra have larger C/N than our cut with a typical value per pixel in the range of 20-30. 
This high C/N assures that the power spectrum measurement at small scales will not be strongly affected by uncertainties in the noise modeling.

\begin{table} 
\caption{\small List of quasars used for this analysis. For each object we report the name (column 1) based on the J2000 coordinates of the object and the emission redshift (column 2). The redshift intervals associated with the Ly$\alpha$ absorption used for this analysis are also reported with the corresponding median C/N level per pixel (column 3, 4 \& 5). Finally, the instrument with which the spectrum was taken is listed in column 6. }
\centering \begin{tabular}{c c c c c c c} 

\hline
Name\ & $z_{\rm em}$\ & \multicolumn{2}{c}{$z_{\rm Ly\alpha}$} & C/N & Instrument\ \\
\multicolumn{1}{c}{} &
\multicolumn{1}{c}{} &
\multicolumn{1}{c}{$z_{\rm start}$} &
\multicolumn{1}{c}{$z_{\rm end}$} &
\multicolumn{1}{c}{} &
\multicolumn{1}{c}{} \\

\hline J2111$-$0156 & 4.89 & 3.99 & 4.79 & 20 & HIRES\\
J0011$+$1446 & 4.94 & 4.03 & 4.84 & 33 & HIRES\\
J1425$+$0827 & 4.95 & 4.04 & 4.85 & 40 & UVES \\
J1008$-$0212 & 5.04 & 4.11 & 4.94 & 22  & UVES\\
J1101$+$0531 & 5.05 & 4.12 & 4.94 & 23 & UVES\\
J0025$-$0145 & 5.07 & 4.12 & 4.95 & 26 & HIRES\\
J1204$-$0021 & 5.09 & 4.16 & 4.99 & 15  & HIRES\\
J0131$-$0321 & 5.12 & 4.23 & 5.08 & 20  & HIRES\\
J0957$+$0610 & 5.17 & 4.22 & 5.07 & 27 & UVES\\
J0741$+$2520 &5.19 &  4.24 & 5.09  & 17 & HIRES\\
J0915$+$4924 & 5.20 & 4.25 &  5.10 & 16 & HIRES\\
J0747$+$1153 & 5.26 & 4.30 &  5.16 & 18 & HIRES\\
J1659$+$2709 & 5.32 & 4.34 &  5.21 & 25 & HIRES\\
J0306$+$1853 & 5.36 & 4.37 &  5.22 & 42 & HIRES\\
J0231$-$0728&  5.42  & 4.43 &  5.31  &  31 & HIRES\\
\hline 
\end{tabular}		
\label{table:QSO}
\end{table}

\section{THE SIMULATIONS}
\label{sec:sim}
To test systematics associated with the observed power spectrum and to interpret our observational results, we used synthetic spectra derived from hydrodynamical simulations and processed to closely match the characteristics of the real data. 
We ran a large set of hydrodynamical simulations that span a range of thermal histories at $z > 4$.
The simulations run following the Sherwood simulations suite \citep{Bolton17} which uses a modified version of the parallel smoothed particle hydrodynamics code P-GADGET-3, an updated and extended version of GADGET-2 \citep{Springrl05}. The models adopt the cosmological parameters $\Omega_{m}=0.308$, $\Omega_{\Lambda}=0.692$, $h=0.678$, $\Omega_{b}=0.0482$, $\sigma_{8} =0.829$ and $n_{s}=0.961$, consistent with the cosmic microwave background constraints of \cite{Planck14}. Initial conditions were obtained using transfer functions generated by CAMB \citep{Lewis00}. Because the vast majority of the absorption systems probed by the Ly$\alpha$ forest at $z>4$ corresponds to overdensities $\Delta=\rho/\bar{\rho}\lesssim10$ our analysis will not be affected by the star formation prescription \citep{Viel13b}. Therefore, to increase the computational speed, gas particles with temperature $T<10^{5}$ K and overdensity $\Delta>10^{3}$ are converted to collisionless particles \citep{Viel04a}.

The bulk of our simulations uses a box size of 10 $h^{-1}$cMpc and $2\times512^{3}$ gas and dark matter particles, corresponding to a gas particle mass of $9.97\times10^{4}h^{-1}M_{\odot}$. In addition, we use runs with larger box size and different mass resolution to test numerical convergence (see Appendix \ref{sec:convergence}). 

We note that our simulations are not intended to self-consistently model reionization. Instead, we employ models with a wide variety of thermal histories so that our ultimate constraints on the temperature and integrated heating of the IGM are as general as possible. The gas in our models becomes optically thin at a redshift $z_{\rm OT}$, after which it is photo-ionized and heated by a uniform ultraviolet background (UVB), which is a scaled version of the background from \cite{HM12}. The thermal history of a given simulation is therefore determined by the choice of $z_{\rm OT}$ and UVB scaling factor.

The photo-heating rates from \cite{HM12} ($\epsilon^{HM12}_{i}$) for the different species ($i$=[H{\sc \,i}, He{\sc \,i}, He{\sc \,ii}]) have been rescaled proportionally by a constant factor $\zeta$ using the relation $\epsilon_{i}=\zeta\epsilon^{HM12}_{i}$ (see Table \ref{table:sim}). The combination of $z_{\rm OT}$ and $\zeta$ will determine both the instantaneous temperature and the total integrated heating per unit mass at the epoch where the power spectrum is measured. Models with larger $z_{\rm OT}$ and/or $\zeta$ will tend to have higher values of $u_{0}$.

A summary of the simulations used in this work is listed in Table \ref{table:sim}. For each model we selected the simulation outputs between $4.0\lesssim z\lesssim5.4$ with a redshift step $\Delta z=0.1$. At each redshift, synthetic spectra of Ly$\alpha$ forest were produced by choosing 5000 ``lines of sight'' throughout the simulation box. In Section \ref{sec:syntheticLOS} we describe how these lines of sight were combined to create realistic mock spectra.

Following N16, the integrated thermal history in our simulations is parametrized using $u_{0}$, the cumulative energy deposited per unit mass into the gas at the mean density. At each redshift $u_{0}$ is defined as:
\begin{equation}\label{eq:u0}
u_{0}(z)=\int_{z}^{z_{\rm OT}} \frac{\sum_{i} n_{i}\epsilon_{i}}{\bar{\rho}}\frac{dz}{H(z)(1+z)} 
\end{equation}
 where $\bar{\rho}$ is the mean mass density and $n_{i}$ and $\epsilon_{i}$ represent, respectively, the number density and the photo-heating rates for the species $i$=[H{\sc \,i}, He{\sc \,i}, He{\sc \,ii}].
 As shown in N16 (see their Figure 4), this parameter correlates with the density power spectrum of the cosmic gas in the simulations, with larger $u_{0}$ corresponding to a smoother distribution of gas for overdensities $\Delta<10$. These are the overdensities at which the Ly$\alpha$ forest is sensitive at $z>4$ \citep[e.g.,][]{Becker11}, suggesting that, at these redshifts, $u_{0}$ serves as a useful parametrization for the prior IGM thermal history. In Section \ref{sec:u0Analysis} we will further consider the redshift range of integration over which $u_{0}$ optimally correlates with the flux power spectrum.
 
\begin{figure*} 
\begin{center}
\includegraphics[width=1.5\columnwidth]{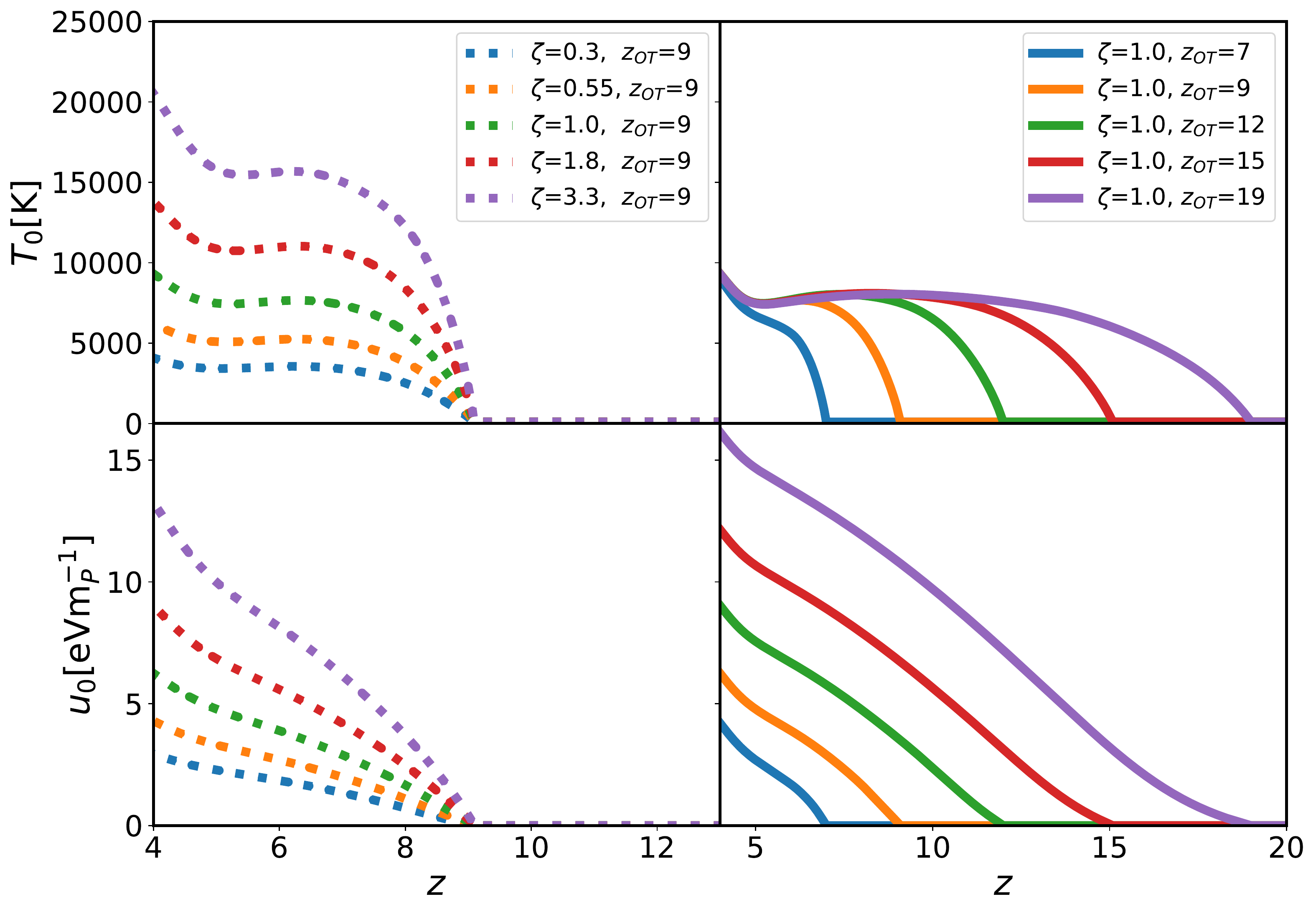} 
\caption{\small Examples of the evolution of parameters governing the thermal state of the IGM in our simulations. Left panels: evolution as a function of redshift of the temperature (top) and the cumulative energy per unit mass at the mean density (bottom) for models in which heating begins at the same $z_{\rm OT}$ but the photo-heating rates are changed. Right panels: models with the same photo-heating rates and different $z_{\rm OT}$. While increasing the photo-heating rate produces larger values in both the temperature and $u_{0}$, models with different $z_{\rm OT}$ converge to the same value of $T_{0}$ provided sufficient time has elapsed after the onset of heating. The values of these quantities at $z=5$ for all our simulations are also listed in Table \ref{table:sim}. The full suite of thermal histories is plotted in Figure \ref{fig:TotThermalH}.}
\label{fig:evolUo}
\end{center}
\end{figure*}

 Examples of the evolution of $u_{0}$ in our models are presented in Figure \ref{fig:evolUo} along with the corresponding evolution of the temperature at the mean density, $T_{0}$ (for the complete set of models, see Appendix \ref{sec:TotT} ). The left panels show how increasing the photo-heating rate in the simulations produces larger values in both the temperature and $u_{0}$. The right--hand panels show models with the same photo--heating rate but different $z_{\rm OT}$. These converge to the same value of $T_{0}$ provided sufficient time has elapsed after the onset of heating ($\Delta z\sim 1-2$; e.g., \citealt{McQuinn16}); however, they remain distinct in terms of $u_0$ values, reflecting differences in the total integrated thermal history and therefore in the amount of pressure smoothing.
\begin{table*} 
\caption{\small List of hydrodynamical simulations used in this work. Entries in bold correspond to the Sherwood simulations first introduced in \cite{Bolton17}; the model names used in that work are given in brackets. For each simulation we report the name (column 1), box size (column 2), mass resolution (column 3), the redshift at which the gas becomes optically thin (column 4), and the constant factor used to rescale the photo-heating rates for different thermal histories (column 5). The thermal parameters that describe the T--$\rho$ relation at $z=5$ are also listed: the temperature of the gas at the mean density (column 6) and the power-law index $\gamma$ (column 7). Finally, the cumulative energy per unit mass deposited into the IGM at the mean density by $z=5$ is given in column 8 (see text for details). Further details on the simulation methodology are presented in \cite{Bolton17}. }
\centering \begin{tabular}{c c c c c c c c} 
\hline
Model\ & $L[h^{-1}$cMpc]\ & $M_{gas}[h^{-1}M_{\odot}]$\ & $z_{\rm OT}$\ & $\zeta$\ & $T^{z=5}_{0}$[K]\ & $\gamma^{z=5}$\ & $u^{z=5}_{0}$[eV $m_{\rm p}^{-1}$ ]\\ 
\hline
 S10$_{-}$0.3z7 & 10 & $9.97\times10^{4}$ & 7 & 0.3& 3162 & 1.52 & 1.3\\
 S10$_{-}$0.3z9 & 10 & $9.97\times10^{4}$ & 9 & 0.3& 3388 & 1.49 & 2.3\\
 S10$_{-}$0.3z15 & 10 & $9.97\times10^{4}$ & 15 & 0.3& 3388 & 1.51 & 5.0\\
 S10$_{-}$0.55z7 & 10 & $9.97\times10^{4}$ & 7 & 0.55& 4553 & 1.51 & 1.9\\
 S10$_{-}$0.55z9 & 10 & $9.97\times10^{4}$ & 9 & 0.55& 5086 & 1.48 & 3.3\\
 S10$_{-}$0.55z12 & 10 & $9.97\times10^{4}$ & 12 & 0.55& 5110 & 1.51 & 5.2\\
 S10$_{-}$0.55z15 & 10 & $9.97\times10^{4}$ & 15 & 0.55& 5093 & 1.52 & 7.4\\
 S10$_{-}$0.55z19 & 10 & $9.97\times10^{4}$ & 19 & 0.55& 5074 & 1.52 & 10.0\\
 S10$_{-}$1z7 & 10 & $9.97\times10^{4}$ & 7 & 1.0& 6607 & 1.50 & 2.7\\
 S10$_{-}$1z9 & 10 & $9.97\times10^{4}$ & 9 & 1.0& 7413 & 1.51 & 4.7\\
 S10$_{-}$1z12 & 10 & $9.97\times10^{4}$ & 12 & 1.0& 7510 & 1.51 & 7.6\\
 \textbf{S10$_{-}$1z15 (10--512)} & \textbf{10} & \textbf{9.97$\times$10$^{\textbf{4}}$} & \textbf{15} & \textbf{1.0}& \textbf{7413} & \textbf{1.50} & \textbf{10.6}\\
 S10$_{-}$1z19 & 10 & $9.97\times10^{4}$ & 19 & 1.0& 7457 & 1.52 & 14.7\\
 S10$_{-}$1.8z7 & 10 & $9.97\times10^{4}$ & 7 & 1.8& 9725 & 1.49 & 3.9\\
 S10$_{-}$1.8z9 & 10 & $9.97\times10^{4}$ & 9 & 1.8& 10866 & 1.50 & 6.8\\
 S10$_{-}$1.8z12 & 10 & $9.97\times10^{4}$ & 12 & 1.8& 10900 & 1.51 & 10.9\\
 S10$_{-}$1.8z15 & 10 & $9.97\times10^{4}$ & 15 & 1.8& 10865 & 1.51 & 15.5\\
 S10$_{-}$1.8z19 & 10 & $9.97\times10^{4}$ & 19 & 1.8& 10827& 1.52 & 21.4\\
 S10$_{-}$3.3z7 & 10 & $9.97\times10^{4}$ & 7 & 3.3& 13803 & 1.48 & 5.5\\
 S10$_{-}$3.3z9 & 10 & $9.97\times10^{4}$ & 9 & 3.3& 15488 & 1.50 & 9.9\\
 S10$_{-}$3.3z12 & 10 & $9.97\times10^{4}$ & 12 & 3.3& 15821 & 1.48 & 16.3\\
 S10$_{-}$3.3z15 & 10 & $9.97\times10^{4}$ & 15 & 3.3& 15488 & 1.52 & 23.2\\
\hline 
S10$_{-}$1z9$_{-}$g1 & 10 & $9.97\times10^{4}$ & 9 & 1.0& 7413 & 1.00 & 4.7\\
\textbf{S20$_{-}$1z15 (20--1024)} & \textbf{20} & \textbf{9.97$\times$10$^{\textbf{4}}$} & \textbf{15} & \textbf{1.0}& \textbf{7413}& \textbf{1.50} & \textbf{10.6}\\
\textbf{S40$_{-}$1z15 (40--2048)} & \textbf{40} & \textbf{9.97$\times$10$^{\textbf{4}}$} & \textbf{15} & \textbf{1.0}& \textbf{7413} & \textbf{1.50} & \textbf{10.6}\\
\textbf{S40$_{-}$1z9 (40--2048--zr9)}  & \textbf{40} & \textbf{9.97$\times$10$^{\textbf{4}}$} & \textbf{9} & \textbf{1.0}& \textbf{7413} & \textbf{1.51} & \textbf{4.7}\\
S10$_{-}$1z15$_{-}$256 &10&$7.97\times10^{5}$ &15 &1.0 & 7413 & 1.50 & 10.6\\
S10$_{-}$1z15$_{-}$768 & 10 & $2.95\times10^{4}$&15 & 1.0 & 7413 & 1.50 & 10.6 \\
\hline 
\end{tabular}		
\label{table:sim}
\end{table*}

\section{The Lyman-$\alpha$ flux power spectrum }
\begin{figure*} 
\begin{center}
\includegraphics[width=1.5\columnwidth]{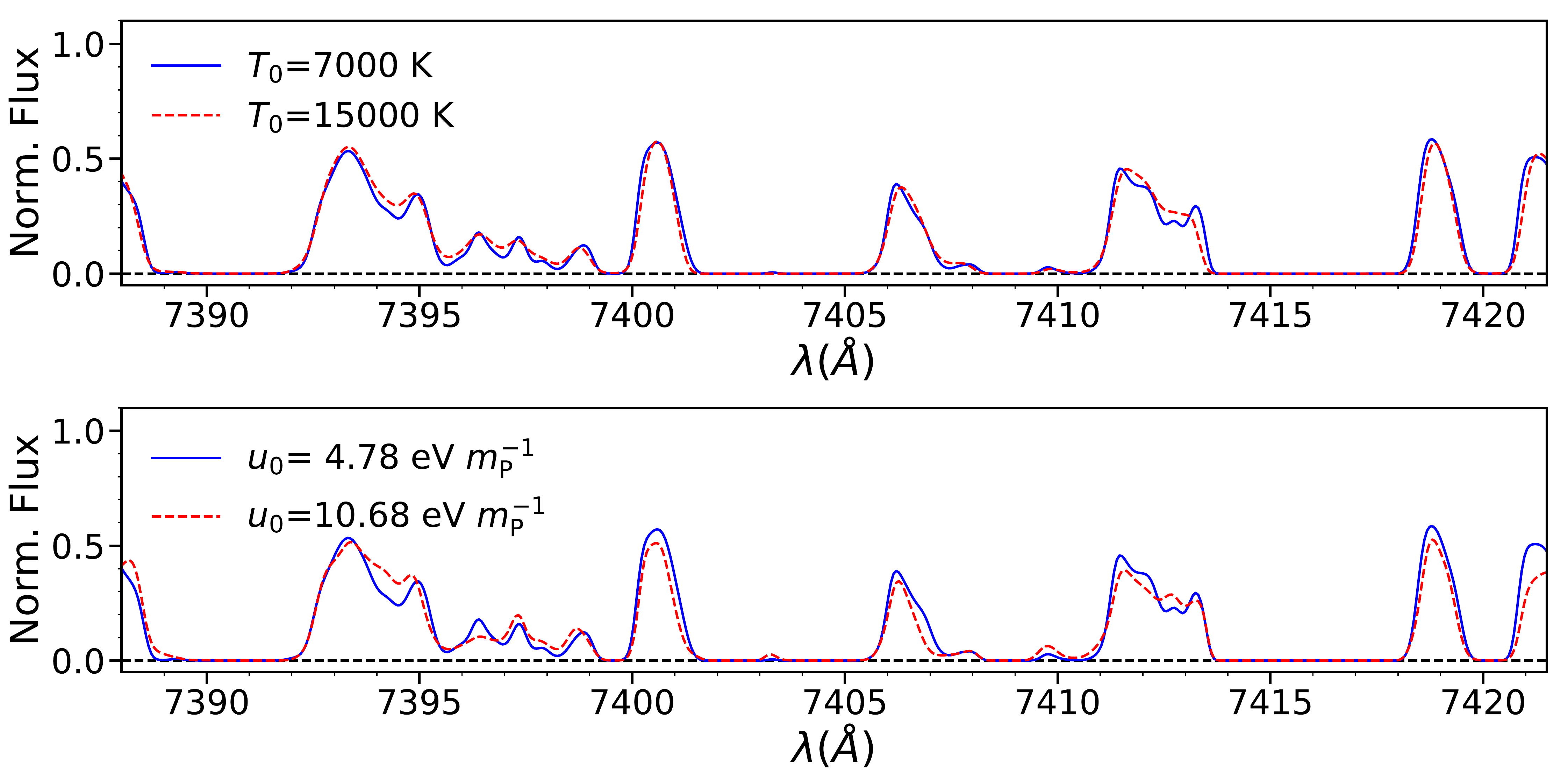} 
\caption{\small The effect of thermal broadening and integrated heating on simulated Ly$\alpha$ forest spectra at $z=5$. Top panel: the effect of thermal broadening on the absorption features in models characterized by the same integrated heating ($u_{0}=4.78$ eV $m_{\rm p}^{-1}$) but post--processed to different instantaneous temperatures. Bottom panel: the effect of pressure smoothing on the Ly$\alpha$ absorption for models with the same temperature ($T_{0}=7000$ K) but different thermal histories.  }
\label{fig:LOS}
\end{center}
\end{figure*}

Both thermal broadening and pressure smoothing tend to reduce the amount of small--scale structure in the forest. Figure \ref{fig:LOS} shows the effect of thermal broadening (top panel) and pressure smoothing (bottom panel) on simulated Ly$\alpha$ forest spectra at $z=5$. While the impact of $T_{0}$ and $u_{0}$ are visually similar, the scale dependences of these effects makes it possible to break the degeneracy (e.g., N16, \citealt{Onorbe17}).

\begin{figure*} 
\centering 
\subfigure
{ 
\includegraphics[width=3.3in]{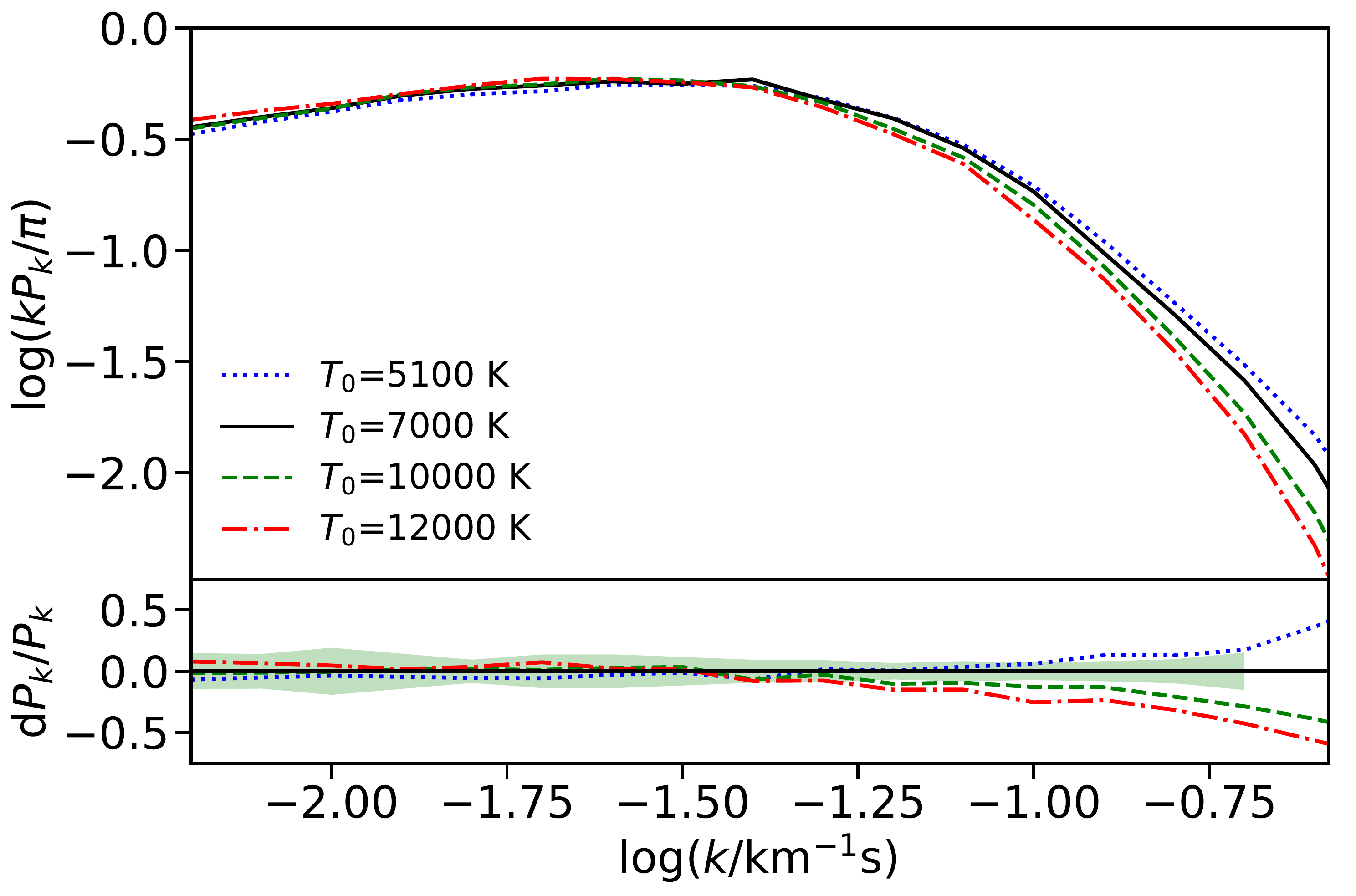}
 \label{fig:T0}
}
\subfigure
{ 
\includegraphics[width=3.3in]{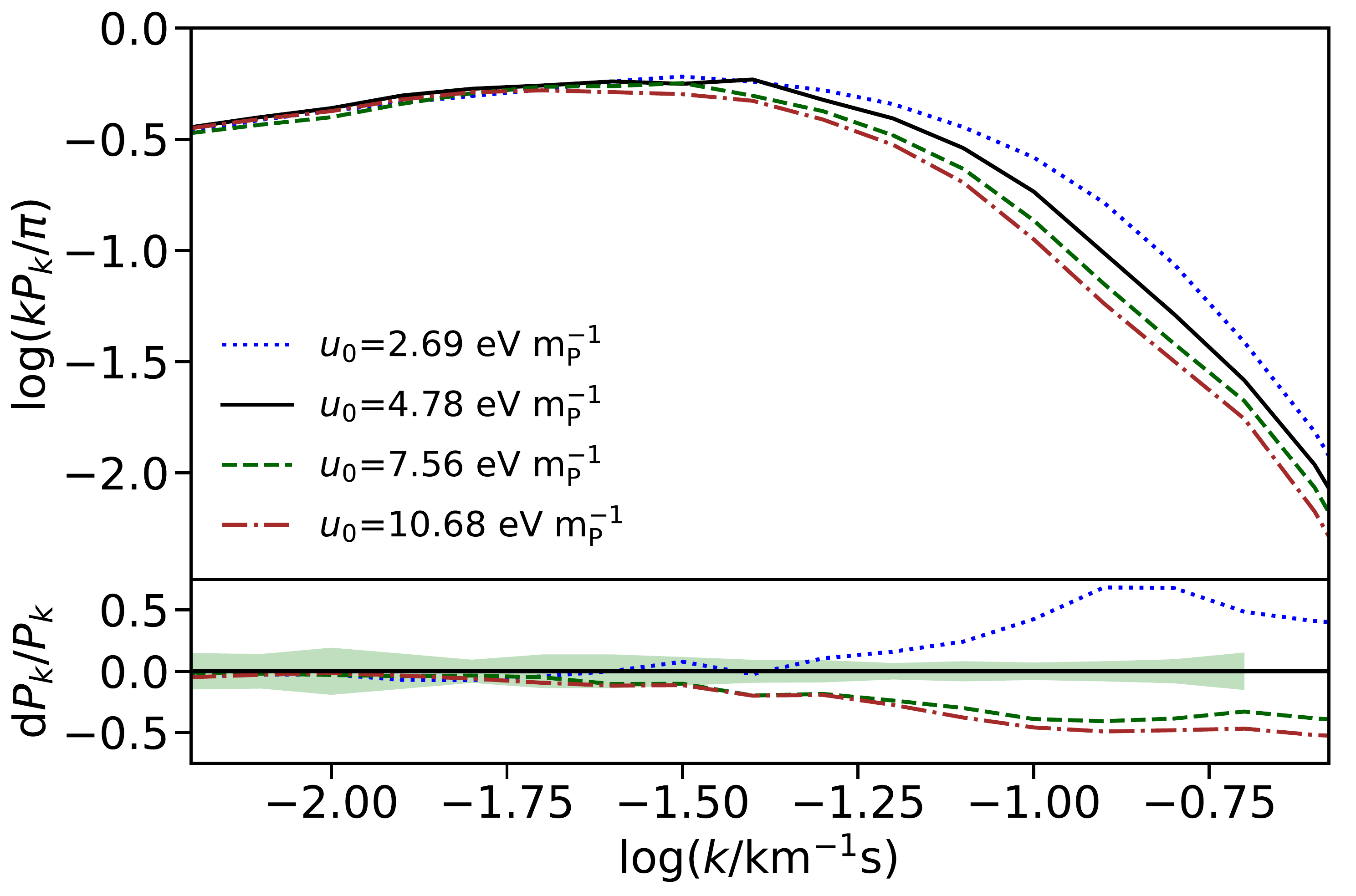} 
\label{fig:u0}
} 
\subfigure
{ 
\includegraphics[width=3.3in]{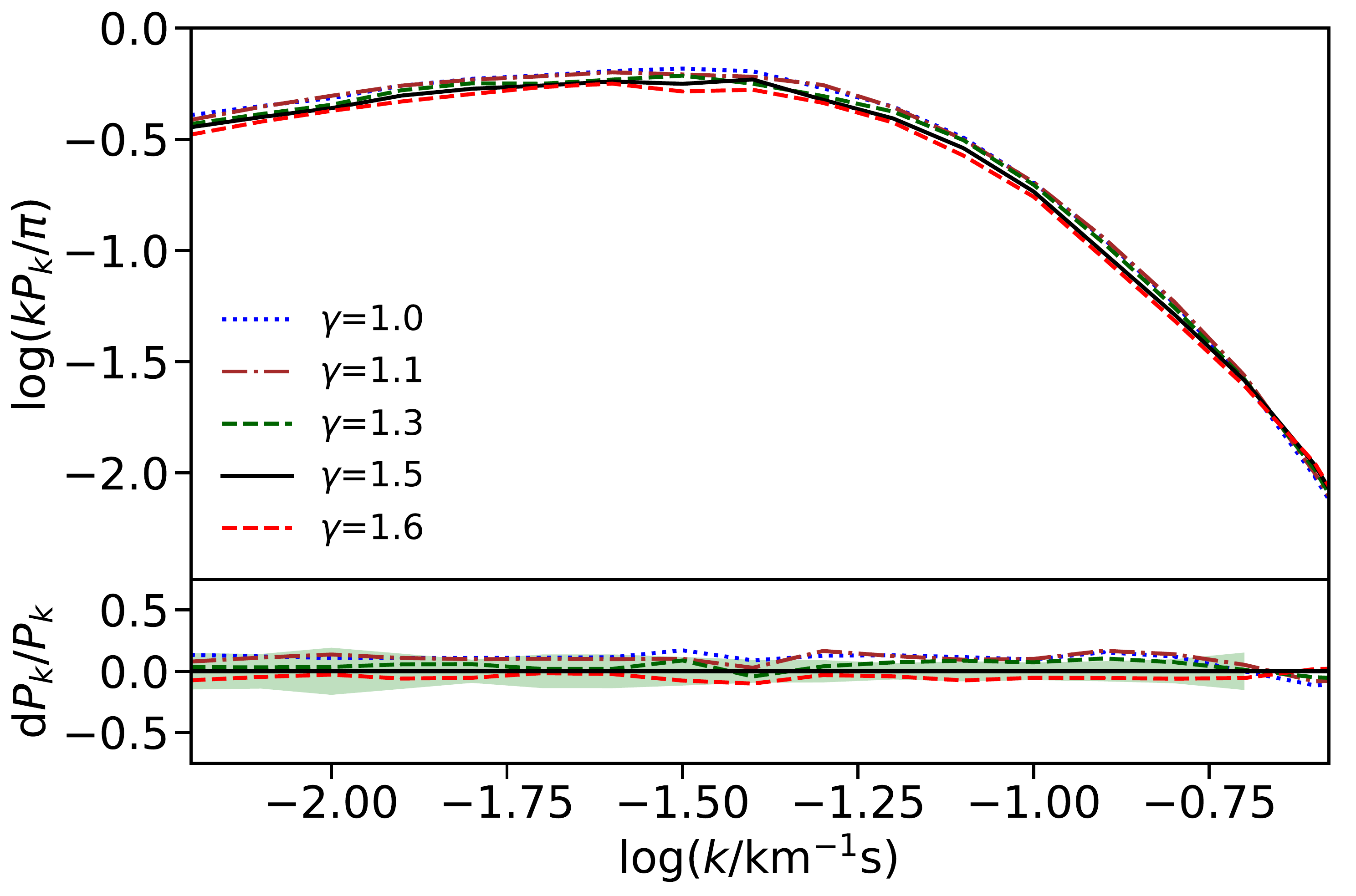}
 \label{fig:gamma}
}
\subfigure
{ 
\includegraphics[width=3.3in]{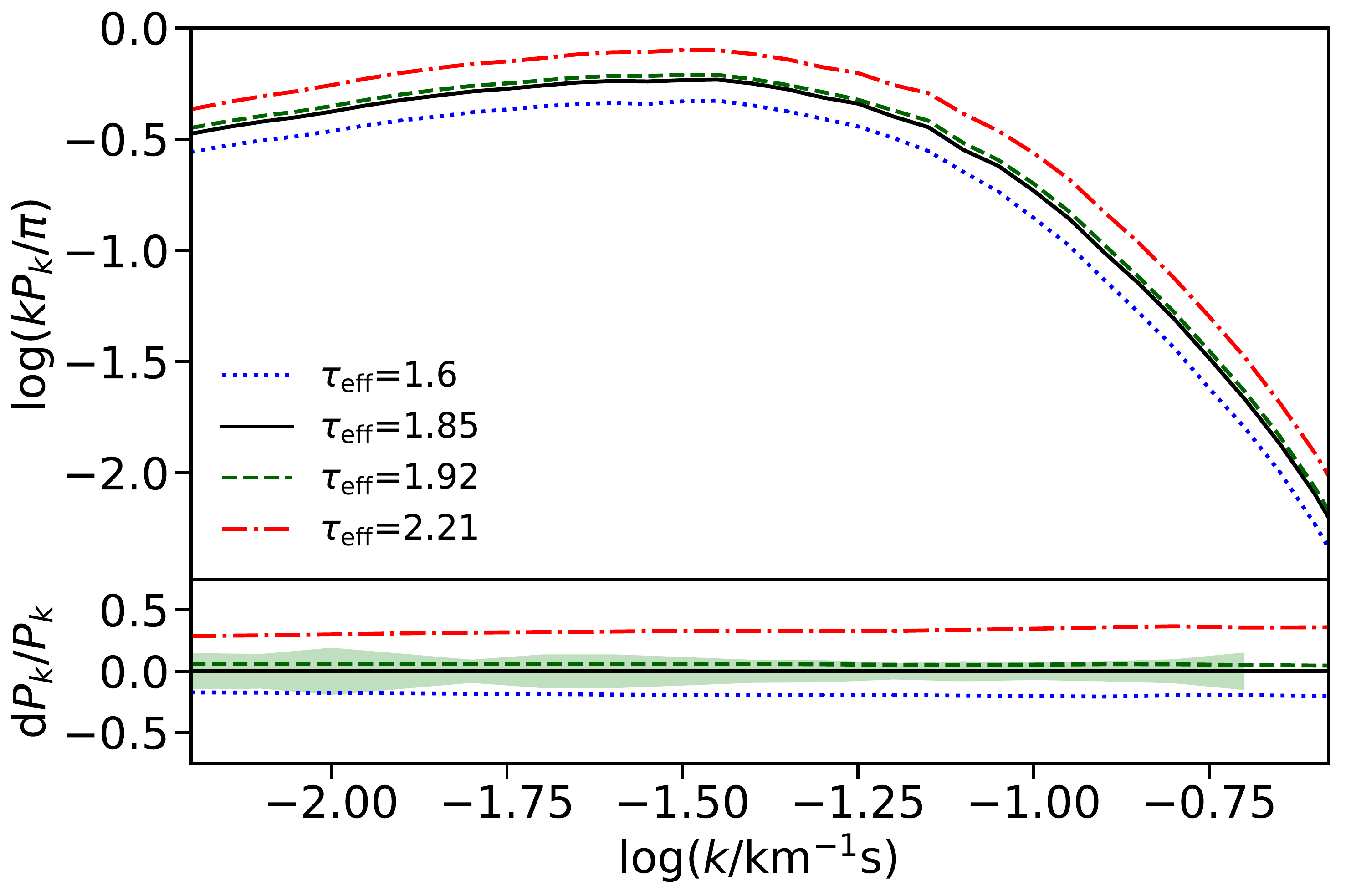} 
\label{fig:tau0}
} 
 \caption{\small The effects of varying our model parameters on the 1D flux power spectrum at $z=5.0$. In all panels we plot a fiducial model with $T_{0}=7000$ K, $\gamma=1.5$, $u_{0}=4.78$ eV $m_{\rm p}^{-1}$ and $\tau_{\rm eff}=1.85$ using a black solid line. The four parameters are varied separately as indicated in each panel.
Residuals d$P_{k}/P_{k}$ relative to the fiducial value are displayed for each scale. For comparison, the $68\%$ errors relative to the observational power spectrum computed in this work at $z=5$ are also shown (shaded green region). Models with higher temperature show decreasing power towards smaller scales with the most prominent effect at scales $\log(k/$km$^{-1}$s$)>-1$ (top left). Changes in the integrated thermal history (top right) produce variations in the pressure smoothing experienced by the gas. This effect has a somewhat different scale dependence than pure thermal broadening.
The power spectrum at this redshift is not highly sensitive to variations in $\gamma$ (bottom left) although decreasing $\gamma$ tends to increase the power at $\log(k/$km$^{-1}$s$)\lesssim-0.8$. Differences in the effective optical depth (i.e., mean flux) create changes in the normalization of the power spectrum (bottom right).}
 \label{fig:ThermalEffect}
\end{figure*}

The top row of Figure \ref{fig:ThermalEffect} demonstrates how the shape of the 1D Ly$\alpha$ flux power spectrum at $z=5$ varies for models with different instantaneous temperature (left panel) and integrated thermal histories (right panel). Similar results were shown in N16, but are expanded here to include a broader range of thermal histories. As described in Section \ref{sec:T0var}, we use post--processing to vary the $T_{0}$ for a fixed $u_{0}$ (top left) or to impose the same $T_{0}$ for models with different $u_{0}$ (top right). We also demonstrate the impact of varying $\gamma$ (bottom left) and the effective optical depth, $\tau_{\rm eff}$ (bottom right). As noted by N16, the scale dependence of $T_{0}$ and $u_{0}$ differ somewhat.  While the impact of pure thermal broadening increases continuously towards smaller scales, the effect of changing $u_{0}$ peaks near $\log(k/$km$^{-1}$s$)\sim -0.9$ to $-0.8$.

Comparing the two panels of the first row of Figure \ref{fig:ThermalEffect}, it is clear that in order to distinguish models characterized by an early reionization (large $u_{0}$) from those with high $T_{0}$ values, it is necessary to probe the power spectrum down to $\log(k/$km$^{-1}$s$)\sim-0.7$.
Our effort to measure the power spectrum down to these scales is described in the following section.

\section{Data analysis }
\label{sec:DataAnalysis}
In this Section we describe our procedure for measuring the flux power spectrum from the observed spectra. The following strategies have been tested using synthetic spectra for two reasons: first, to detect and quantify systematic effects in the calculation of the power spectrum, and second, to guarantee a fair comparison between simulated models and the observed data. 

\subsection{Rolling mean}
\label{sec:Rolling}  
\begin{figure} 
\centering
\subfigure
{
\includegraphics[width=1.0\columnwidth]{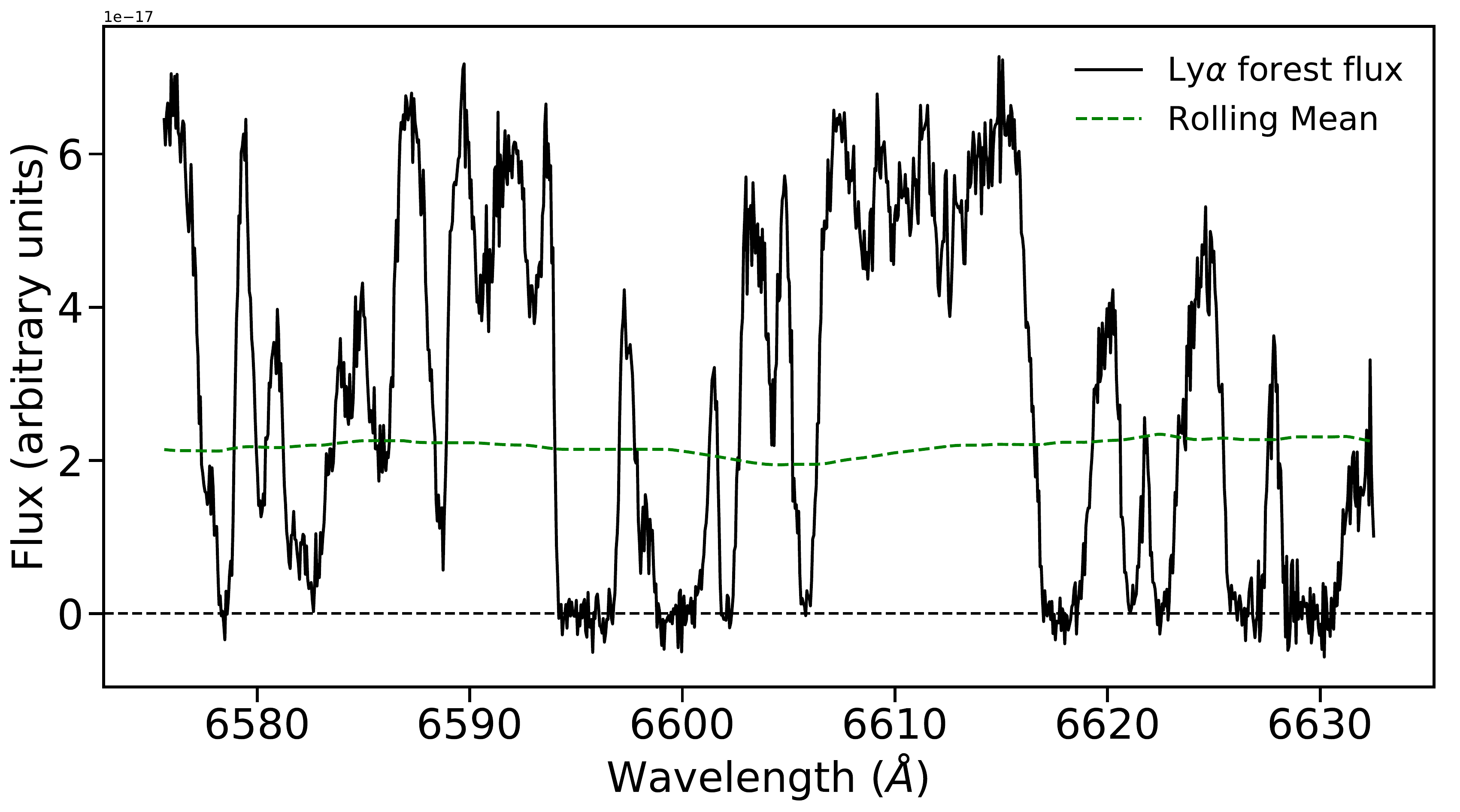} 
}
\subfigure
{ 
\includegraphics[width=1.0\columnwidth]{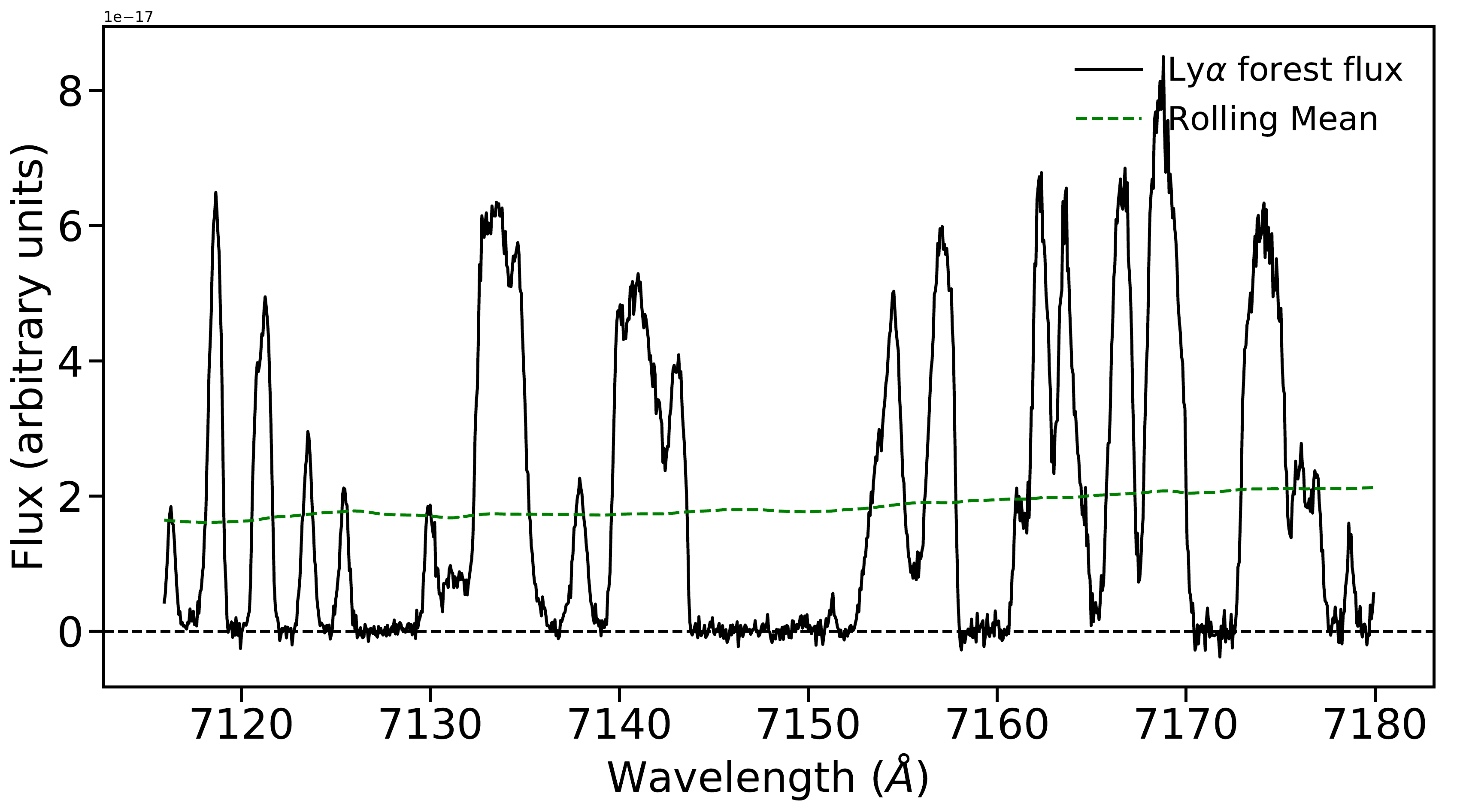}
}

\caption{\small Examples of observed 20 $h^{-1}$cMpc Ly$\alpha$ forest sections extracted from our sample. Top panel: Ly$\alpha$ forest at $z\simeq 4.4$ extracted from the spectrum of J0741+2520 with a C/N per pixel $\sim17$. Bottom panel: Ly$\alpha$ forest at $z\simeq 4.9$ extracted from the spectrum of J0306+1853 with a C/N per pixel of $\sim40$. The corresponding values of the of the boxcar rolling mean, measured within a 40 $h^{-1}$cMpc window, are also shown (green dashed line).  }
\label{fig:LySec}
\end{figure}

We performed the power spectrum measurement on the flux contrast estimator
  \begin{equation}\label{eq:df}
\delta_{F}= \frac{F-\bar{F}}{\bar{F}} ,
\end{equation}
where $F$ is the transmission in the Ly$\alpha$ forest and $\bar{F}$ is the mean flux.
When computing $\delta_{F}$ we need to first divide out the intrinsic shape of the quasar spectrum, which can impact the power spectrum at large scales ($\log(k/$km$^{-1}$s$) \lesssim-2$; e.g., \citealt{Kim04,Viel13,Irsic17}). However, directly estimating the continuum is difficult at $z\gtrsim 4$ due to the high levels of absorption in the Ly$\alpha$ forest.
We therefore used a rolling mean approach, wherein $\bar{F}$ is estimated locally by smoothing the observed spectrum using a boxcar average. 
We used a boxcar window of 40 $h^{-1}$cMpc, which was chosen to be short enough to roughly capture relevant features in the quasar continua over the forest (see Appendix \ref{sec:RollSys} for details). Examples of this approach are presented in Figure \ref{fig:LySec}.
\vspace{5mm}
\subsection{Proximity regions}
Regions near to a quasar are subjected to the local influence of its UV radiation field and are therefore expected to show lower Ly$\alpha$ absorption with respect to the cosmic mean. To avoid any proximity effect bias we excluded these regions from the analysis. The UV flux of a bright quasar, is thought to affect regions $\lesssim$10 proper Mpc along its line of sight \citep[e.g.,][]{Scott00, Worseck06}. 
We conservatively masked 30 proper Mpc bluewards of the quasar redshift Ly$\alpha$ emission line. Moreover, to exclude possible blueshifted Ly$\beta$ absorption we also masked a velocity interval corresponding to 10 proper Mpc redwards of the Ly$\beta$ emission line. 
Excluding the proximity regions moderately changes the power (by $\gtrsim$5$\%$) only for the highest redshift bin at $z=5$, although the correction is always well within the statistical error.  
 \subsection{DLAs}
 \label{sec:Masking1}
 We excluded damped Lyman-$\alpha$ (DLA) systems from our spectra. DLAs were identified visually and masked prior to computing the power spectrum. This step changes the power up to $\sim$5--$10\%$ which is within the statistical uncertainties at all scales.
 \subsection{Bad pixels}
 \label{sec:Masking2}
 We masked bad pixels characterized by negative or zero flux errors. We also masked discrete regions affected by sky emission line residuals, which tend to be noisy. These features mainly impact smaller scales than the ones we want to compute ($\log(k/$km$^{-1}$s$)\gtrsim-0.5$), but they may affect the evaluation of the correct noise power (see Section \ref{sec:Noise}) and therefore need to be removed. 
\subsection{Ly$\alpha$ sections and redshift sub-samples}
\label{sec:sections}
We compute the flux power spectrum on sections of 20 $h^{-1}$cMpc (comoving distance). This scale was chosen to be small enough that we would have enough sub--samples ($N>30$) to evaluate the statistical uncertainty in the flux power via bootstrapping, yet large enough to avoid significant windowing affects (see Appendix \ref{sec:Windowing}). Each of the Ly$\alpha$ sections have been examined by eye to avoid sections containing too many masked pixels. 
The power spectrum results from the useful sections are then collected and averaged in redshift bins of $\Delta z=0.4$ centered at $z=4.2$, 4.6 and 5.0.
\subsection{Measuring the power spectrum}
For each of the 20 $h^{-1}$cMpc forest regions we calculate the power spectrum from the flux contrast $\delta_{F}$ defined in Eq. \ref{eq:df}. Our spectra are unevenly sampled because they are masked so we use a Lomb--Scargle algorithm \citep{Lomb76, Scargle82} to compute the raw power of each region ($P_{\rm masked}(k)$). In all of our calculations we use $k$-bins logarithmically spaced with $\Delta$$\log$$k=0.1$.
To obtain the final power spectrum values, $P_{\rm F}(k)$, for each section we first correct the raw $P_{\rm masked}(k)$ for the effect of masking. Secondly, we subtract from the corrected $P_{\rm data}(k)$ an estimate of the contribution to the power from noise, $P_{\rm N}(k)$. All these steps are described in the following sections.
\subsubsection{Masking correction function}
\label{sec:MF}
The masking procedure described in Sections \ref{sec:Masking1} and \ref{sec:Masking2}, and in particular the masking of sky line residuals impacts the power spectrum due to the application of a complex window function. In order to correct for this we apply a masking correction function, $C_{m}(k)$, to the raw power obtained from each of the 20 $h^{-1}$cMpc Ly$\alpha$ forest sections,
\begin{equation}\label{eq:maskingCorr }
P_{\rm data}(k)=P_{\rm masked}(k)\times C_{m}(k) ,
\end{equation} 
where $P_{\rm data}(k)$ is the corrected quantity used to infer the final power and $P_{\rm masked}(k)$ is the raw power initially computed from masked spectra.

We determine the effect of masking for each of the Ly$\alpha$ forest sections contributing to the analysis using the following procedure. First, we create hundreds of synthetic spectra with the same characteristics (i.e. size, noise, redshift) of each of the real 20 $h^{-1}$cMpc sections with and without the same masking applied. The final correction is then obtained from the average of the ratio between the power of the unmasked ($P_{sim}$) and masked ($P^{mask}_{sim}$) simulated spectra,
\begin{equation}
C_{m}(k)=\Bigg<\frac{P_{sim}(k)}{P^{mask}_{sim}(k)}\Bigg> .
\end{equation}
Because the impact of masking on the power spectrum in principle depends on its underlying shape, possible systematics may arise from choosing a particular simulation run to compute $C_{m}(k)$. We use the $20 h^{-1}$cMpc run of Table \ref{table:sim} for the final correction, therefore, we quantify these possible uncertainties in Appendix \ref{sec:mfunction}.
\subsubsection{Noise subtraction}
\label{sec:Noise}

\begin{figure} 
\centering
\includegraphics[width=0.9\columnwidth]{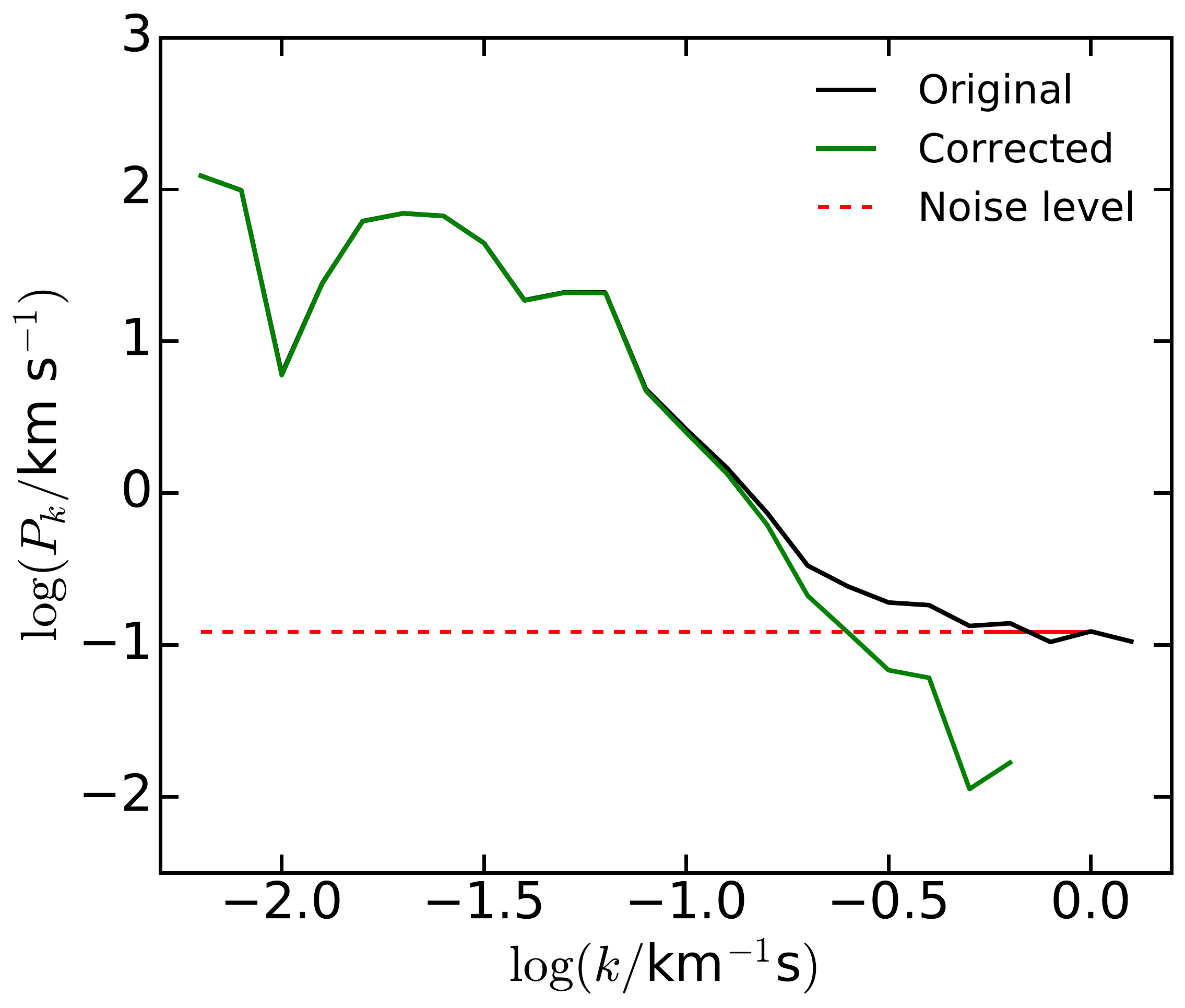} 
\caption{\small A demonstration of the subtraction of the noise power from the raw power spectrum of an observed 20 $h^{-1}$cMpc section of Ly$\alpha$ forest. The original power computed directly from the spectra (black solid line) shows a flattening towards the smallest scale becoming roughly constant for $\log(k/$km$^{-1}$s$)\gtrsim-0.2$ when the noise starts to dominate. Assuming white noise, we fit the noise power with a constant value at the smallest scale (red dashed line) and then subtract it from the total power spectrum obtaining the corrected, ``noiseless" version (green solid line). }
\label{fig:LyNoise}
\end{figure}

In principle the noise power can be directly computed from the flux error array output by the data reduction pipeline \citep[e.g.,][]{Irsic17,Walther18}. This approach, however, relies on the precision of the pipeline; underestimating or overestimating these uncertainties could significantly impact on the final power, especially at the small scales we are interested in. 
We therefore estimate the amount of noise for each forest section directly from the raw power spectrum of the data. At the smallest scales ($\log(k/$km$^{-1}$s$)\gtrsim-0.2$ ) the power is dominated by noise fluctuations and, assuming that the noise in adjacent wavelength bins is uncorrelated, can be fitted with a constant value. We then assume that $P_{\rm N}$ is constant over all scales and subtract it from the total power obtaining the ``noiseless'' power spectra. This method is illustrated in Figure \ref{fig:LyNoise} and has been tested on synthetic data after adding the observational noise arrays to the simulated lines of sight (see Appendix \ref{sec:NoiseTreat}).
\subsubsection{Resolution correction}
\label{sec:resolutionCorr}
In this work we forward model the synthetic spectra generated from simulations to match the instrumental resolution and pixel size of the data. For reference, however, we include a version of the observed flux power spectrum that has been corrected for resolution as
\begin{equation}\label{eq:PSf}
P_{\rm F}(k)= \frac{P_{\rm data}(k)-P_{\rm N}(k)}{W^{2}_{R}(k,R,dv_{p})} ,
\end{equation}
using the window function adopted in \cite{Delabouille15b},
\begin{equation}\label{eq:Wr}
W_{R}(k,R,dv_{p})=\exp\Big(-\frac{1}{2}(kR)^{2}\Big) \frac{\sin(k dv_{p}/2)}{(kdv_{p}/2)}.
\end{equation}
Assuming our nominal resolution $R=2.55$ km s$^{-1}$ ($\rm FWHM=6$ km s$^{-1}$) and pixel size $dv_{p}=2.5$ km s$^{-1}$, the correction for the smallest scale considered in this work ($\log(k/$km$^{-1}$s$)=-0.7$ ) is $W^{2}_{R}\sim0.76$. 

We note that the actual spectral resolution of the data will depend on the seeing of the observation and may be different from the nominal one. A possible error in the power spectrum due to uncertainties in the spectral resolution, even when forward modeling the simulations, must therefore be taken into account. We estimate an error of $10\%$ in the spectral resolution, corresponding to an uncertainty in the power of $\lesssim$5$\%$ at $\log(k/$km$^{-1}$s$)\lesssim -0.7$. This correction is smaller than our statistical error, so we do not expect that uncertainties in the resolution will significantly affect the measurements (see Appendix \ref{sec:ResCorr}). 

\subsection{Metals}
\label{sec:Metals}
 The flux power spectrum measured directly from the observational spectra contains both the power coming from the Ly$\alpha$ forest and a small contribution from intervening metal lines. These lines tend to show individual components significantly narrower than Ly$\alpha$ ($b\lesssim15$ km s$^{-1}$), which will increase the power on small scales \citep[e.g.,][]{Lidz10}. This Section describes our approach to quantifying and removing the effect of metals on our final power spectrum measurements.  

 The high level of Ly$\alpha$ absorption at high redshift makes it very challenging to directly identify all metal lines in the forest. We therefore estimate the metal power spectrum directly from regions of quasar spectra redwards of the Ly$\alpha$ emission line, where only metal absorption systems are present \citep[e.g.,][]{McDonald05,Delabrouille13}. The metal power measured in this way will not take into account transitions with rest--frame wavelength shorter than the Ly$\alpha$ line. Correlation features like the one observed for Si{\sc \,iii} ($\lambda$ 1206) in \cite{McDonald06}, however, will tend to affect the power spectrum on scales larger than the ones considered in this work ($\log(k/$km$^{-1}$s$)\lesssim -2.5$).
 
 We measured the metal power spectrum from two samples of high--resolution quasar spectra. First, we use a sub-set of the spectra listed in Table \ref{table:QSO} with emission redshift $4.5\lesssim z_{\rm em}\lesssim{5.3}$. Second, we use a sample of spectra of quasars with emissions redshifts $3.4 \lesssim z_{\rm em} \lesssim4.1$ from \cite{Boera14,Boera16} (Table \ref{table:QSOmetals}). The latter sample allows us to measure the metal power spectrum over observed wavelengths similar to those spanned by the Ly$\alpha$ forest at $4.0\lesssim z\lesssim4.4$. While metals redwards of Ly$\alpha$ are not a perfect estimate of those that appear in the forest at higher redshifts, analyzing multiple samples allows us to check for redshift evolution in the metal power spectrum.

\begin{figure} 
\begin{center}
\includegraphics[width=1.0\columnwidth]{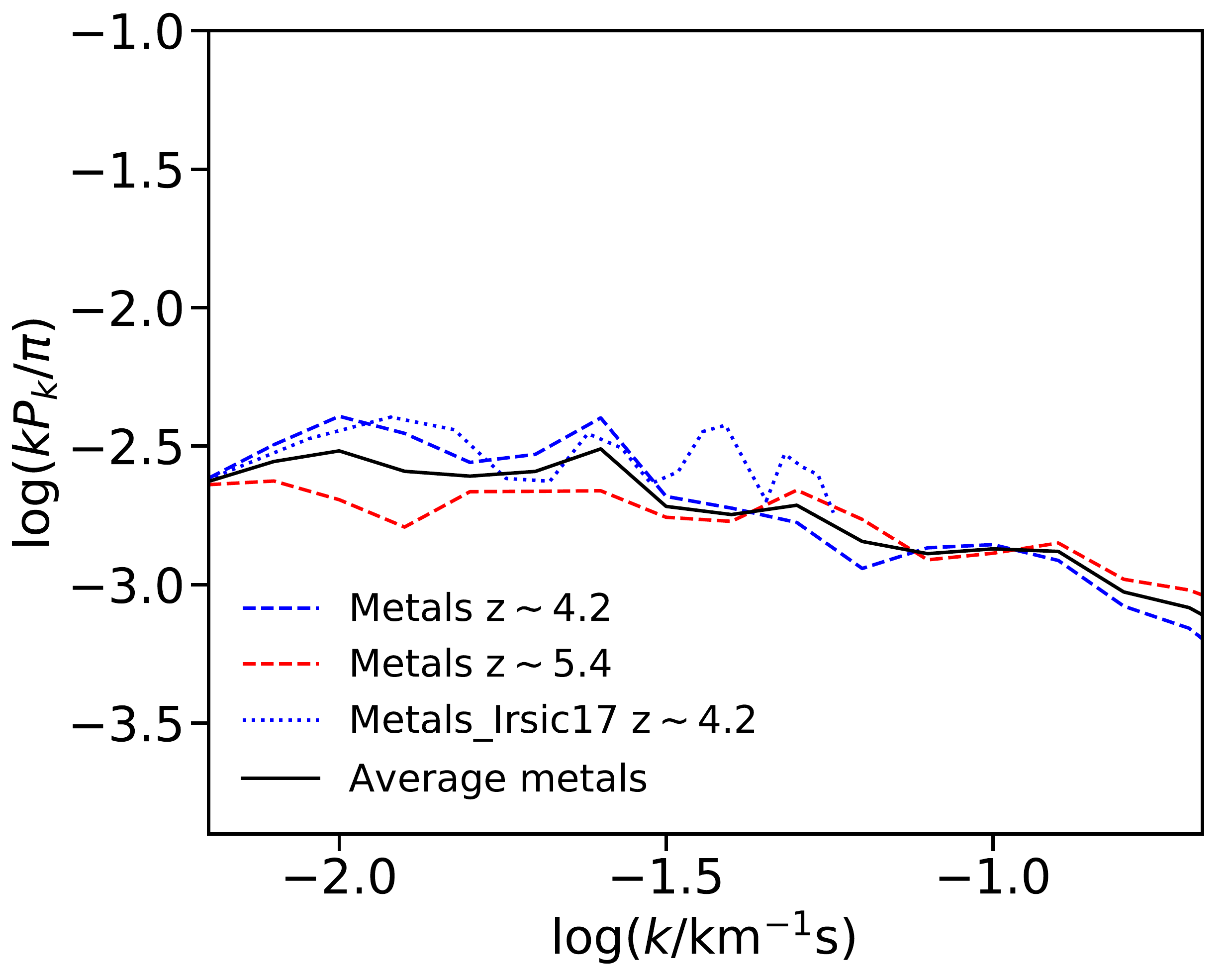} 
\caption{\small The metal power spectra measured redwards of the Ly$\alpha$ emission line. The metal contribution at $z\sim4.2$ (blue dashed line) computed from the 
quasars with emissions redshifts $3.4\lesssim z_{\rm em} \lesssim4.1$ of Table \ref{table:QSOmetals} is compared with the metal power measured at $z\sim5.4$ (red dashed line) from the quasars subsample with $4.5\lesssim z_{\rm em}\lesssim{5.3}$ of Table \ref{table:QSO}. The average is given by the black solid line. For comparison, the metal power spectrum computed from the XQ-100 data sample at $z\sim4.2$ \citep{Irsic17} is also shown (blue dotted line).}
\label{fig:Metals}
\end{center}
\end{figure}

Figure \ref{fig:Metals} shows the comparison between our measurements of the metal power spectrum at $z\sim4.2$ (blue dashed line) and at $z\sim5.4$ (red dashed line), obtained using the same data analysis procedure described in the previous sections. The most significant difference in the metal power between these two redshifts is at large scales ($\log(k/$km$^{-1}$s$)\lesssim-1.5$) where the contribution of metals to the final flux power spectrum measurement is less relevant. 
Note that our measurement of the metal power spectrum at $z\sim4.2$ is also in good agreement with the one computed from the XQ-100 data sample at the same redshift \citep{Irsic17} (blue dotted line) even if the latter is slightly nosier.

Given the weak evolution in redshift of the metals power (already noted in previous works, e.g., \citealt{Delabrouille13}) we corrected our final power spectrum measurements assuming a metal contribution constant with redshift and equal to the average between our two measured metal power spectra (black solid line in Figure \ref{fig:Metals}). 
The effect of the metal correction on the final flux power spectrum is shown in Appendix \ref{sec:met}.

\begin{table} 
\caption{\small List of quasars used for the analysis of the metal power spectrum at $z\sim4.2$. For each object we report the name (column 1) based on the J2000 coordinates of the quasar and the emission redshift (column 2). All the spectra have been taken with the UVES spectrograph (see \citealt{Boera14} and \citealt{Murphy18} for details). }
\centering \begin{tabular}{c c } 
\hline
Name\ & $z_{\rm em}$\  \\
\hline J010604$-$254651& 3.36500 \\
J162116$-$004250&   3.70270 \\        
J132029$-$052335&   3.70000 \\        
J124957$-$015928 &  3.63680\\                   
J014049$-$083942 &  3.71290 \\         
J115538$+$053050  & 3.47520 \\           
J014214$+$002324  & 3.37140 \\              
J123055$-$113909 &  3.52800 \\             
J110855$+$120953 &  3.67160 \\          
J005758$-$264314 &  3.65500 \\ 
\hline 
\end{tabular}		
\label{table:QSOmetals}
\end{table}

\subsection{The new power spectrum measurements}
\label{sec:PSResults}
\begin{figure*} 
\begin{center}
\includegraphics[width=1.6\columnwidth]{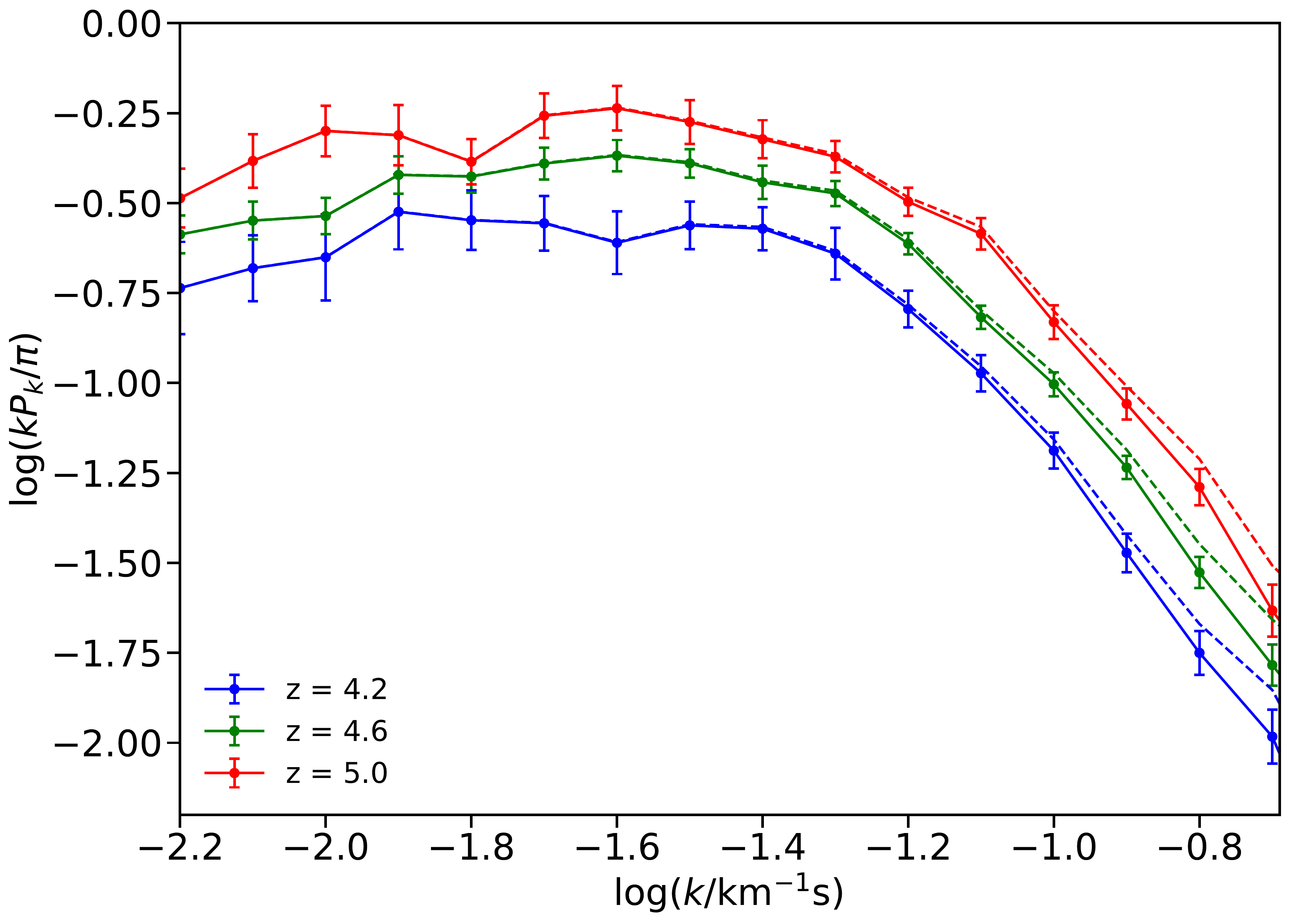} 
\caption{\small Our measurements of the Ly$\alpha$ flux power spectrum for the $\Delta z =0.4$ bins centered at $z=4.2$ (blue), 4.6 (green) and 5.0 (red). Higher effective optical depths determine an increase in the power towards higher redshifts. Values are obtained following the steps presented in Section \ref{sec:DataAnalysis} with (dashed lines) and without (solid line and data points) instrumental resolution correction. Vertical error bars are 1$\sigma$ errors taken from the corrected and regularized covariance matrix (Section \ref{sec:CovMatrix}). Note that the larger number of spectra contributing to the $z=4.6$ bin (roughly double the number of spectra of the other two bins) are reflected in the significantly smaller errors.}
\label{fig:ObsPower}
\end{center}
\end{figure*}

The main observational results of this work are presented in Figure \ref{fig:ObsPower}, where we plot the final Ly$\alpha$ flux power spectrum measured from our data. The values are tabulated in Appendix \ref{sec:Values} and reported as a function of scale for the three redshift bins centered at $z=4.2,4.6$ and 5.0.
Solid colored lines and data points represents the power spectrum results without the resolution correction described in \ref{sec:resolutionCorr}, while the corresponding dashed lines are the measurements corrected for finite resolution and pixel size. 
The 1$\sigma$ errors are estimated from the bootstrap covariance matrix of the data, corrected and regularized following the procedure described in Section \ref{sec:CovMatrix}.

Our measurements at scales $\log(k/$km$^{-1}$s$)\gtrsim-1$ are the first ones made at these redshifts (see \citealt{Walther18} for an analysis at $z<4$). At larger scales, however, we can compare with the results derived from high--resolution spectra by \cite{Viel13}, and medium--resolution data by \cite{Irsic17}. These comparisons are presented in Appendix \ref{sec:VielComp} and \ref{sec:IrsicComp}.

\subsection{Covariance matrix}
\label{sec:CovMatrix}
As demonstrated in previous works \citep[e.g.,][]{Viel13,Irsic17}, the covariance matrix obtained via bootstrapping of a limited data set is necessarily noisy.
We therefore regularized the observed covariance matrix using the correlation coefficients estimated from the simulated spectra following an approach similar to the one used by \cite{Lidz06}.
We first used the simulations to verify the ability of the bootstrapped errors to reproduce the real statistical variance. For this test, using the 40 $h^{-1}$cMpc box simulation S40$_{-}$1z15, we created hundreds of samples of simulated lines of sight  that closely reproduce the characteristics of the observational data  (see Section \ref{sec:syntheticLOS} for details) and we compare the variance computed directly from these realizations with the uncertainty obtained from the bootstrapping of only one synthetic sample randomly chosen. We verify that, as already shown by previous studies \citep[e.g.,][]{Kim04,Irsic13,Delabrouille13}, the bootstrapping technique underestimates the cosmic variance by up to $\sim$25$\%$, with the discrepancy level increasing towards smaller scales. We therefore increased the observational bootstrapped error at all scales by $\sim$15-25$\%$, where the correction has been computed separately for each redshift. 

The elements of the final covariance matrix, $C_{ij}$, are then computed as:
\begin{equation}
C_{ij}=R_{ij}\sqrt{C^{data}_{ii}C^{data}_{jj}},
\end{equation}
with
\begin{equation}
R_{ij}=\frac{C^{sim}_{ij}}{\sqrt{C^{sim}_{ii}C^{sim}_{jj}}}, 
 \end{equation}
where $C^{data}_{ii}$ are the diagonal elements of the bootstrapped observational covariance matrix, corrected as previously described, and $C^{sim}$ are the elements of the simulated covariance matrix obtained from the multiple realizations of synthetic lines of sight. 

\begin{figure*} 
\centering
\includegraphics[width=1.65\columnwidth]{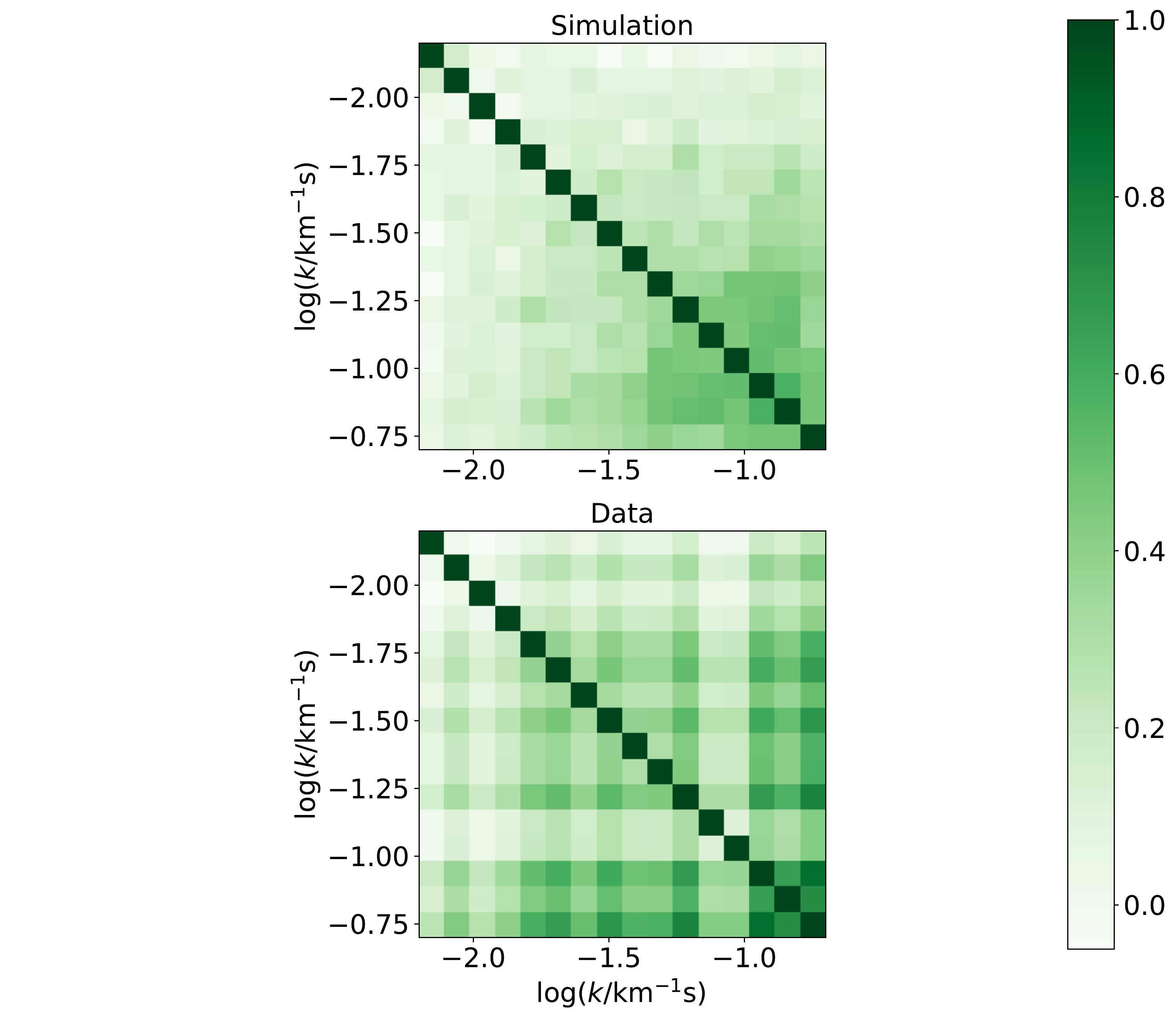} 
\caption{\small Example of the Ly$\alpha$ flux power spectrum covariance matrix computed by directly bootstrapping the data (bottom panel) and from many realizations of our 40  $h^{-1}$cMpc simulation model (top panel) for the redshift bin at $z=5$. The bootstrapped matrix is visibly nosier than the simulated one due to the limited sample size; however, the correlation structures are similar between the two panels, with stronger correlations towards smaller scales.}
\label{fig:CovMat}
\end{figure*}

For comparison, Figure \ref{fig:CovMat} shows the simulated and observational covariance matrices for the redshift bin at $z=5$. As expected, the bootstrapped matrix is noisier but both the matrices show a similar structure, with correlations increasing towards the smallest scales, $\log(k/$km$^{-1}$s$)\gtrsim-1.5$. 
This similarity gives us confidence that the simulation model used for the covariance matrix regularization is reasonably capturing the data properties. We note that 
the off--diagonal correlation structure will depend somewhat on the precise shape of the power spectrum and therefore on the thermal parameters characterizing the model.
In Appendix \ref{sec:cutoffMa} we verify that the particular choice of the simulation S40$_{-}$1z15 for this analysis is not significantly affecting our final constraints.

\section{Simulations analysis} 
In this Section we describe how we calibrate and analyze synthetic spectra to create power spectrum models that will be used to fit the observational measurements in Section \ref{sec:thermalConstr}.
\label{sec:simModel}

\subsection{Constructing mock lines of sight}
\label{sec:LOSsin}
 To ensure the correct comparison between simulations and observational data we need to produce mock lines of sight with the same resolution and redshift coverage as our observed sample. 
 We first resample and smooth the synthetic Ly$\alpha$ spectra produced from the simulations in Table \ref{table:sim} to match the spectral resolution and the pixel size of the real spectra.  
 We then progressively merge multiple synthetic sections randomly selected from the $\Delta$$z=0.1$ simulation snapshot closest to the Ly$\alpha$ forest redshift that we want to cover. 
 We choose an arbitrary starting point along each section, taking advantage of the periodicity of the simulation box.
 We take into account the mild redshift evolution of the mean flux along the line of sight by rescaling the optical depths such that the global effective Ly$\alpha$ optical depth ($\tau_{\rm eff}=-\ln(\bar{F})$) in the simulation box follows the relation
 \begin{equation}
 \tau_{\rm eff}=1.56  \Big(\frac{1+z}{5.75}\Big)^{4.0}.
 \label{eq:tau}
 \end{equation}
 See Appendix \ref{sec:opticalD} for details.
 We note that, while we use Equation \ref{eq:tau} to calibrate the optical depth evolution within each simulated line of sight, the overall Ly$\alpha$ mean flux of each redshift bin will be treated as a free parameter in our models.

 When required for testing purposes (see Appendix \ref{sec:SysTest}), the spectral noise at the same level of the corresponding observational line of sight is added to the synthetic spectra as well as the same pixel masking. Because the power spectrum computed from the real data is corrected for these systematics, the final power spectrum models used for the MCMC fit are computed without noise or pixel masking.
 \subsection{Modeling the power spectrum}
To retrieve the final flux power spectrum for each of the models in Table \ref{table:sim} we average the power computed from hundreds of mock data samples. Each measurements has been obtained following a similar procedure to the one described in Section \ref{sec:DataAnalysis}, with a few necessary expedients:
\begin{itemize}
\item Ly$\alpha$ sections: As for the real data we compute the flux contrast estimator (Equation \ref{eq:df}) using a 40 $h^{-1}$cMpc boxcar rolling mean over the reconstructed line of sight. 
After the rolling mean is applied we re--divide the line of sight into the original 10 $h^{-1}$cMpc sections and use these to compute the power spectra.
We verified that discontinuities in the flux on the border between individual sections do not substantially affect the rolling mean. 
\item Mass resolution and box size corrections: Our 10 $h^{-1}$cMpc models with $2\times512^{3}$ gas and dark matter particles represent a necessary compromise in terms of computational resources \citep{Bolton09b} and need to be corrected for small errors due to box size and resolution convergence. Therefore, we rescaled our models by factors obtained from reference simulations with larger box size (40 $h^{-1}$cMpc box) and higher mass resolution ($2\times768^{3}$ particles) in the convergence tests presented in Appendix \ref{sec:convergence}.
\end{itemize}

\label{sec:syntheticLOS}
\subsection{Varying model parameters }

To be able to fit the power spectrum measurement of Section \ref{sec:PSResults} we need a grid of models that cover the parameter space that we want to explore. In each redshift bin we consider four parameters to describe the power spectrum: the thermal parameters $T_{0}$, $u_{0}$ and $\gamma$ and the effective Ly$\alpha$ optical depth, $\tau_{\rm eff}$. While the large set of simulations, listed in Table \ref{table:sim} spans a wide range of thermal histories, by themselves they are not sufficient to evaluate the power spectrum in all the possible combination of thermal parameters. We therefore use the interpolation scheme described below.

 \subsubsection{Varying $T_{0}$ and $\gamma$}
 \label{sec:T0var}
 In order to separate the impact of thermal broadening and Jeans smoothing in the power spectrum models, we applied a simple post--processing procedure to the simulated spectra. This is achieved by translating and rotating the entire T--$\rho$ plane of the simulations to match the new $T_{0}$ and $\gamma$  values.
 We recompute the optical depths in each of our models over an extended range of power law T--$\rho$ relationships, with $T_{0}$=[3000-15000 K] in steps of 1000 K and $\gamma$=[0.7-1.7] in steps of 0.1. 

 We note that at $z\gtrsim 4$ the Ly$\alpha$ forest is mainly sensitive to gas close to the mean density \citep[e.g.,][]{Becker11}. For this reason we do not expect to place strong constraints on $\gamma$. 
 We nevertheless treat $\gamma$ as a free parameter in our fitting code. The impact of $T_{0}$ and $\gamma$ on the flux power spectrum are demonstrated in Figure \ref{fig:ThermalEffect}. Note that scales $\log(k/$km$^{-1}$s$)\gtrsim-0.8$ seem to be insensitive to variations in $\gamma$ while considerable changes in this parameter create minor shifts in the power for scales $\log (k/$km$^{-1}$s$)\lesssim-0.8$.

 \subsubsection{Varying $\tau_{\rm eff}$} 
 We rescaled the optical depths in our models to span a wide range of $\tau_{\rm eff}$ values. At each redshift the reference value has been obtained from Equation \ref{eq:tau} while the entire range of optical depths covered by our models is 
$\tau_{\rm eff}$=[0.6--2.2] in steps of 0.1. Different mean fluxes within each redshift bin have been obtained by multiplying Equation \ref{eq:tau} by single, fine tuned scalars when calibrating the simulated lines of sight.  The impact of varying $\tau_{\rm eff}$ on the power spectrum is shown in the bottom right panel of Figure \ref{fig:ThermalEffect}.

 \subsubsection{Varying $u_{0}$}
\label{sec:u0Analysis}
\begin{figure*} 
\centering
\subfigure
{ 
\includegraphics[width=1.0\columnwidth]{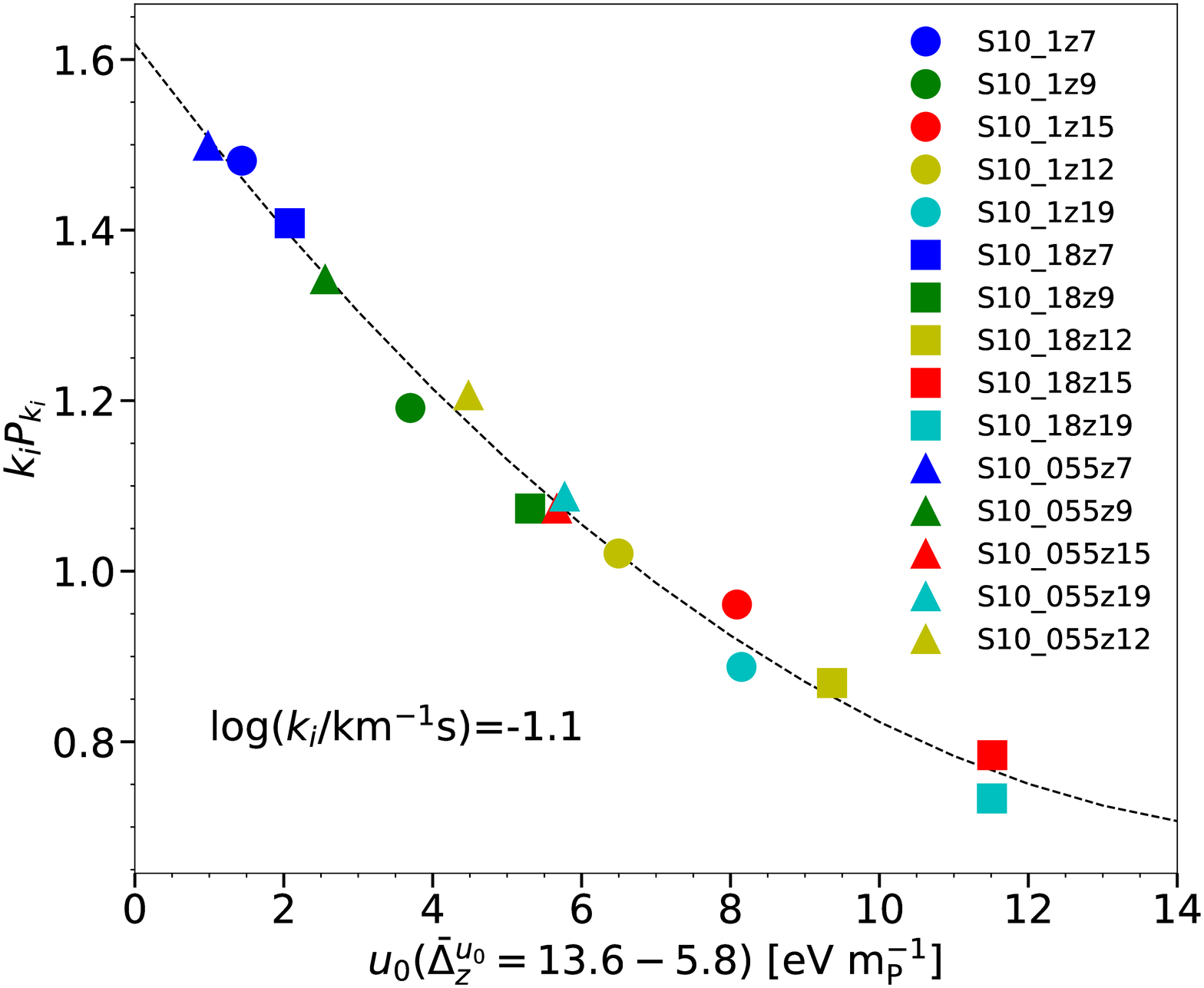}
 
}
\subfigure
{ 
\includegraphics[width=1.0\columnwidth]{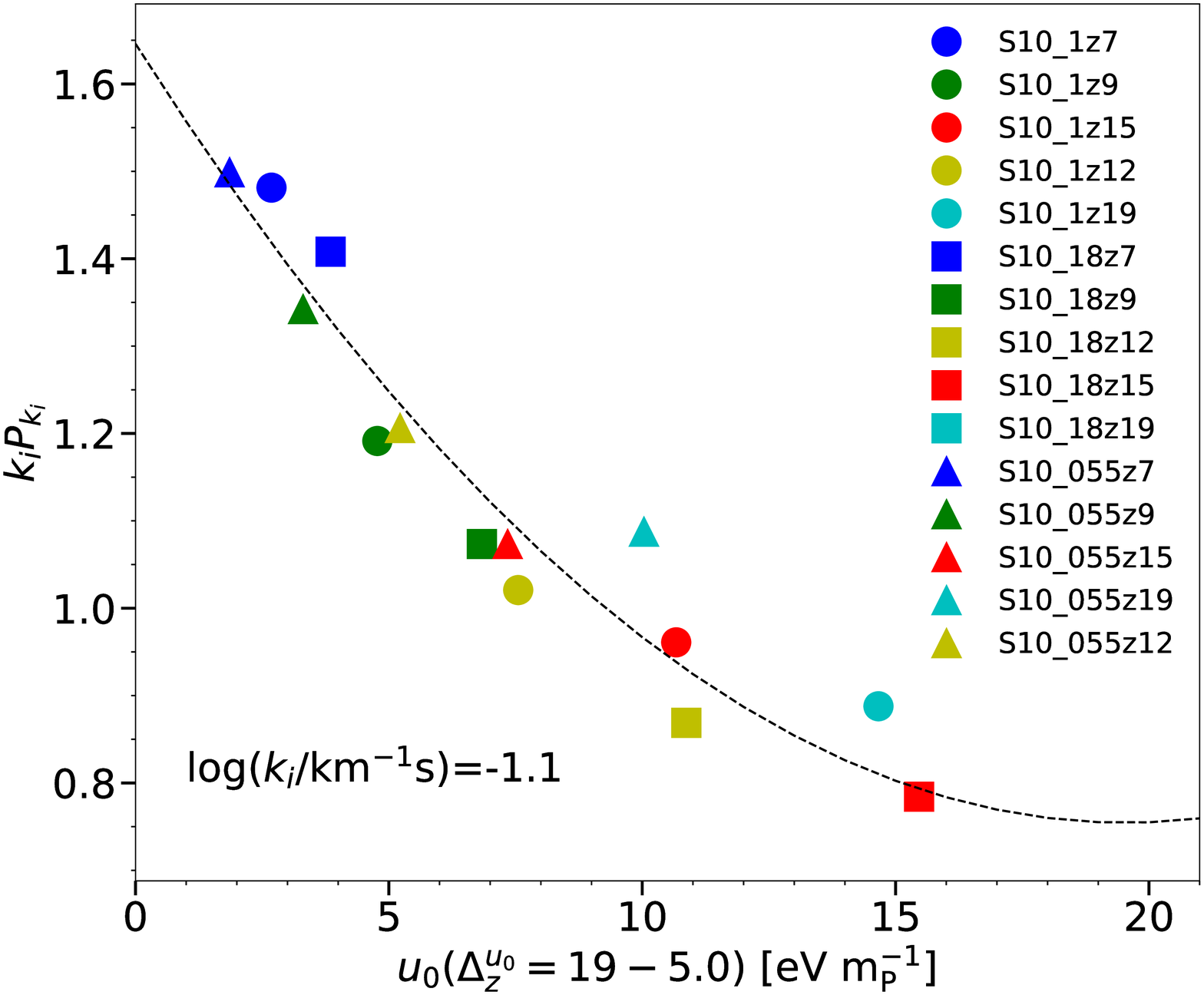} 

} 
 \caption{\small Examples of the relationship between the power spectrum at $z=5.0$ computed for $\log(k/$km$^{-1}$s$)=-1.1$ and $u_{0}$ obtained integrating over the characteristic redshift range $\bar{\Delta z^{u_{0}}}= 13.6-5.8$ (left panel) and integrating over a non-optimal redshift range $\Delta z^{u_{0}}= 19-5.0$ (right panel). Different colors correspond to different simulations post--processed to the same value of $T_{0}=10000$ K, $\gamma=1.5$, and $\tau_{\rm eff}=1.85$.}
 \label{fig:u0Rel}
\end{figure*}

By post--processing our simulations to a common set of thermal parameters we can isolate how the power spectrum depends on the integrated heating. N16 demonstrated that the flux power spectrum at $z=5.0$, (averaged over scales $-1.5\lesssim\log(k/$km$^{-1}$s$)\lesssim-0.8$) correlates with $u_{0}$. They further argued that the correlation is strongest when $u_{0}$ is integrated between $z= [12-5]$, reflecting the timescales over which the Jeans smoothing is sensitive to heat injection. Here we re--evaluate this redshift dependence using our more extended suite of models.
 For each of the redshift bins at which we compute the power spectrum and each of the scales sensitive to $u_{0}$ ( $-1.4\lesssim\log(k/$km$^{-1}$s$)\lesssim-0.8$) we empirically determine the ``characteristic'' redshift range of integration ($\bar{\Delta z^{u_{0}}}$) for which the power is closest to a one--to--one function of $u_{0}$. The method is demonstrated in Figure \ref{fig:u0Rel}. We first post--process all of the 10 $h^{-1}$ cMpc simulations of Table \ref{table:sim} to the same values of $T_{0}$, $\gamma$ and $\tau_{\rm eff}$. We then fit a power law to $kP_{k}$ versus $u_{0}$, where $u_{0}$ is integrated using Eq. \ref{eq:u0} over a redshift interval $\Delta z^{u_{0}}$. The preferred interval, $\bar{\Delta z^{u_{0}}}$, is the one that minimizes the $\chi^{2}$ for this fit.
 
\begin{figure} 
\centering
\includegraphics[width=1.0\columnwidth]{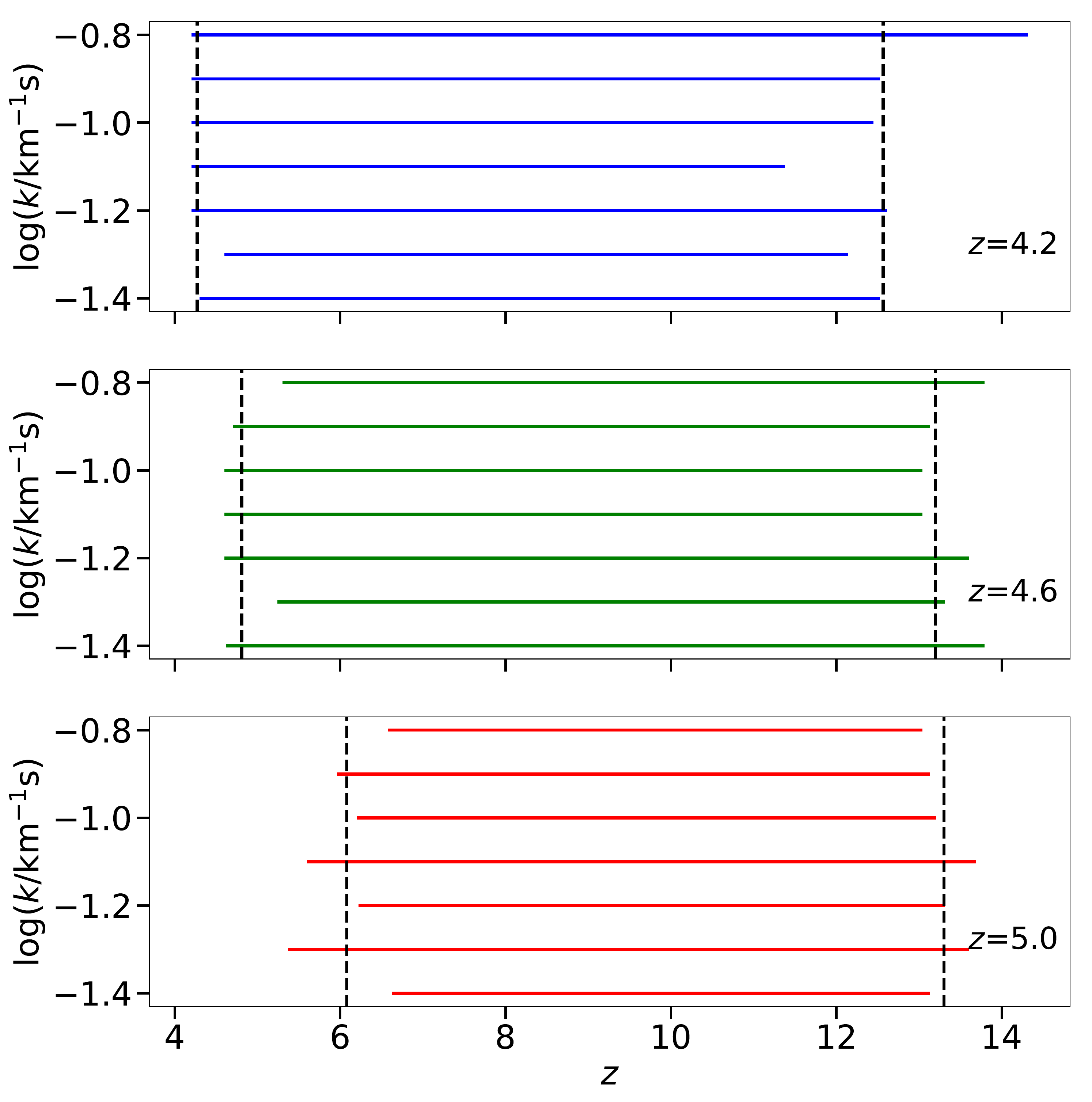} 
\caption{\small Characteristic $\bar{\Delta z^{u_{0}}}$ computed for $-1.4\lesssim\log(k/$km$^{-1}$s$)\lesssim -0.8$. For each $\log(k/$km$^{-1}$s$)$ on the y axis, the colored horizontal line covers the redshift interval that produces the best fit between the power spectrum and the integrated heating. Each panel shows the results for a different redshift bin. While there is a mild evolution with redshift, the $\bar{\Delta z^{u_{0}}}$ values are generally constant within the same redshift bin. We use the average $\bar{\Delta z^{u_{0}}}$ indicated by the black dashed vertical lines to compute the fiducial $u_{0}$ parameters at each redshift.}
\label{fig:CDz}
\end{figure}

The characteristic $\bar{\Delta z^{u_{0}}}$ computed for the scales $-1.4\lesssim\log(k/$km$^{-1}$s$)\lesssim-0.8$ are reported in Figure \ref{fig:CDz} for the different redshift bins. 
As expected, the sensitivity of the power spectrum to the previous thermal history varies slightly with the redshift at which the power spectrum is measured. The Ly$\alpha$ structures observed at progressively lower redshifts seem to slowly lose sensitivity to earlier epochs; while the power spectrum measured at $z=5.0$ still maintains sensitivity up to $z\gtrsim13$, at $z=4.2$ the forest traces the thermal history of the gas mainly for $z\lesssim12$. Interestingly, we find that the power spectrum at $z=5.0$ is less sensitive to heating happening at $z\lesssim 6$, even though the power spectra at $z=4.6$ and 4.2 retain sensitivity all the way down to their respective redshifts.
We generally expect that the gas density distribution will exhibit some delay in responding to changes in gas pressure. Further investigation revealed that a delay did appear for all three redshifts when peculiar velocities were turned off. The delay increased with the redshift at which the power spectrum was measured, with a delay at $z=5.0$ that was larger than the one found with peculiar velocities turned on.
This suggests that peculiar velocities may play a role by decreasing the delay between heat injection and a change in the power spectrum. Presumably this occurs because, as the gas is heated, redshift distortions created by accelerating the gas precede changes in the density field. This effect may partly explain the lack of a gap at $z=4.2$ and 4.6. For now we adopt these relations as ampirical, and leave more detailed physical insights to future work.

Because the $\bar{\Delta z^{u_{0}}}$ values are generally constant among different scales within the same redshift bin, we adopt average values (black dashed vertical lines in Figure \ref{fig:CDz}) as integration bounds in Eq. \ref{eq:u0}. The fiducial redshift range over which $u_{0}$ is integrated is given in Table \ref{table:u0Ranges}.
We note that our $u_{0}$--$kP_{k}$ relationship, while remarkably tight over scales sensitive to $u_{0}$, do exhibit scatter. In the final MCMC analysis therefore, the amount of scatter about the $u_{0}$--$kP_{k}$ fit at each scale has been included as systematic uncertainty. 
We note that, while we chose our fiducial redshift ranges to maximize the sensitivity of the power spectrum to $u_{0}$, in principle we could constrain this parameter integrated within any reasonable redshift range if properly accounting for the systematic uncertainty in the $u_{0}$ versus $kP_{k}$ fit.

 \begin{table} 
\caption{\small Fiducial redshifts ranges used to compute the $u_{0}$ parameters for fitting the flux power spectrum at different redshifts. Redshift bins are indicated in column 1 with the corresponding fiducial $u_{0}$ redshift intervals in column 2.   }
\centering \begin{tabular}{c c c} 
\hline
$z$\  &  $\Delta z_{fid}^{u_{0}}$\ \\
\hline
4.2 & 4.2--12\\
4.6 & 4.6--13\\
5.0 & 6.0--13\\
\hline 

\end{tabular}		
\label{table:u0Ranges}
\end{table}

\section{Thermal state constraints}
 \label{sec:thermalConstr}
 To obtain constraints on the IGM temperature and integrated thermal history from the observational power spectrum measurements obtained in Section \ref{sec:DataAnalysis}, we adopted a Bayesian MCMC approach to measure $T_{0}$, $u_{0}$, $\gamma$ and $\tau_{\rm eff}$ for each of the three redshift bins independently. 
In this Section we present the method and the main findings of this analysis.
\subsection{The MCMC analysis }
\label{sec:MCMC}
We constructed a grid of power spectrum models following the post--processing approach given above, where for a given choice of $T_{0}$, $\gamma$, and $\tau_{\rm eff}$ the dependence of the power spectrum on $u_{0}$ is derived from the fits described in Section \ref{sec:u0Analysis}.
We then perform a multi-linear interpolation among the grid points of the four dimensional parameter space. We implemented the interpolation scheme using a Bayesian MCMC approach. At each redshift, applying flat priors for all variables, we obtain the set of parameters that maximize a Gaussian multivariate likelihood function:
\begin{equation}
\ln \mathcal{L}=-\frac{1}{2}\vec{\Delta}^{T}C^{-1}\vec{\Delta} -\frac{1}{2}\ln \det(C)- \frac{N}{2}\ln 2\pi, 
\end{equation}
 where $\vec{\Delta}$ is the residual vector between the power spectrum values of the data and the model and $C$ is the $N\times N$ data covariance matrix (where $N$ is the number of data points).
 
We tested the interpolation scheme by removing one of the models used for the interpolation and using it to generate mock data (see Appendix \ref{sec:Interpolation}). We found that the parameters $u_{0}$ and $T_{0}$ are recovered accurately. Small biases (within the $68\%$ uncertainties) appear in the recovered values of $\gamma$ and $\tau_{\rm eff}$ due to their intrinsic degeneracy at large scales and the poor sensitivity of the high redshift power spectrum to $\gamma$. Fortunately, however, the relatively weak constraints on these parameters do not bias our results for $T_{0}$ and $u_{0}$.

To test the reliability of the best fitting values, for each redshift we ran three independent chains of $2\times10^{5}$ iterations (half of which are discarded as burn--in) from different randomly chosen initial parameters. We verify that all the chains were converged by comparing the between--chain and within--chain variances for each parameter using the Gelman--Rubin test.
 
\subsection{Results}
\label{sec:ResultsThermal}

\begin{figure*} 
\centering 
\includegraphics[width=4.2in]{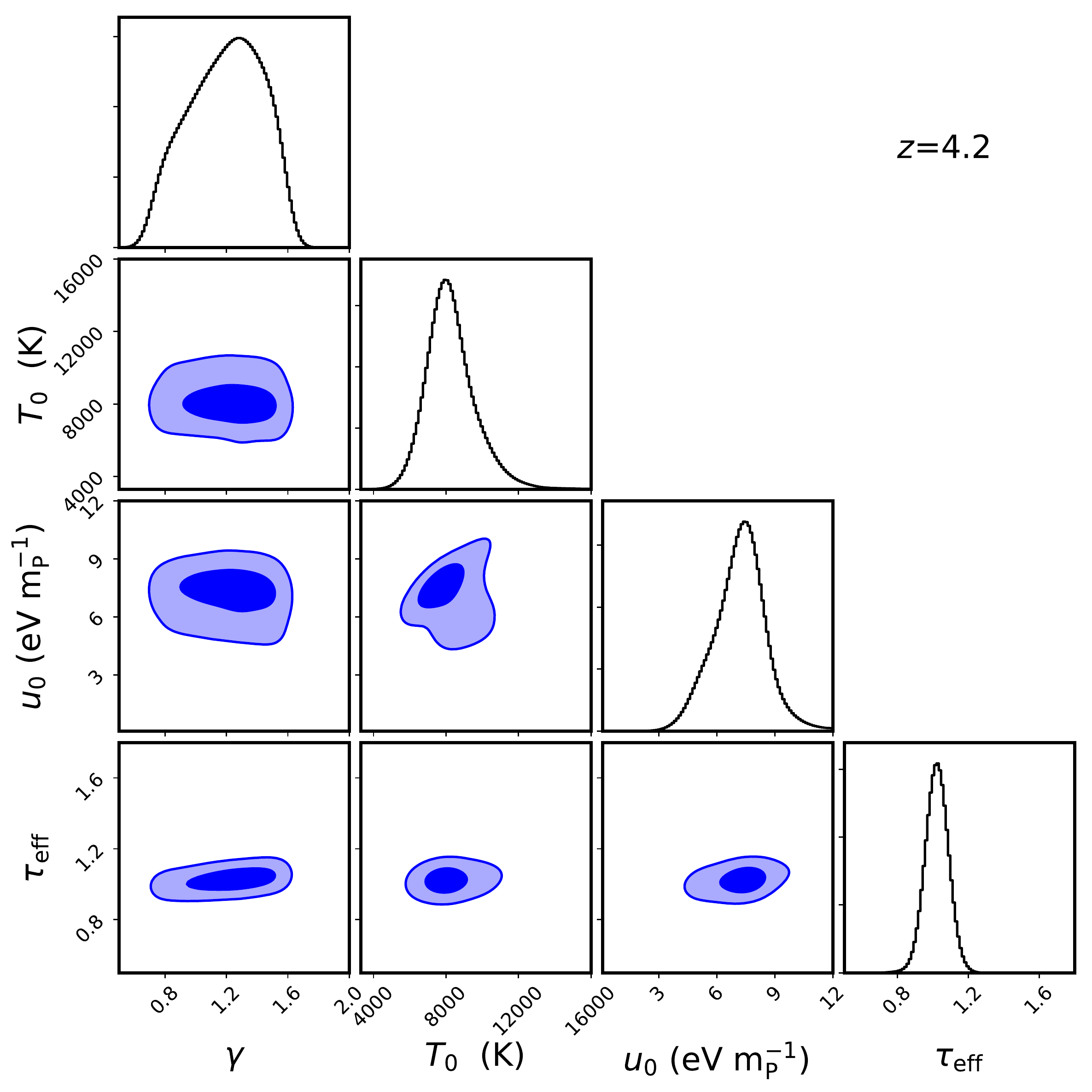}
\caption{\small Probability distributions for the parameters $T_{0}$, $u_{0}$, $\gamma$ and $\tau_{\rm eff}$ obtained from the MCMC analysis of the power spectrum at $z=4.2$. Contours show the 68$\%$ and $95\%$ marginalized two--dimensional probability distributions while the black histograms display the one--dimensional marginalized posterior distributions for each parameter.  }
 \label{fig:MCMCz4.2}
 \end{figure*}

\begin{figure*} 
\centering 
\includegraphics[width=4.2in]{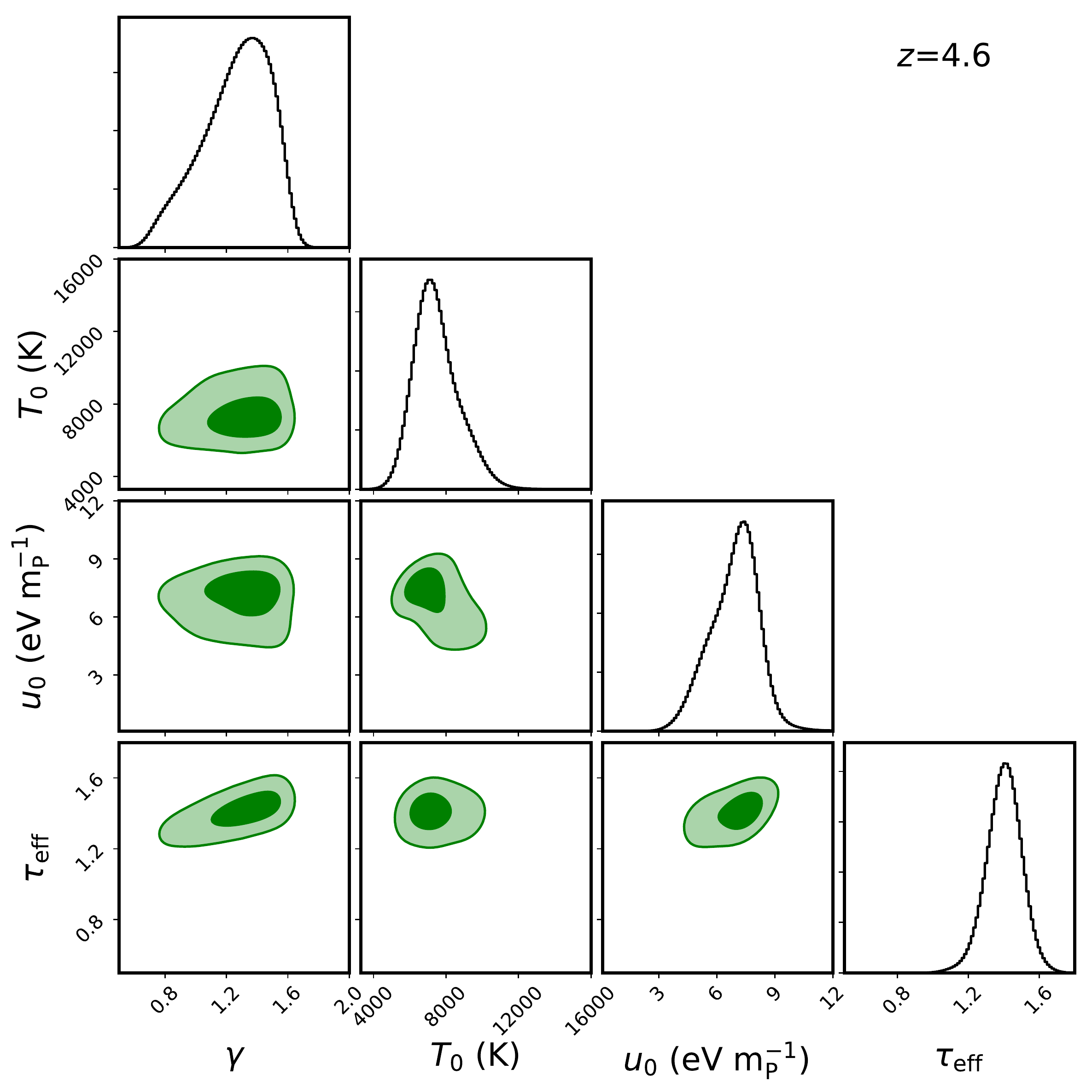} 
\caption{\small As in Figure \ref{fig:MCMCz4.2} but for the power spectrum at $z=4.6$.}
\label{fig:MCMCz4.6}
\end{figure*}

\begin{figure*}
\centering 
\includegraphics[width=4.2in]{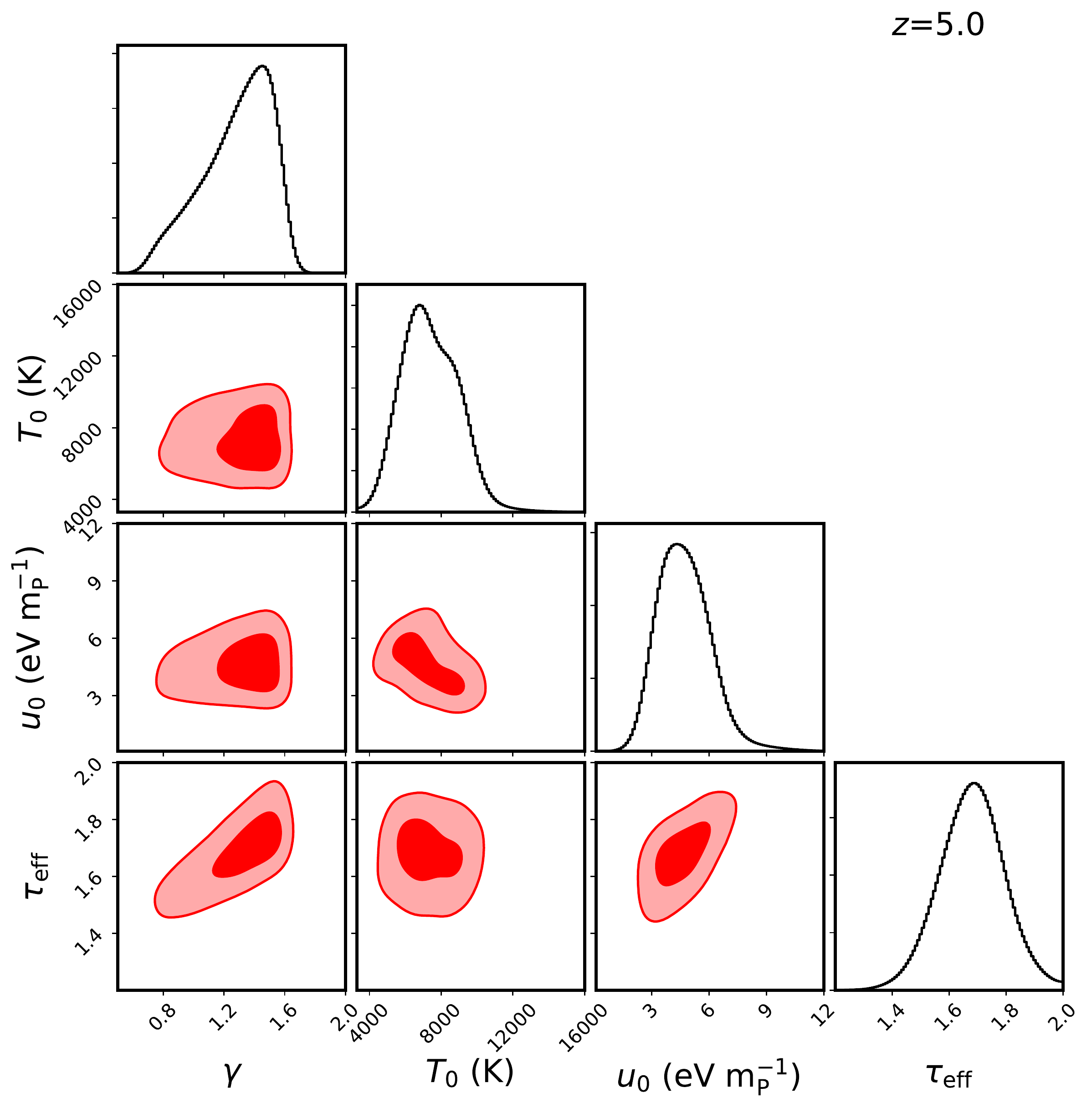} 
\caption{\small As in Figure \ref{fig:MCMCz4.2} but for the power spectrum at $z=5.0$.}
 \label{fig:MCMCz5.0}
\end{figure*}

Figures \ref{fig:MCMCz4.2}, \ref{fig:MCMCz4.6} and \ref{fig:MCMCz5.0} display the posterior likelihood distributions for the parameters $T_{0}$, $u_{0}$, $\gamma$ and $\tau_{\rm eff}$ at redshifts 4.2, 4.6, and 5.0 respectively.
While the inclusion of small scales ($\log(k/$km$^{-1}$s$)\gtrsim-1.0$) in the power spectrum allows relatively tight constraints on both $T_{0}$ and $u_{0}$, some degeneracy between these two variables is still noticeable at all redshifts. This is expected since both of these parameters act on intermediate scales in a similar way.  
Degeneracies between $\gamma$ and $\tau_{\rm eff}$ increase with redshift with slightly weaker constraints on $\tau_{\rm eff}$ obtained towards higher redshifts. As expected, $\gamma$ shows broad bounds at all redshift (with the 1$\sigma$ contours covering almost the entire parameter space), reaffirming that the Ly$\alpha$ forest at high--redshifts mainly probes gas around the mean density and is not highly sensitive to the slope of the T--$\rho$ relation. Figure \ref{fig:MCMCz4.2}, \ref{fig:MCMCz4.6} and \ref{fig:MCMCz5.0} demonstrate that our measurements of $u_{0}$ and $T_{0}$ are not highly affected by degeneracies with $\gamma$ and $\tau_{\rm eff}$.

The final results of the MCMC analysis are summarized in Table \ref{table:Tprob}. The temperatures are constrained with $\sim$15$\%$ uncertainties at all redshifts, while the error on $u_{0}$ varies from $\sim$18$\%$ for the $z=4.2$ and $z=4.6$ redshift bins up to $\sim$30$\%$ at the highest redshift, in good agreement with the forecast presented by N16. Our results for $\tau_{\rm eff}$ are highly consistent with the measurements of \cite{Becker13a} at $z=4.2$. We are somewhat higher at $z=4.6$, but all together our constraints appear to bridge the evolution of $\tau_{\rm eff}$ at $z\lesssim4$ measured by \cite{Becker13a} and at $z\gtrsim5$ from \cite{Bosman18} (see Appendix \ref{sec:opticalD}).

\begin{figure} 
\centering
\includegraphics[width=1.0\columnwidth]{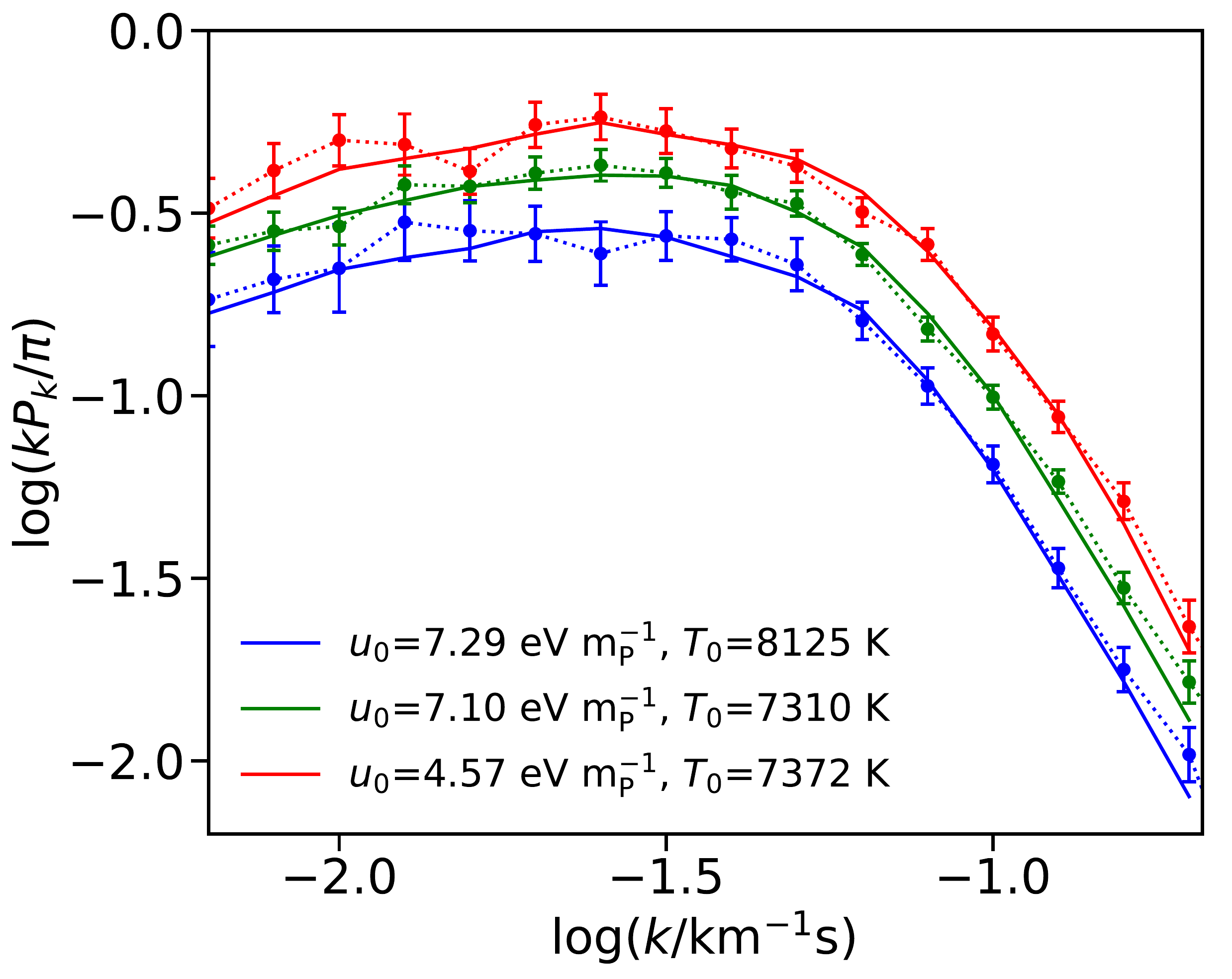} 
\caption{\small  The best fitting models for our high--resolution power spectrum measurements. The best fit models at $z=5.0$ (red solid line), $z=4.6$ (green solid line) and $z=4.2$ (blue solid line) are superimposed on the corresponding observational measurement (color--coded data points and dotted lines). The corresponding best fitting parameters are also reported. }
\label{fig:BestFits}
\end{figure}

\begin{table} 
\caption{\small Best fitting values and marginalized $68\%$ confidence intervals for the fits to our power spectrum measurements. The power spectrum redshift (column 1) is reported along with the best fitting values of $T_{0}$ (column 2), $u_{0}$ (column 3), $\gamma$ (column 4) and $\tau_{\rm eff}$ (column 5).}
\centering \begin{tabular}{c c c c c} 
\hline
$z$\  &  $T_{0}/10^{3} [K]$\ & $u_{0}$[eV $m_{p}^{-1}]$ \ & $\gamma$\ & $\tau_{\rm eff}$ \\
\hline
4.2 & $8.13^{+1.34}_{-0.97}$  &  $7.29 ^{+0.98}_{-1.35}$ & $1.21 ^{+0.23}_{-0.28}$  & $1.02 ^{+0.04}_{-0.04}$\\
4.6 & $7.31^{+1.35}_{-0.88}$  & $7.10 ^{+0.83}_{-1.45}$ & $1.29 ^{+0.19}_{-0.26}$  & $1.41 ^{+0.08}_{-0.09}$ \\
5.0 & $7.37^{+1.67}_{-1.39}$ & $4.57 ^{+1.37}_{-1.16}$ & $1.33 ^{+0.18}_{-0.27}$  & $1.69 ^{+0.10}_{-0.11}$  \\
\hline 

\end{tabular}		
\label{table:Tprob}
\end{table}
In Figure \ref{fig:BestFits} we show the best power spectrum models compared with the measurement. Visually there is good agreement with the data at all redshifts. 

\begin{figure} 
\centering
\includegraphics[width=1.0\columnwidth]{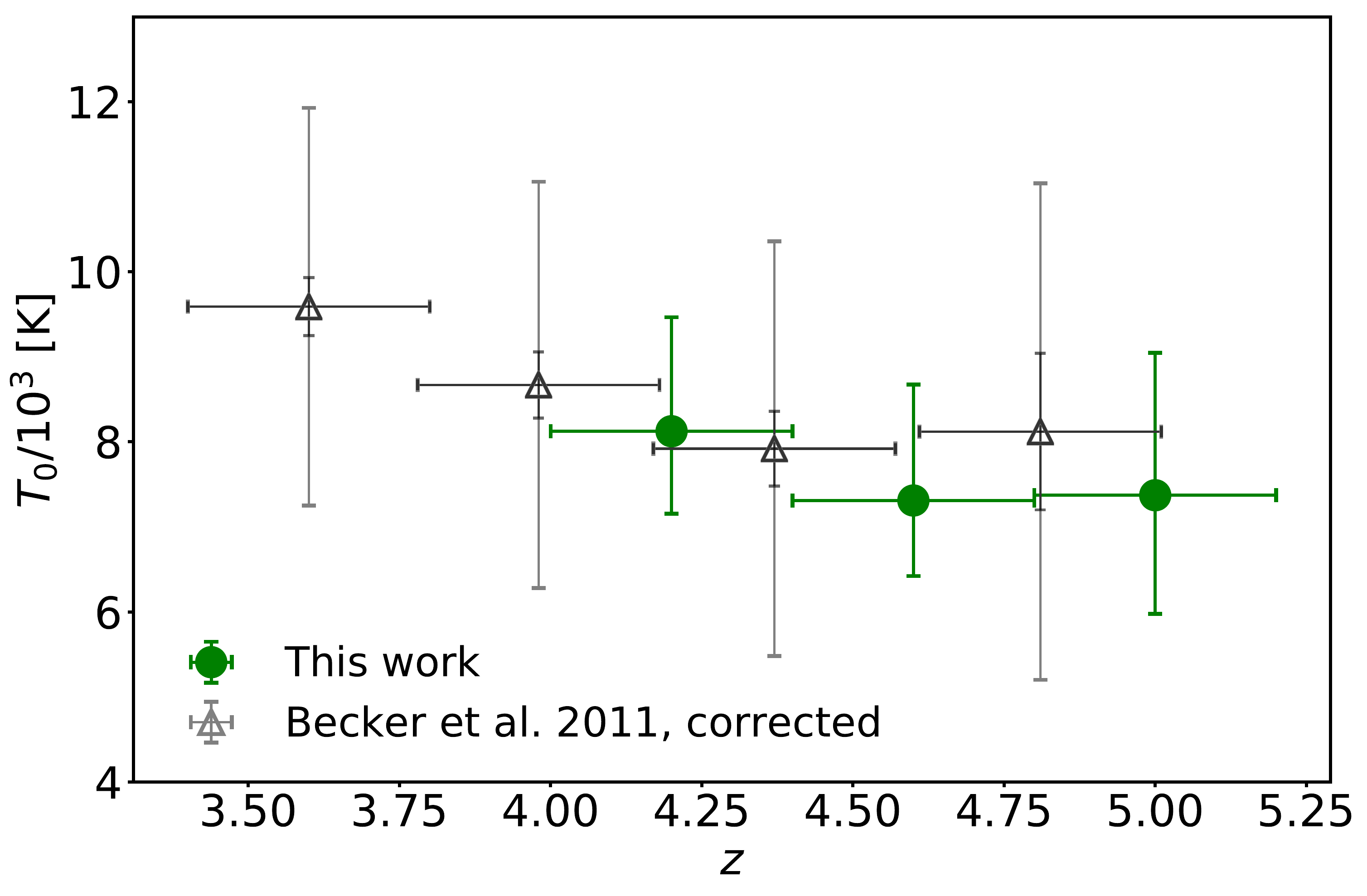} 
\caption{\small Temperature at the mean density of the IGM obtained in this work (green points) and from the curvature measurement of \cite{Becker11} (gray triangles) at $z\gtrsim3.5$. The Becker et al. $T_{0}$ values have been inferred assuming $\gamma\sim1.5$. Vertical error bars are $68\%$ confidence intervals for this work. For Becker et al. the small error bars are the $68\%$ statistical uncertainties, while the extensions in lighter gray include the Jeans smoothing uncertainty estimated by those authors.}
\label{fig:Tcomp}
\end{figure}

The final temperature constraints at the IGM mean density are presented in Figure \ref{fig:Tcomp}. Our new results (green points) are compared with the previous measurements of \citealt{Becker11} (gray triangles) at $z>3.5$ obtained with the curvature method. We have added the systematic uncertainty for Jeans smoothing estimated by Becker et al. to those data. See also Appendix \ref{sec:GlobalComp} for a more comprehensive comparison of recent temperature measurements at the redshifts covered by our analysis.
Our temperature measurements show good agreements with this previous work in the overlapping redshift bins. This accord is significant because we analyzed a largely independent set of quasar spectra with a different method and using a new suite of hydrodynamical simulations.
Most significantly, we now explicitly fit for $u_{0}$, removing the systematic uncertainty in $T_{0}$ related to Jeans smoothing.
Overall our $T_{0}$ values are consistent with little evolution over $4.2\lesssim z\lesssim 5.0$. Given the known trend of increasing temperatures at $z\lesssim4$ \citep[e.g.,][]{Becker11,Boera14,Walter18} our measurement at $z=4.2$ may include a contribution from the initial phase of IGM reheating due to the He{\sc \,ii} reionization (e.g., \citealt{Worseck11, Syphers14}). We consider this possibility in the final part of our analysis, when we use the new thermal constraints to evaluate hydrogen reionization scenarios.

\begin{figure} 
\centering
\includegraphics[width=1.0\columnwidth]{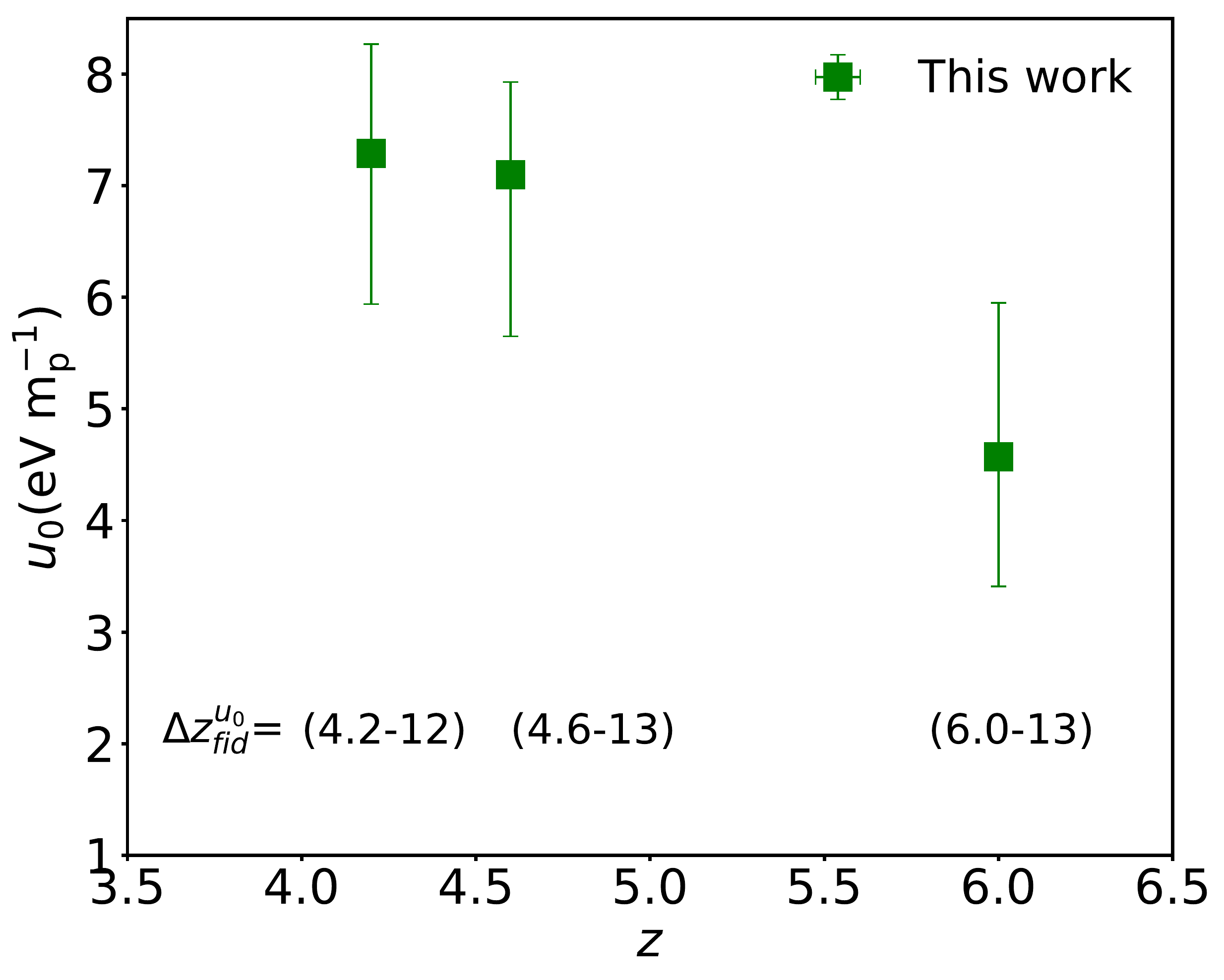} 
\caption{\small Our constraints on the integrated heat input into the IGM, computed for the fiducial redshift range indicated below each point. Error bars are marginalized $68\%$ confidence intervals. Note that, for each of the observational power spectra, the fiducial redshift range used to compute $u_{0}$ has been chosen to maximize the sensitivity of the  power spectrum to this parameter and, therefore, minimize the measurement's  errors (see Section \ref{sec:u0Analysis} for details). }
\label{fig:u0comp}
\end{figure}

Finally, in Figure \ref{fig:u0comp} we present our first constraints on the integrated thermal history of the IGM. Our $u_{0}$ measurements, are plotted at the minimum redshift of the fiducial ranges given in Table \ref{table:Tprob}. As expected, $u_{0}$ increases from $z=6$ to $z=4.2$, reflecting ongoing heat injection after reionization.

\section{Reionization constraints}
\label{sec:reionizationModels}
Our observational constraints on $T_{0}$ and $u_{0}$ can, in principle, be used to test any reionization model for which a thermal history can be calculated. While we leave for future work the analysis of extended and more realistic reionization scenarios, in this Section we demonstrate the potential of this approach using semi-analytical models of instantaneous reionization.

\subsection{Modeling instantaneous reionization}
We model the thermal history of instantaneous hydrogen reionization using a semi-analytical approach similar to the one adopted in \cite{Sanderbeck16}. 
To obtain $T_{0}$ as a function of redshift we solve the equation describing the temperature evolution of a Lagrangian fluid element at the cosmic mean density, i.e., with $\Delta=1$ \citep[e.g.,][]{MiraldaRees94, HuiGnedin1997, McQuinn16},
\begin{equation}
\frac{dT}{dt}=-2HT +\frac{2T}{3\Delta}\frac{d\Delta}{dt} +\frac{2}{3k_{B}n_{tot}}\frac{dQ}{dt},
\label{eq:Tev}
\end{equation}
where $H$ is the Hubble parameter and $n_{tot}$ is the total number density of particles (electrons and ions). Equation \ref{eq:Tev} is valid in the approximation that the number of particles remains fixed, describing well the post--reionization gas. The first term on the right side of Equation \ref{eq:Tev} takes into account the cooling due to adiabatic expansion while the second term gives the adiabatic heating and cooling due to structure formation. The thermal history at the mean density has been shown to depend weakly on this second term \citep{McQuinn16}; we will therefore ignore it in our calculation. The differences among the models will depend instead upon the third term, which encodes photo-heating (the only heating source considered in these calculations) and radiative cooling processes. 
We can expand this term as:
\begin{equation}
\frac{dQ}{dt}=\sum_{X}{\frac{dQ_{photo,X}}{dt}}+\frac{dQ_{Compton}}{dt}+\sum_{i}\sum_{X}{R_{i,X}n_{e}n_{X}},
\label{eq:balance}
\end{equation}
where  $\frac{dQ_{photo,X}}{dt}$ is the photo-heating rate of ion $X$, $\frac{dQ_{Compton}}{dt}$ is the Compton cooling rate and $R_{i,X}$ is the cooling rate coefficient for the ion $X$ and cooling mechanism $i$. Because we are modeling the temperature of the gas at the end of hydrogen reionization, in Equation \ref{eq:balance} we will consider only the H{\sc \,i} (dominant) and He{\sc \,i} photo-heating. As for the cooling term, we include Compton and H{\sc \,ii} recombination in our calculations. As discussed in \cite{McQuinn16}, these represent the relevant cooling processes that shape the temperature evolution. We compute these cooling terms using the rate coefficients provided by \cite{HuiGnedin1997}. 
The optically thin photo-heating after reionization is modeled as \citep[e.g.,][]{Sanderbeck16}
\begin{equation}
\frac{dQ_{photo,X}}{dt}\approx\frac{h\nu_{X}}{\gamma_{X}-1+\alpha_{bk}}\alpha_{A,X}n_{\tilde{X}}n_{e} ,
\label{eq:phot}
\end{equation}
where $\nu_{X}$ is the frequency associated with the ionization potential of species $X$, and $\gamma_{X}$ is the corresponding approximate power--law index of the photo-ionization cross section, for which we assume $\gamma_{X}=2.8$ for H{\sc \,i} and $\gamma_{X}=1.7$ for He{\sc \,i}.
Equation \ref{eq:phot} is valid in the approximation of photo-ionization equilibrium with an ionizing background that has a power--law specific intensity of the form $J_{\nu}\propto \nu^{-\alpha_{\rm bk}}$. In detail, the photo-heating rate will also depend on $\alpha_{A,X}$, the case A recombination coefficient associated with the transition from $X${\sc \,i}$\rightarrow X$ for species $X\in$ [H{\sc \,i}, He{\sc \,i}]; on the number density of the species $\tilde{X}\in$ [H, He]; and on the electron number density $n_{e}$.

To compute the total energy deposited into the gas by photo-heating, $u_{0}$, we just need to consider the third term of Equation \ref{eq:Tev} (and the first term of Equation \ref{eq:balance}).
The equation to solve for $u_{0}$ will then be
\begin{equation}
\frac{dT}{dt}= +\frac{2}{3k_{B}n_{tot}}\sum_{X}{\frac{dQ_{photo,X}}{dt}}.
\label{eq:u0base}
\end{equation}
Because the specific internal energy can be expressed as $u=\frac{3}{2}k_{B}T\frac{1}{m}$, Equation \ref{eq:u0base} can be solved as
\begin{equation}
\frac{du_{0}}{dt}=\frac{1}{\bar{\rho}}\sum_{X}{\frac{dQ_{photo,X}}{dt}},
\label{eq:u0sec}
\end{equation}
where $\bar{\rho}$ is the mean mass density.

\subsection{Instantaneous reionization parameters}
\begin{figure*} 
\begin{center}
\includegraphics[width=2.012\columnwidth]{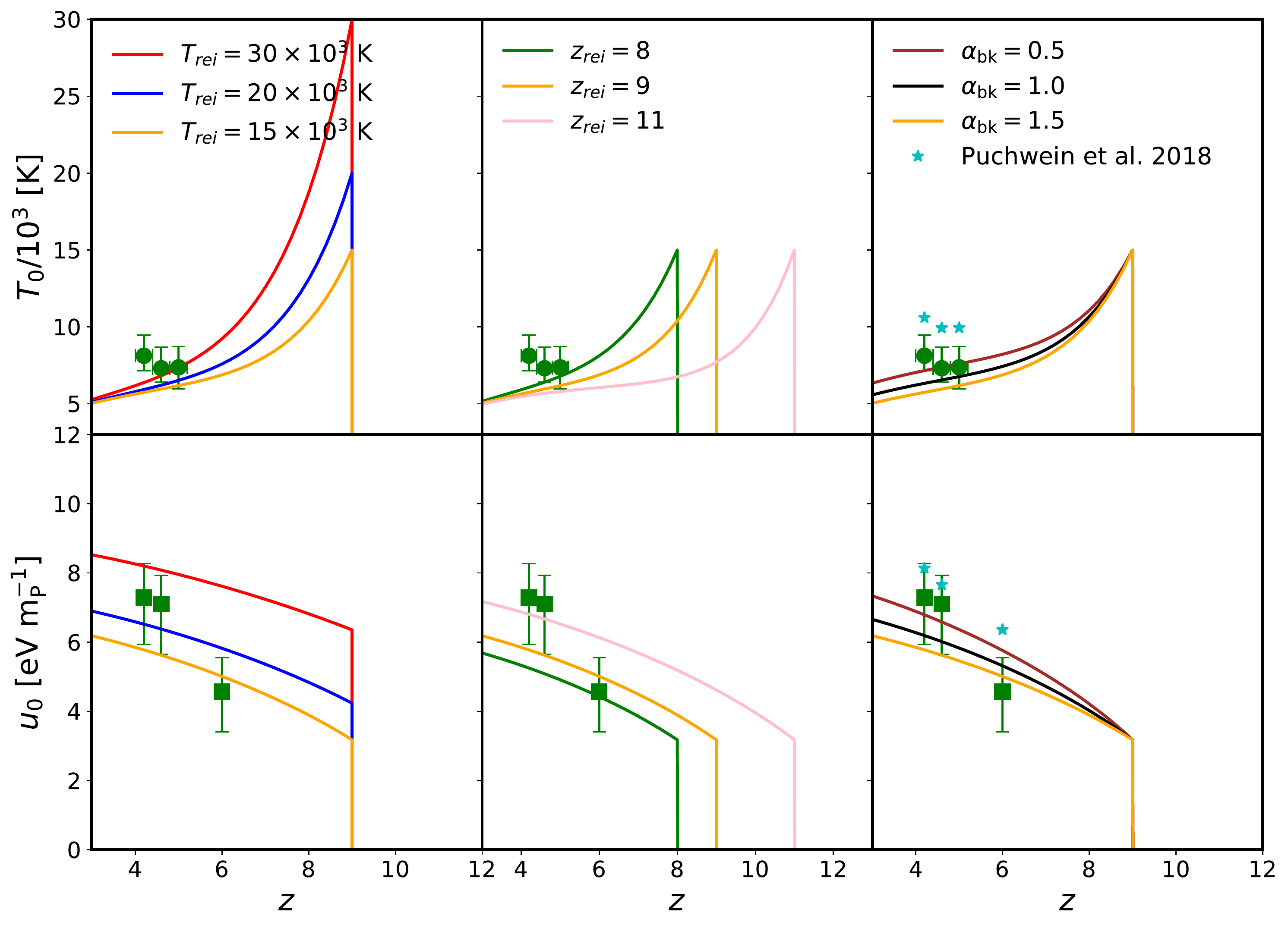} 
\caption{\small Evolution of the temperature of the IGM (top row) and the total energy injected per unit mass (bottom row) at the mean density, for toy models of instantaneous hydrogen reionization. A fiducial model with $T_{\rm rei}=15,000$ K, $z_{\rm rei}=9$ and $\alpha_{\rm bk}=1.5$ is plotted in orange in each panel. Column 1 shows the effect of varying the reionization temperature. Column 2 shows how the evolution of $T_{0}$ and $u_{0}$ is influenced by the redshift of instantaneous reionization. Column 3 varies the spectral index of the post--reionization ionizing background. In all panels, the green points with $68\%$ error bars are the observational constraints obtained in this work. For comparison, in column 3 we also report the values of temperature and heating extracted from the empirically calibrated UV background model of \citealt{Puchwein18} (cyan stars).}
\label{fig:reiModels}
\end{center}
\end{figure*}

We parametrize our models using three numbers: the redshift of instantaneous reionization, $z_{\rm rei}$, the temperature reached by the IGM during hydrogen reionization, $T_{\rm rei}$, and the spectral index of the post--reionization ionizing background, $\alpha_{\rm bk}$. 
 Figure \ref{fig:reiModels} presents the effects on the evolution of $T_{0}$ (top row) and $u_{0}$ (bottom row) of these three parameters.
The first column shows models with the same $z_{\rm rei}$ and $\alpha_{\rm bk}$ but different reionization temperatures. Radiative transfer calculations suggest that temperatures during reionization should reach 17,000 K $\lesssim T_{\rm rei}\lesssim$25,000 K \citep[e.g.,][]{MiraldaRees94,Daloisio18}; however, we explored $T_{\rm rei}$ down to 10,000 K and up to 30,000 K, where the upper range is similar to the hottest scenario of short and late reionization considered in \citealt{Daloisio18}. 
The second column in Figure \ref{fig:reiModels} demonstrates how changing the timing of reionization influences the histories of $T_{0}$ and $u_{0}$. We tested models spanning a range of redshifts from $z_{\rm rei}=5.5$ up to $z_{\rm rei}=12$. Finally, the third column shows the effect of changing the spectral index of the post--reionization ionizing background. The value of $\alpha_{\rm bk}$ can be connected to the intrinsic spectral index of the sources, $\alpha_{s}$, via the expression $\alpha_{\rm bk}\approx\alpha_{s}-3(\beta-1)$, where $\beta$ is the logarithmic slope of the column density distribution of intergalactic hydrogen absorbers \citep{Sanderbeck16}, which
 is valid at $z\gtrsim3$ when the physical mean free path of 1 Ry photons $\lambda_{MFP}\ll cH^{-1}$. The value of $\beta$ may vary, but for this analysis we adopt $\beta=1.3$ from \cite{Songaila10}.%
 
Here we focus on two cases: reionization driven by star--forming galaxies with a soft $\alpha_{\rm bk}$=1.5 ($\alpha_{s}\sim2.4$, within the commonly adopted range between 1 and 3; e.g., \citealt{Bolton07, Kuhlen12}), and models of quasar--driven reionization with $\alpha_{\rm bk}=0.5$ (corresponding to $\alpha_{s}\sim1.4$; e.g., \citealt{Telfer02,Shull12}). We note that radiative transfer calculations have shown that the temperature increase from the passage of an ionization front may not depend strongly on the spectrum of the ionizing sources \citep{Daloisio18}. We therefore constrain $T_{\rm rei}$ independently from $\alpha_{\rm bk}$.

\subsection{Method}

To be conservative, the constraints on instantaneous reionization models presented in this paper will be obtained using the thermal parameters in the lowest redshift bin ($z=4.2$) only as upper limits because they may be affected by the extra heating due to the He{\sc \,ii} reionization. 
For each combination of parameters ($\alpha_{\rm bk}$, $z_{\rm rei}$, $T_{\rm rei}$) we obtain three likelihood values corresponding to the redshifts of the observational constraints: $L_{model}[{<T^{z=4.2}_{0},<u^{z=4.2}_{0}}]$, $L_{model}[{T^{z=4.6}_{0},u^{z=4.6}_{0}}]$ and $L_{model}[{T^{z=5.0}_{0},u^{z=6.0}_{0}}]$. These probabilities describe how well a model can simultaneously fit the observed values of $T_{0}$ and $u_{0}$ at each redshift.
A given $L_{model}$ is obtained by associating the model's  $T_{0}$, and $u_{0}$ values to the probability derived from the corresponding full posterior distribution in Figures \ref{fig:MCMCz4.2}, \ref{fig:MCMCz4.6}, or \ref{fig:MCMCz5.0} (central panels).

The final probability of each model is computed by multiplying the independent likelihood values obtained at each redshift :
\begin{equation}
\begin{split}
L_{model}=L_{model}[{<T^{z=4.2}_{0},<u^{z=4.2}_{0}}]\times \\
L_{model}[{T^{z=4.6}_{0},u^{z=4.6}_{0}}]\times L_{model}[{T^{z=5.0}_{0},u^{z=6.0}_{0}}]
\end{split}
\end{equation}

\subsection{Results}
\begin{figure} 
\centering
\subfigure
{ 
\includegraphics[width=3.3in]{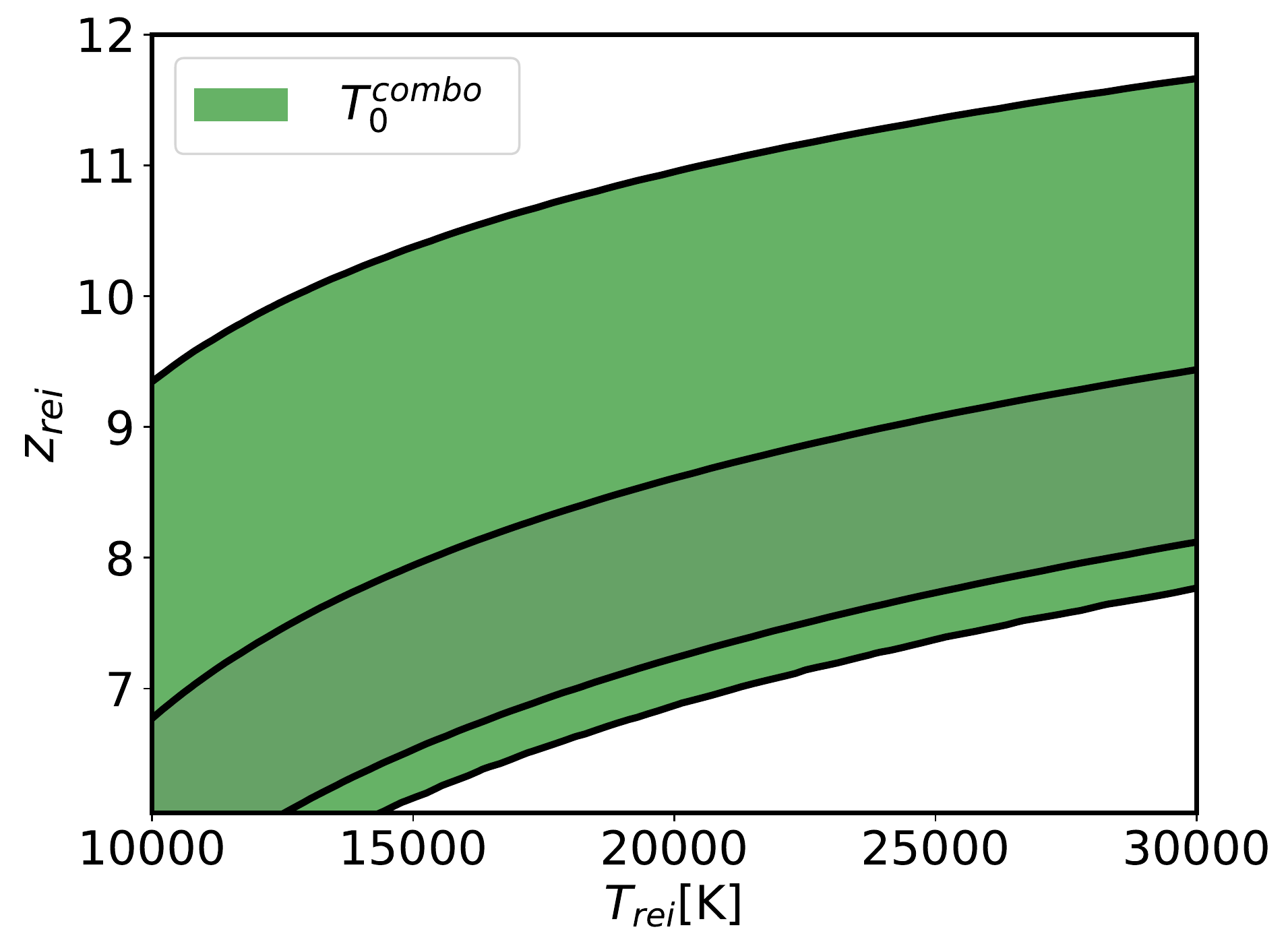}
 
}
\subfigure
{ 
\includegraphics[width=3.3in]{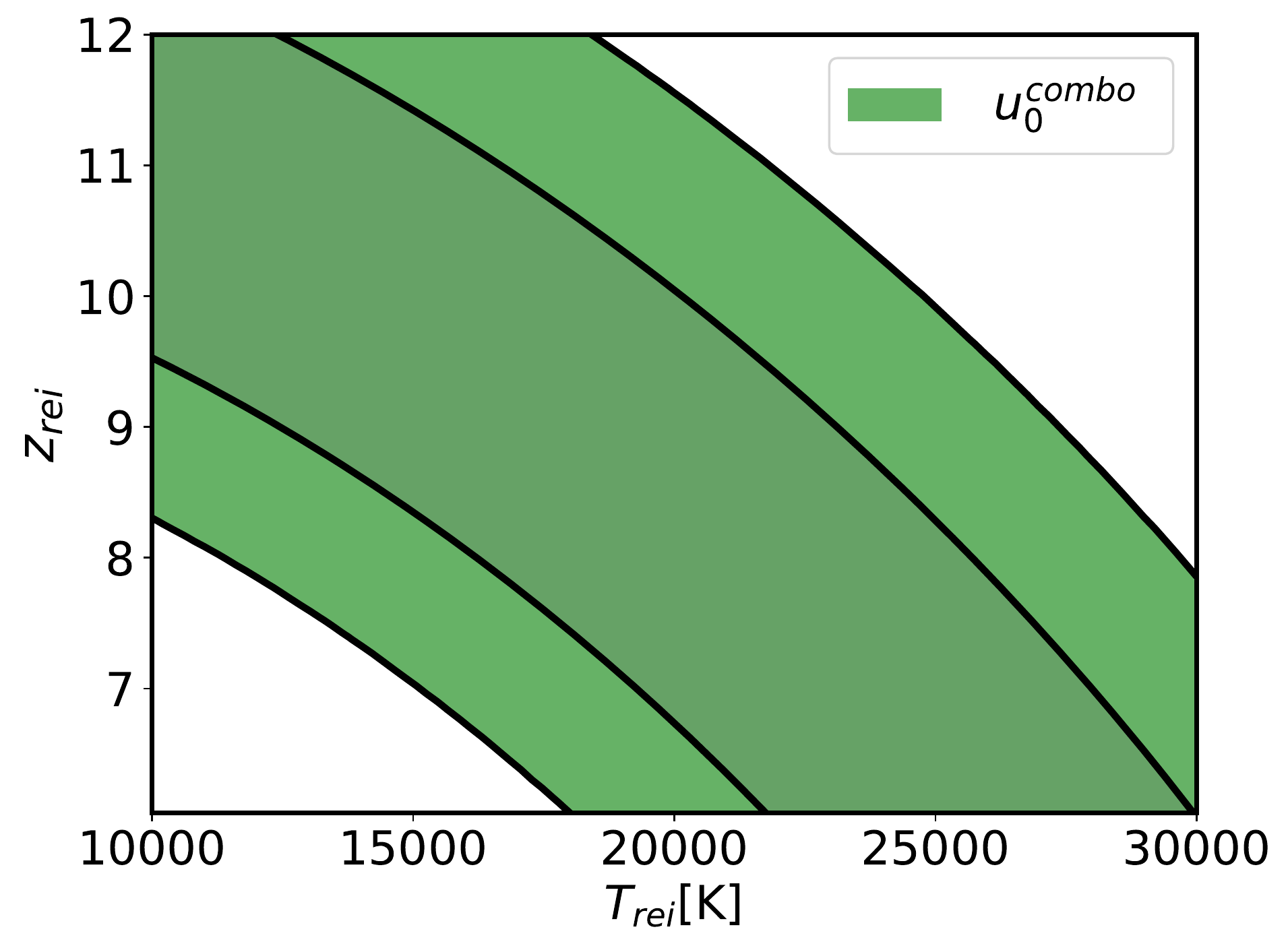} 

} 
\subfigure
{ 
\includegraphics[width=3.3in]{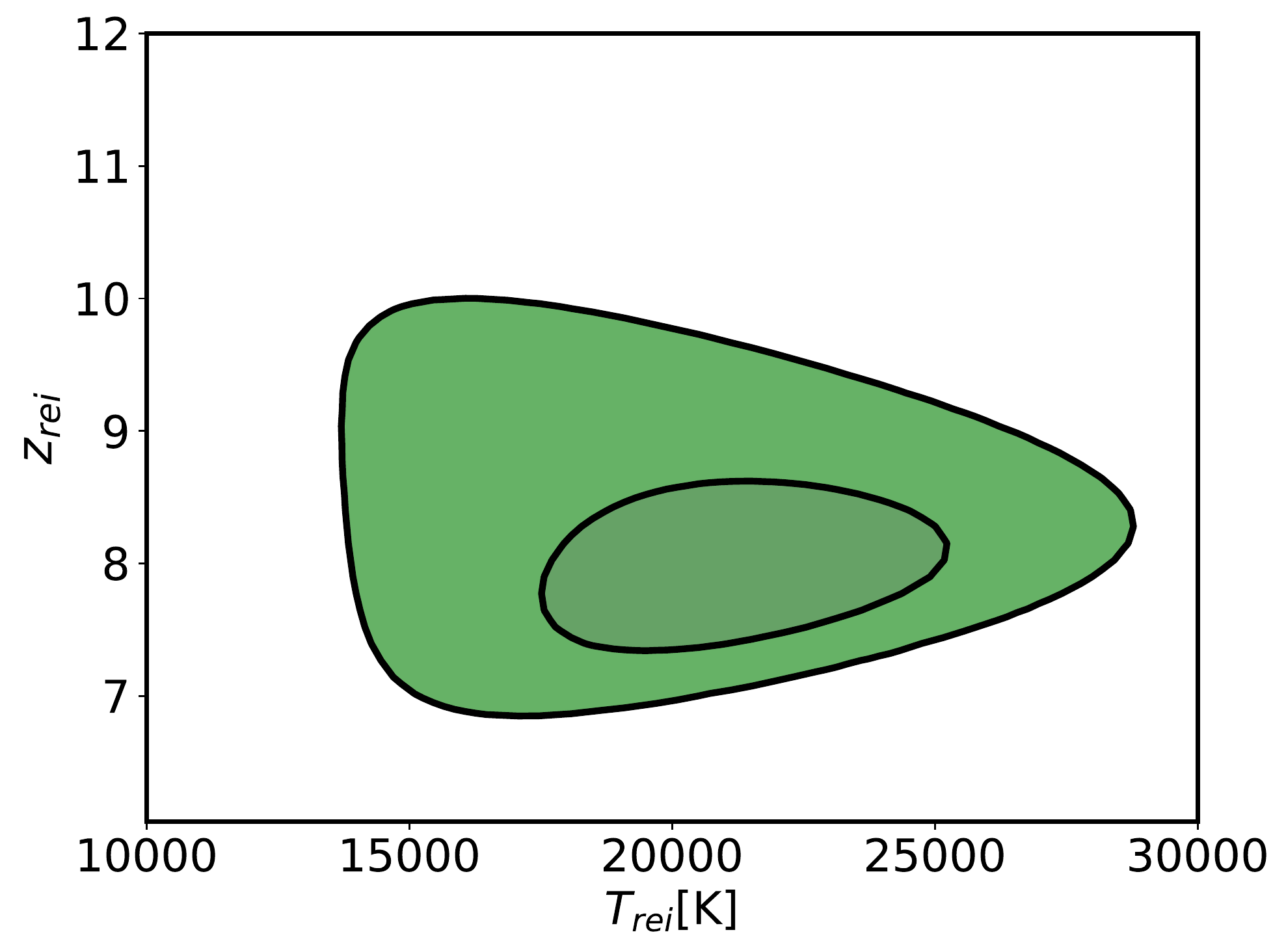} 

} 
 \caption{\small Constraints on instantaneous reionization parameters from our $T_{0}$ and $u_{0}$ measurements. The three panels show the $68\%$ and $95\%$ probability contours in the $T_{\rm rei}$ vs $z_{\rm rei}$ parameter space for $\alpha_{\rm bk}=1.5$, obtained when considering only the temperature results (top panel), only the $u_{0}$ measurements (middle panel) and both together (bottom panel).}

 \label{fig:ConstrainingApproach}
\end{figure}

Before presenting the final results, we stress that, unlike previous attempts to constrain reionization using the instantaneous temperature alone \citep[e.g.,][]{Theuns02,Raskutti12} the power of our approach relies on the simultaneous use of measurements of both $T_{0}$ and $u_{0}$. Figure \ref{fig:ConstrainingApproach} demonstrates how these measurements separately constrain the likelihood contours for our galaxy--driven reionization models. In the top panel the $68\%$ and $95\%$ probability contours are shown for the temperature constraints only, while in the middle panel they are given for the $u_{0}$ constraints only. The models that better fit the $T_{0}$ and $u_{0}$ data cover two different but intersecting regions in the $T_{\rm rei}$ vs $z_{\rm rei}$ parameter space. Applying both constraints
 simultaneously therefore, reduces the allowed parameter space considerably (bottom panel of Figure \ref{fig:ConstrainingApproach}). 

\begin{figure*} 
\centering
\includegraphics[width=1.7\columnwidth]{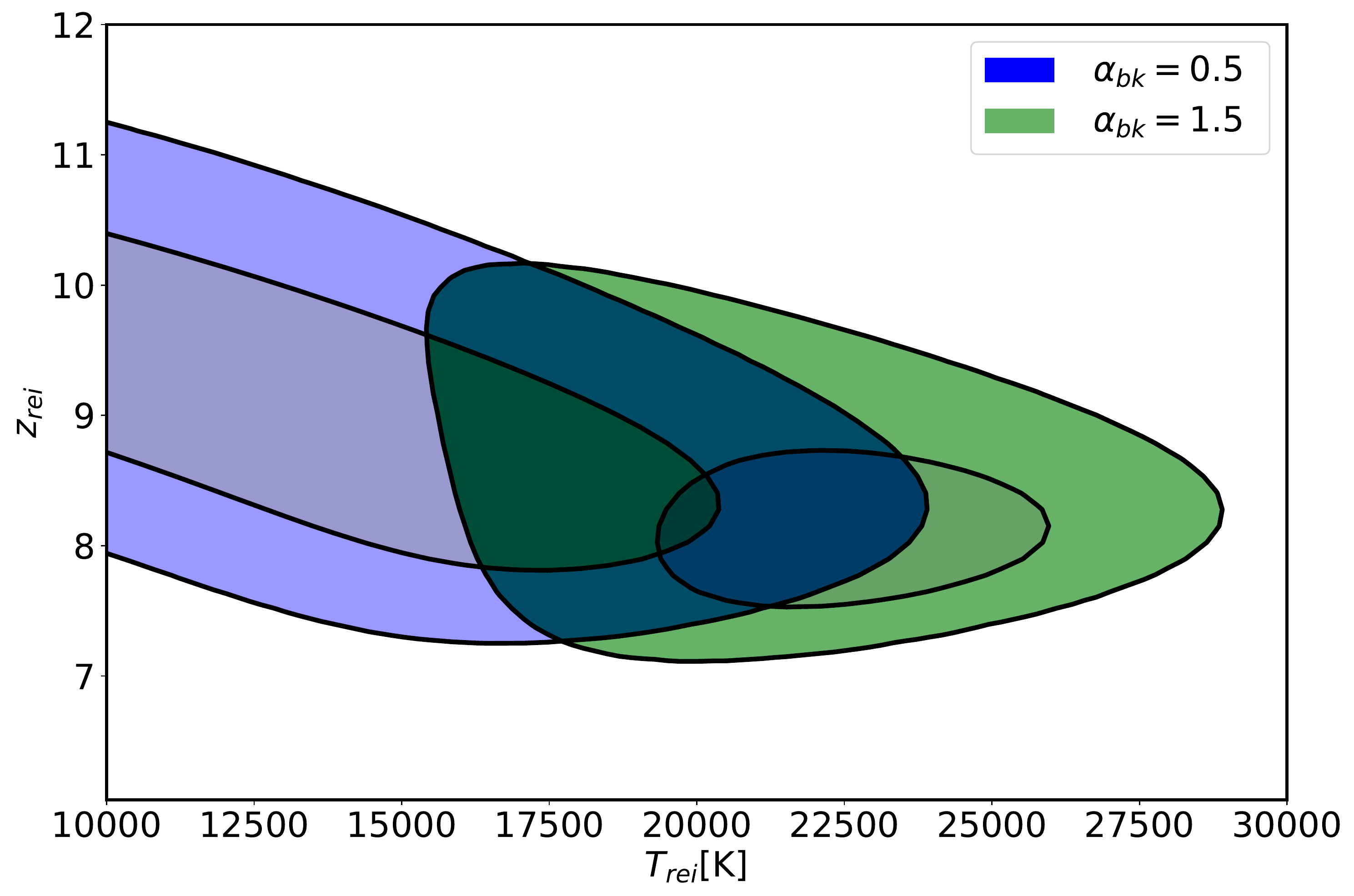} 
\caption{\small Constraints on instantaneous reionization models for two choices of the post--reionization ionizing background spectrum. The two--dimensional $68\%$ and $95\%$ probability contours in the $T_{\rm rei}$ vs $z_{\rm rei}$ parameter space are reported for softer ($\alpha_{\rm bk}=1.5$; green contours) and harder ionizing ($\alpha_{\rm bk}=0.5$; blue contours) backgrounds. }
\label{fig:Resultsno42}
\end{figure*}

Figure \ref{fig:Resultsno42} shows the final $68\%$ and $95\%$ two--dimensional probability contours for models of instantaneous reionization driven by softer ($\alpha_{\rm bk}=1.5$; green contours) and harder sources ($\alpha_{\rm bk}=0.5$; blue contours).
 For softer sources the favored models are the ones with $z_{\rm rei}\sim 8$ and reionization temperature of 20,000 K $\lesssim T_{\rm rei}\lesssim25,000$ K. These temperatures are consistent with the values predicted by radiative transfer models \citep[e.g.,][]{MiraldaRees94,McQuinn12,Daloisio18}.
For harder sources the thermal data prefer earlier reionizations and lower temperatures ($T_{\rm rei}\lesssim 20,000$ K). These results are driven by the fact that for harder post--reionization ionizing backgrounds the IGM temperature needs more time to cool in order to match the relatively low values observed at $z\lesssim5$. We note that even lower $T_{\rm rei}$ would be needed to fit the observations if we included the contribution of He{\sc \,ii} photo-heating, which has been conservatively excluded.  
Radiative transfer calculations may disfavor $T_{\rm rei}\lesssim 17,000$ K, as this would imply reionization front speeds unexpectedly low even for the early stages of reionization \citep{Daloisio18}. Some of the parameter space in Figure \ref{fig:Resultsno42} preferred by harder sources may therefore be disfavored on physical grounds.

In Figure \ref{fig:reiModels}, we also compare our constraints to the empirically calibrated UV background model recently presented by \citealt{Puchwein18}. Their predictions for the instantaneous temperature are larger than our measurements, and are inconsistent with our new
constraints at around the $2\sigma$ level. Furthermore, the cumulative
energy input into the IGM at mean density also exceeds our constraint
at $z=6$, and is again inconsistent at around $2\sigma$. This suggests
that there is slightly too much IGM heating at $z>6$ in the fiducial
\cite{Puchwein18} model, under the assumption of a $\Lambda$-CDM
cosmology. This difference will be exacerbated further for models
where AGN provide a substantial contribution to the photon budget for
reionization.

\begin{figure*} 
\centering
\includegraphics[width=1.7\columnwidth]{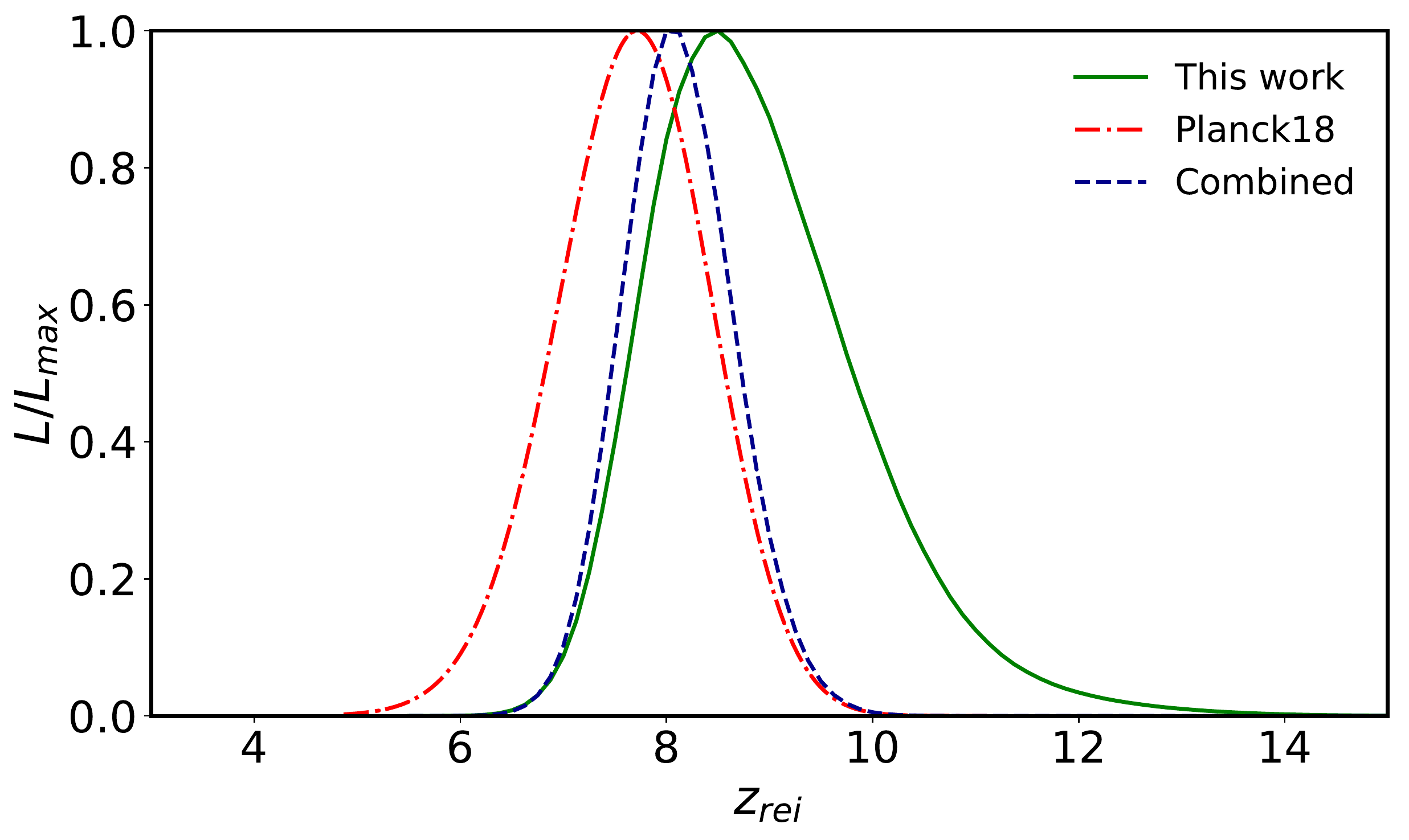} 
\caption{\small The 1D probability distribution for $z_{\rm rei}$ obtained in this work after marginalizing over $T_{\rm rei}$ and $\alpha_{\rm bk}$ (green solid line) and from the Thompson optical depth constraints of \citealt{Planck18} (red dash-dotted line), assuming an instantaneous reionization in both cases. The combined probability distribution distribution is plotted as a blue dashed line.}
\label{fig:PlanckComp}
\end{figure*}

Finally, we compare our results to constraints on the redshift of instantaneous reionization derived from the most recent Planck measurements of the Thompson scattering optical depth \citep{Planck18}.
We marginalized over $T_{\rm rei}$ and $\alpha_{\rm bk}$ (where $\alpha_{\rm bk}$ was allowed to vary from 0.5 to 1.5) to obtain the 1D probability distribution on $z_{\rm rei}$ from the IGM thermal history.
Figure \ref{fig:PlanckComp} compares our marginalized constraints on $z_{\rm rei}$ (green solid line) to those derived for an instantaneous reionization from the Planck baseline optical depth constraint $\tau_{\rm e}=0.0544\pm0.0073$ (based on $Plank$ TT,TE,EE+lowE+lensing; \citealt{Planck18}) (red dot-dashed line).
The two distributions are broadly consistent and have comparable constraining power, with $z_{\rm rei}\simeq 8.5^{+1.1}_{-0.8}$ from the thermal history and $z_{\rm rei}\simeq 7.7^{+0.7}_{-0.7}$ from the Planck results. The combined probability distribution (blue dashed line) give $z_{\rm rei}\simeq8.1 ^{+0.5}_{-0.5}$.
While we emphasize that the instantaneous reionization adopted in this Section is simplistic, the proof of concept presented here demonstrates the potential of our observational constraints to inform reionization scenarios.


\section{CONCLUSIONS}
\label{sec:conclusions}
In this work we have presented the first simultaneous constraints on the instantaneous temperature and integrated thermal history of the IGM at $z>4$ and demonstrated how these results can be used to test different scenarios of hydrogen reionization.
We have utilized a sample of 15 Keck/HIRES and VLT/UVES high--resolution and high--C/N spectra to obtain new measurements of the Ly$\alpha$ forest flux power spectrum over redshifts $4.0\lesssim z\lesssim5.2$, for the first time pushing the measurement down to the smallest scales currently accessible to high--resolution quasar spectra at these redshifts ($\log(k/$km$^{-1}$s$)\le -0.7$). 
We fit the new flux power spectra to obtain robust constraints on the instantaneous IGM temperature, $T_{0}$, and integrated energy input per unit mass, $u_{0}$, marginalizing over the slope of the T--$\rho$ relation and the effective optical depth, and assuming a $\Lambda$-CDM cosmology.

In agreement with previous results from the curvature method \citep{Becker11}, we find temperatures of $T_{0}\sim 7000-8000$ K and no strong temperature evolution over $4.2\lesssim z\lesssim5.0$. Our first constraints on $u_{0}$ show a significant increase from $u_{0}\sim4.5$ eV $m_{\rm P}^{-1}$ at $z > 6$ to 7.1 eV $m_{\rm P}^{-1}$ at $z>4.2-4.6$. These results are consistent with a heating from reionization at $z > 6$ and residual photo-ionization heating over $z\sim 6$ to 4.

Our constraints on $T_{0}$ and $u_{0}$ can be used to test any reionization scenario for which the temperature and the energy injection into the IGM can be calculated. 
As a proof of concept we analyzed simplistic, semi-analytical models of instantaneous reionization. These toy models depend on three parameters, the IGM temperature reached during reionization ($T_{\rm rei}$), the redshift of reionization ($z_{\rm rei}$) and the spectral index of the post--reionization UV background ($\alpha_{\rm bk}$), which is related to the sources driving the reionization process. 
We find that our measurements prefer instantaneous reionization redshifts near $z_{\rm rei}\sim8$ with $T_{\rm rei} \sim 20,000$ K for a relatively soft UV--background dominated by ionizing photons from star--forming galaxies. Our fully marginalized constraints on the reionization redshift, $z_{\rm rei}\simeq8.5^{+1.1}_{-0.8}$, are moreover comparable with those from recent Planck results.

While tests of more realistic scenarios of reionization are left for future work, the proof of concept presented here, demonstrates the potential of the IGM thermal history at high redshift to impose tight constraints on the timing --and possibly the sources-- of reionization.

\section*{acknowledgments}
We thank Simeon Bird and Anson D'Aloisio for helpful conversations.
EB and GDB were supported by the National Science Foundation through grant AST-1615814. JSB acknowledges the support of a Royal Society University Research Fellowship. This work is based on observations made at the W.M. Keck Observatory, which is operated as a scientific partnership between the California Institute of Technology and the University of California; it was made possible by the generous support of the W.M. Keck Foundation. It also includes observations made with ESO Telescopes at the La Silla Paranal Observatory under program ID 092.A-0770. EB thanks Michael Murphy and the Centre of Astrophysics and Supercomputing at Swinburne for granting the access to the Swinburne supercomputer facility during the preparation of this work.
The hydrodynamical simulations used in this work were performed with
supercomputer time awarded by the Partnership for Advanced Computing
in Europe (PRACE) 8th Call. We acknowledge PRACE for awarding us
access to the Curie supercomputer, based in France at the Tres Grand
Centre de Calcul (TGCC). This work also made use of the DiRAC High
Performance Computing System (HPCS) at the University of
Cambridge. These are operated on behalf of the STFC DiRAC HPC
facility. This equipment is funded by BIS National E-infrastructure
capital grant ST/J005673/1 and STFC grants ST/H008586/1,
ST/K00333X/1. We thank Volker Springel for making P-GADGET-3
available.

\begin{appendices}
\appendix
\section{Systematic effects}
\label{sec:Systematics}
In this Appendix we review some of the steps of our analysis to check and quantify possible systematic uncertainties arising from the specific strategies adopted. 
\label{sec:SysTest}
\subsection{Rolling mean}
\label{sec:RollSys}
As described in Section \ref{sec:Rolling} we computed the flux contrast $\delta_{F}$ of Equation \ref{eq:df} using a rolling mean along the entire Ly$\alpha$ forest region. 
The application of this technique on both simulated and observational lines of sight guarantees a fair comparison between models and the real measurement when the continuum level is unknown, but it may introduce possible bias when comparing our power spectrum with previous works in which the power was computed from continuum--normalized spectra.
Using the simulations we tested different averaging functions and window sizes in order to minimize the impact of the rolling mean on the power spectrum at the relevant redshifts and to verify the ability of the rolling mean to capture continuum fluctuations.
As demonstrated below, we found that a 40 $h^{-1}$cMpc boxcar rolling mean is able to recover the power at all relevant scales even in the presence of continuum fluctuations. 
 
\begin{figure} 
\centering 
\subfigure
{ 
\includegraphics[width=3.4in]{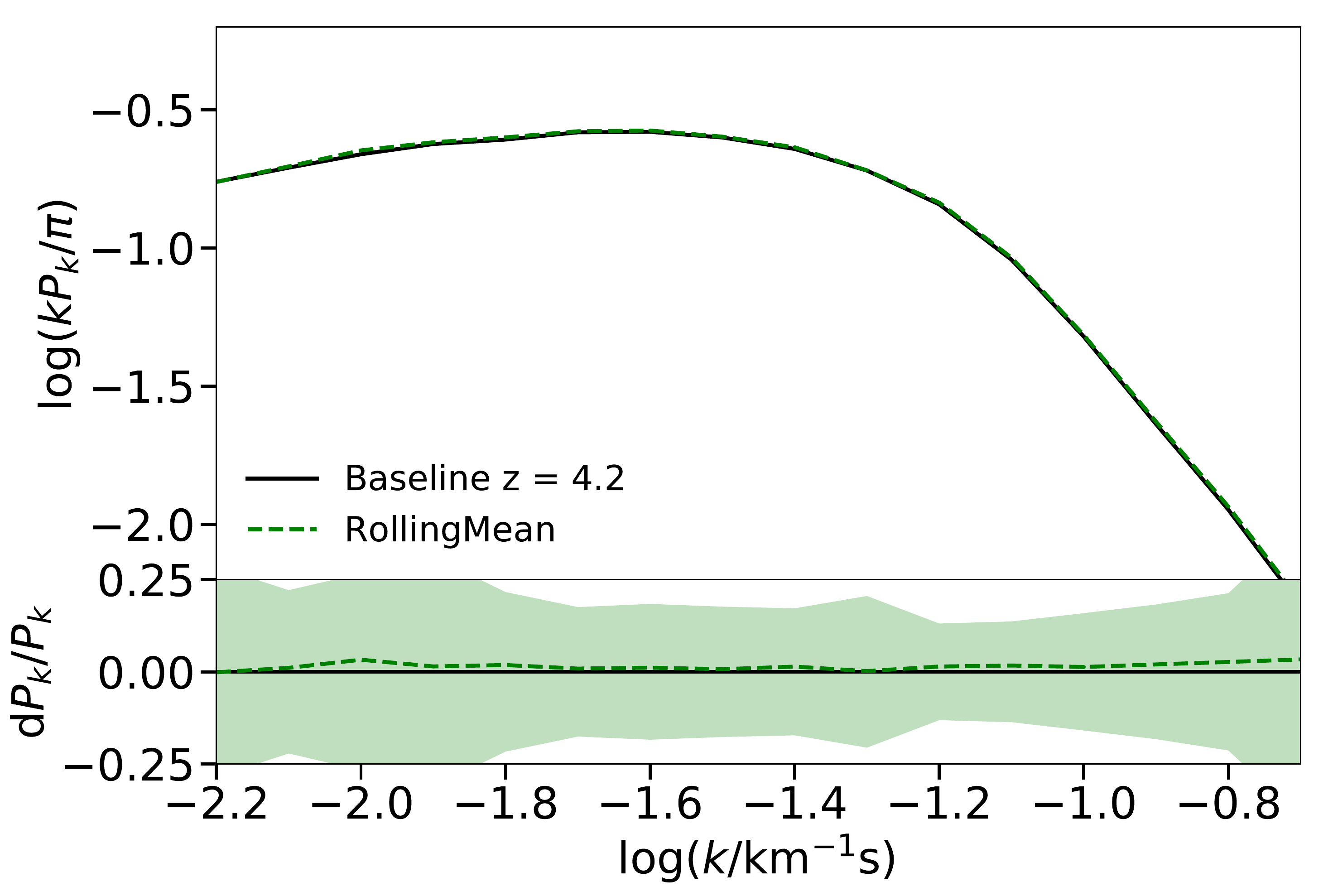}
 \label{fig:z4.2}
}
\subfigure
{ 
\includegraphics[width=3.4in]{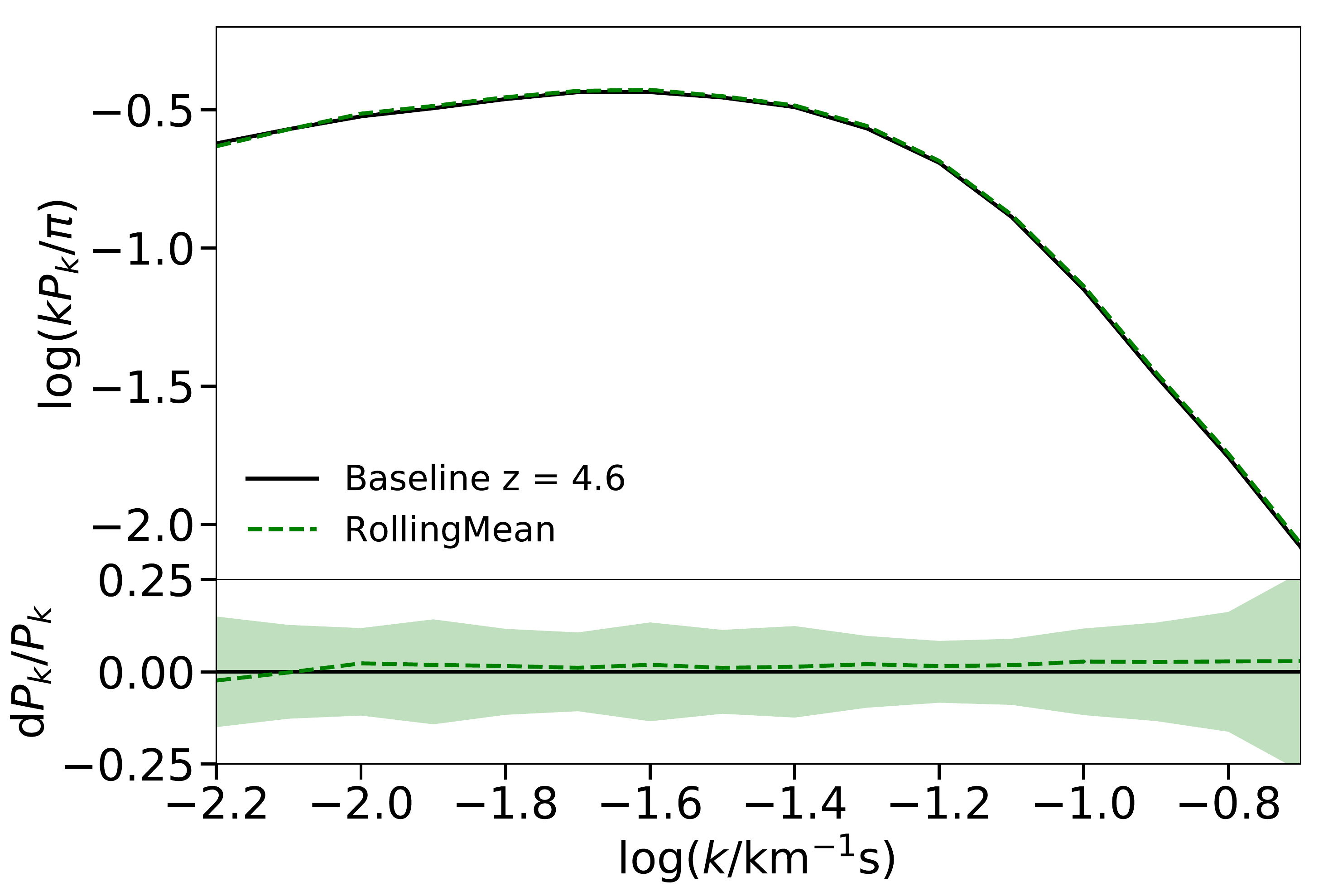} 
\label{fig:z4.6}
} 
\subfigure
{ 
\includegraphics[width=3.4in]{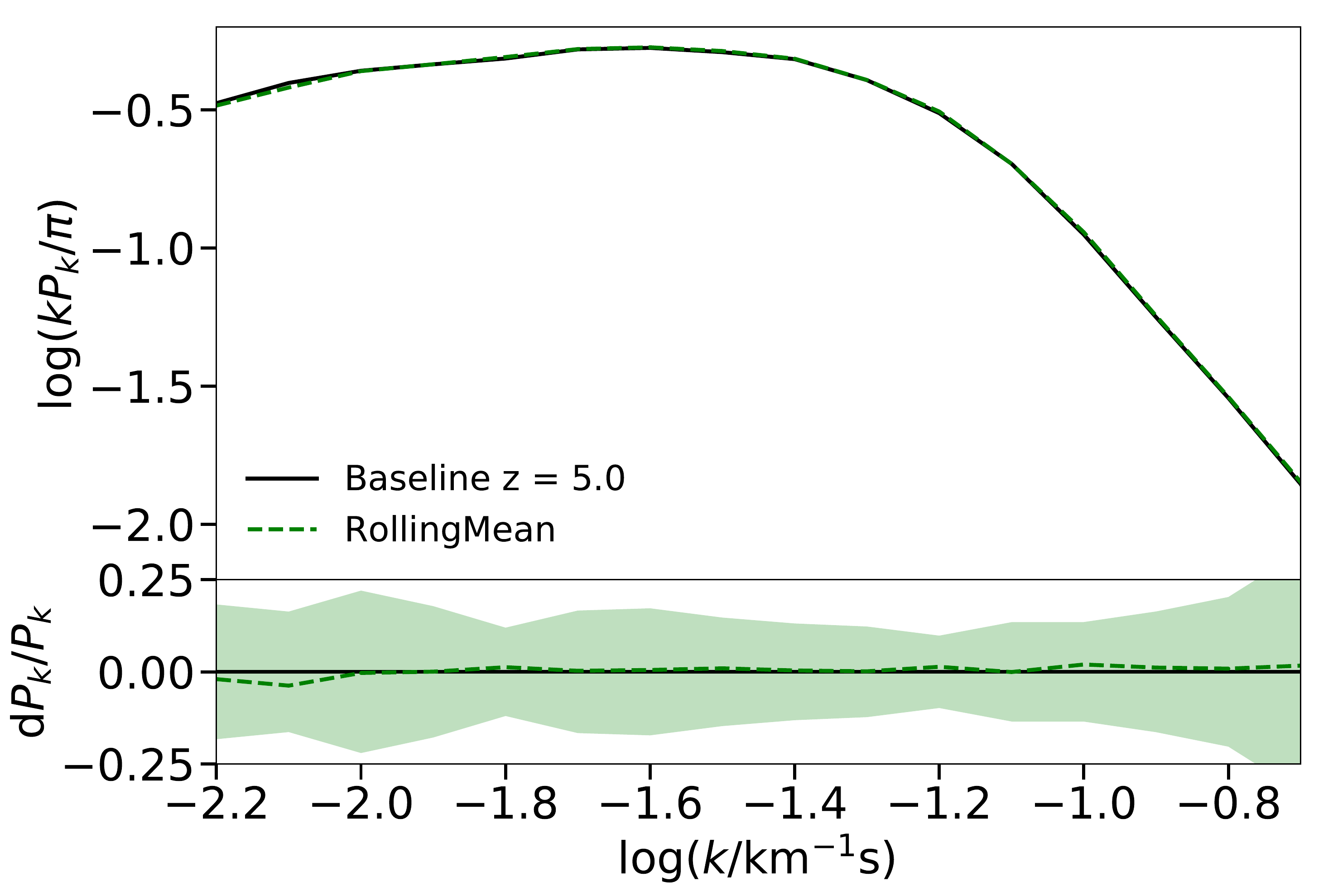} 
\label{fig:z5.0}
} 
 \caption{\small Effect of the rolling mean on the power spectrum measurements. At each of the three redshift bins we show the power spectrum computed using a 40 $h^{-1}$cMpc boxcar rolling mean (green dashed line) compared with the values obtained using a standard global flux average (black solid line). For the rolling mean calculation a random quasar continuum has been imposed on each synthetic line of sight. Differences between the two power spectra are reported in the bottom panel of each plot and are compared with the $68\%$ statistical error for our observed sample (green shaded region).}
 \label{fig:RollingMean}
\end{figure}

Figure \ref{fig:RollingMean} shows, for our three redshift bins, the comparison between the power spectrum computed from simulated data sets using the rolling mean technique (green dashed line) and using a fixed mean flux (black solid line). Both the synthetic samples of lines of sight used in this test have been created following the procedure described in Section \ref{sec:simModel}. 
For the rolling mean model, we first imposed on each of the lines of sight a random continuum selected from the real continua fitted for the XQ-100 survey \citep{Lopez16}. We then run the 40 $h^{-1}$cMpc boxcar rolling mean directly on the total Ly$\alpha$ + continuum flux.
Differences between the two power spectra are shown in the bottom panel of each plot and compared with the statistical error characterizing our observational sample (green shaded region; see Section \ref{sec:CovMatrix}). We note that, at all redshifts, the discrepancies between the two models always lie well within the statistical error, with systematic uncertainties $\sigma_{roll}$ typically $\lesssim 0.20\sigma_{stat}$. We therefore do not expect our results to be sensitive to this averaging choice.

\subsection{Windowing effects}
\label{sec:Windowing}
As explained in Section \ref{sec:sections} we compute the observational power spectrum in 20 $h^{-1}$cMpc sections of Ly$\alpha$ forest. Dividing the spectra into many small regions may introduce artificial excess power at intermediate and small scales due to a windowing effect. This effect does not not appear in the simulations because of their periodicity. 

\begin{figure} 
\centering 
\subfigure
{ 
\includegraphics[width=3.4in]{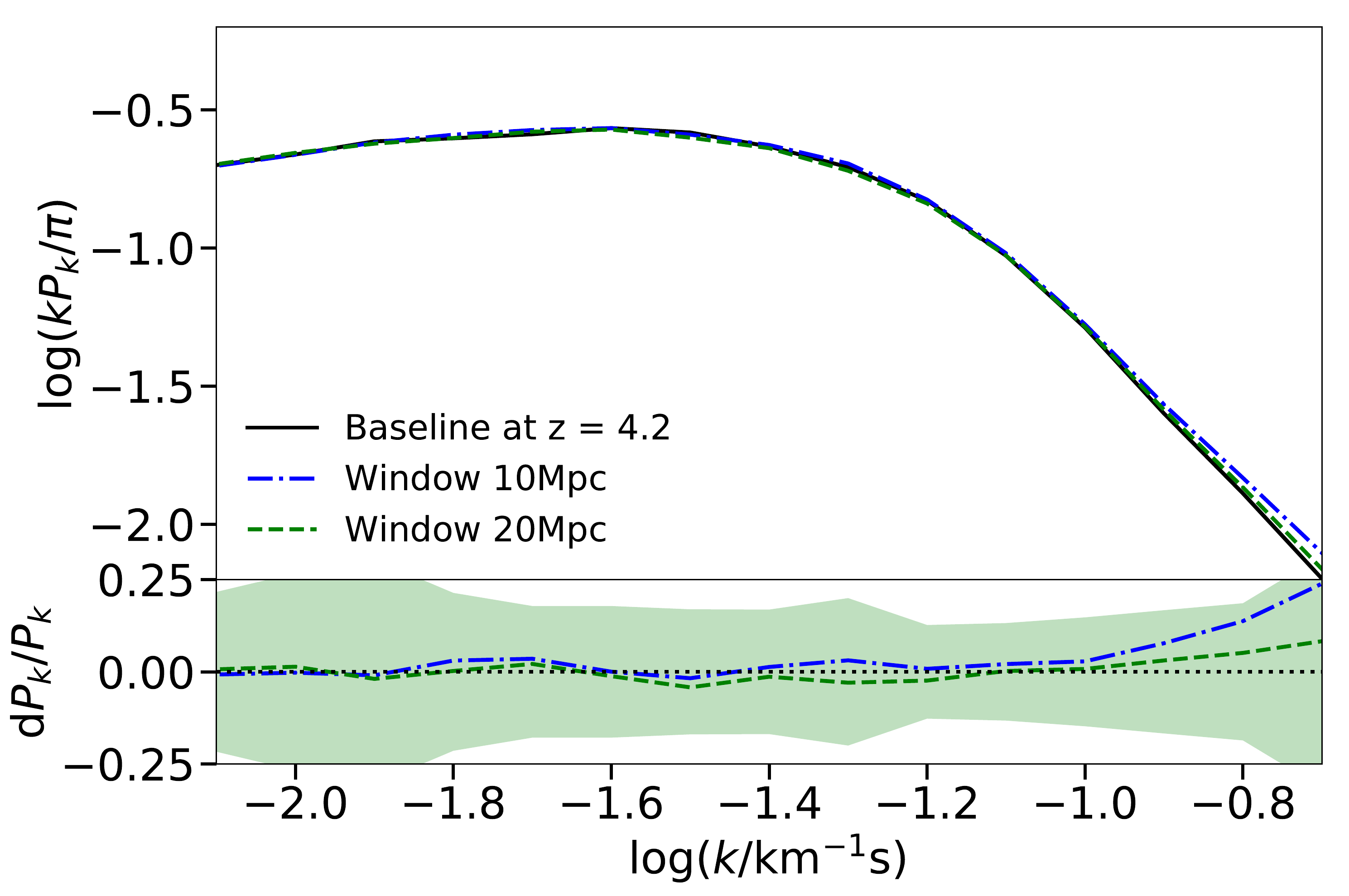}
 \label{fig:z4.2}
}
\subfigure
{ 
\includegraphics[width=3.4in]{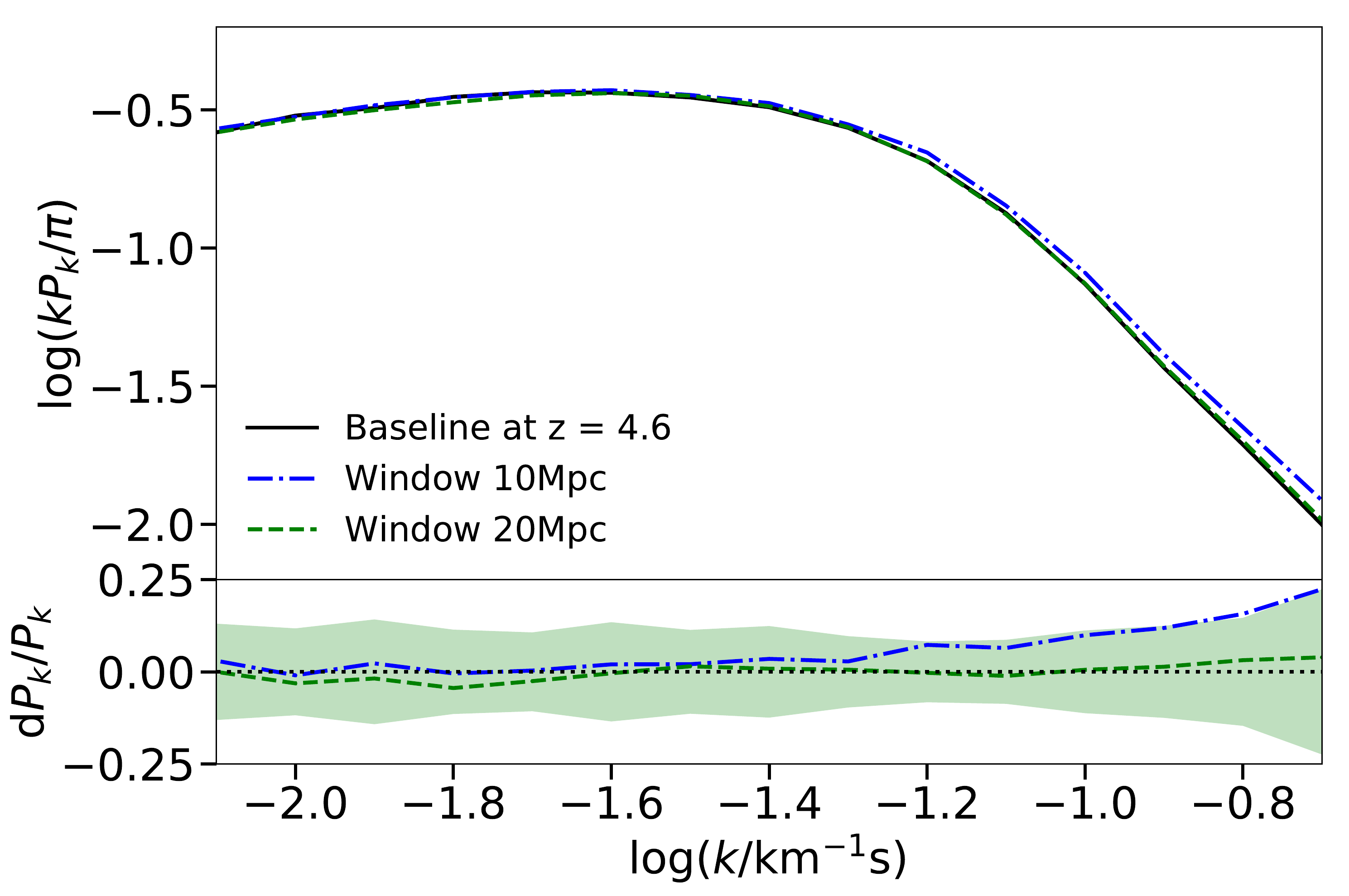} 
\label{fig:z4.6}
} 
\subfigure
{ 
\includegraphics[width=3.4in]{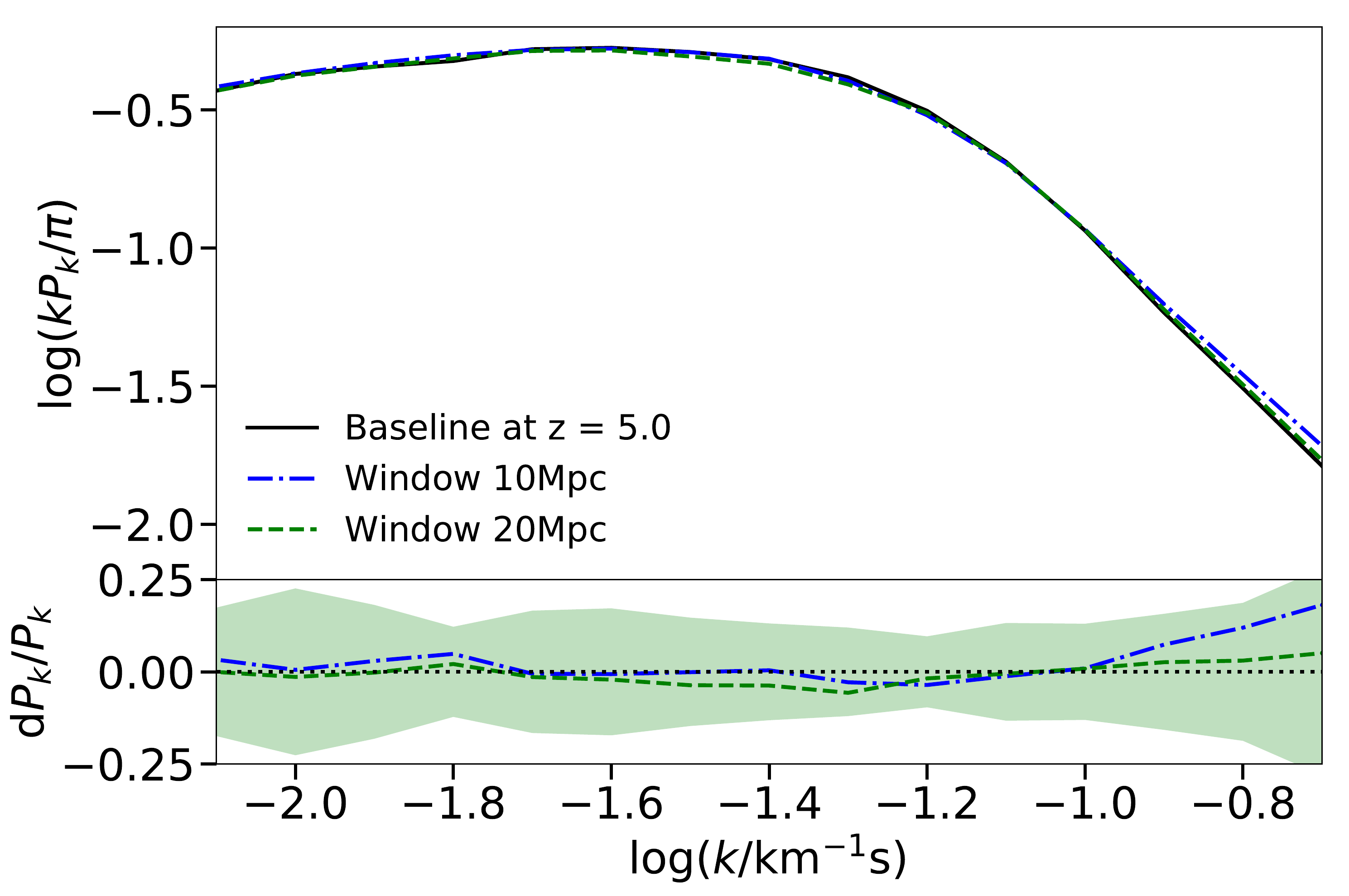} 
\label{fig:z5.0}
} 
 \caption{\small Windowing effect on the flux power spectrum. At each redshift, the baseline model (black solid line) has been computed from the largest simulation box (40 $h^{-1}$cMpc). We also show the power spectrum computed from the same simulation box but after dividing each of the native spectra into smaller sections of 10 $h^{-1}$cMpc (blue dot--dashed line) and 20 $h^{-1}$cMpc (green dashed line). Changes in the power as fractions of the baseline power spectrum are shown in the bottom panel of each plot. Given the small variations in the power for the 20 $h^{-1}$cMpc sections, we opted for this section size when computing the power spectrum from the observed data. For comparison, the green shaded region in the bottom panel shows the $68\%$ statistical errors for our observational results.}
  \label{fig:WindowEffect}
\end{figure}

In Figure \ref{fig:WindowEffect} we present the effect on the flux power spectrum of dividing the spectra into smaller sections. At each redshift bin, the power computed from the largest 40 $h^{-1}$cMpc simulation box (baseline: solid black line) is compared with the power computed from the same simulation but dividing each of the native synthetic spectra into sections of 10 $h^{-1}$cMpc (blue dot--dashed line) and 20 $h^{-1}$cMpc (green dashed line). Changes in the power are reported as fractions of the baseline power spectrum values in the bottom panel of each plot. While cutting the spectra into 10 $h^{-1}$cMpc sections introduce an excess of power at the small scales ($\log(k/$km$^{-1}$s$)\gtrsim-1.1$) of $\sim$20--25$\%$, the windowing effect effect for the 20 $h^{-1}$cMpc sections is less significant, with variations in the power $\lesssim$8$\%$ at all scales. We therefore opted for this latter section size in our analysis.

\subsection{Noise subtraction}
\label{sec:NoiseTreat}
\begin{figure} 
\centering 
\subfigure
{ 
\includegraphics[width=3.4in]{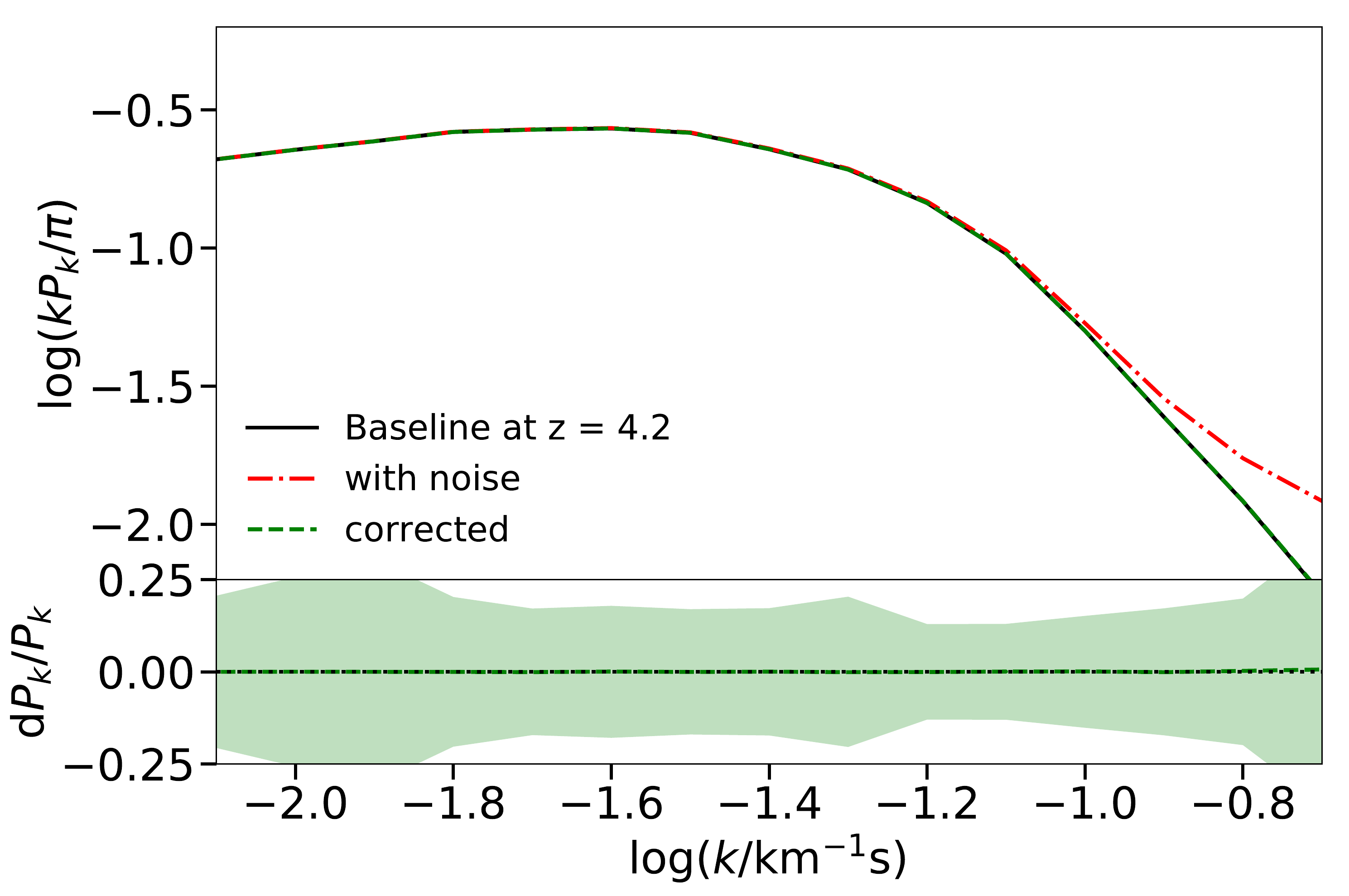}
 \label{fig:z4.2}
}
\subfigure
{ 
\includegraphics[width=3.4in]{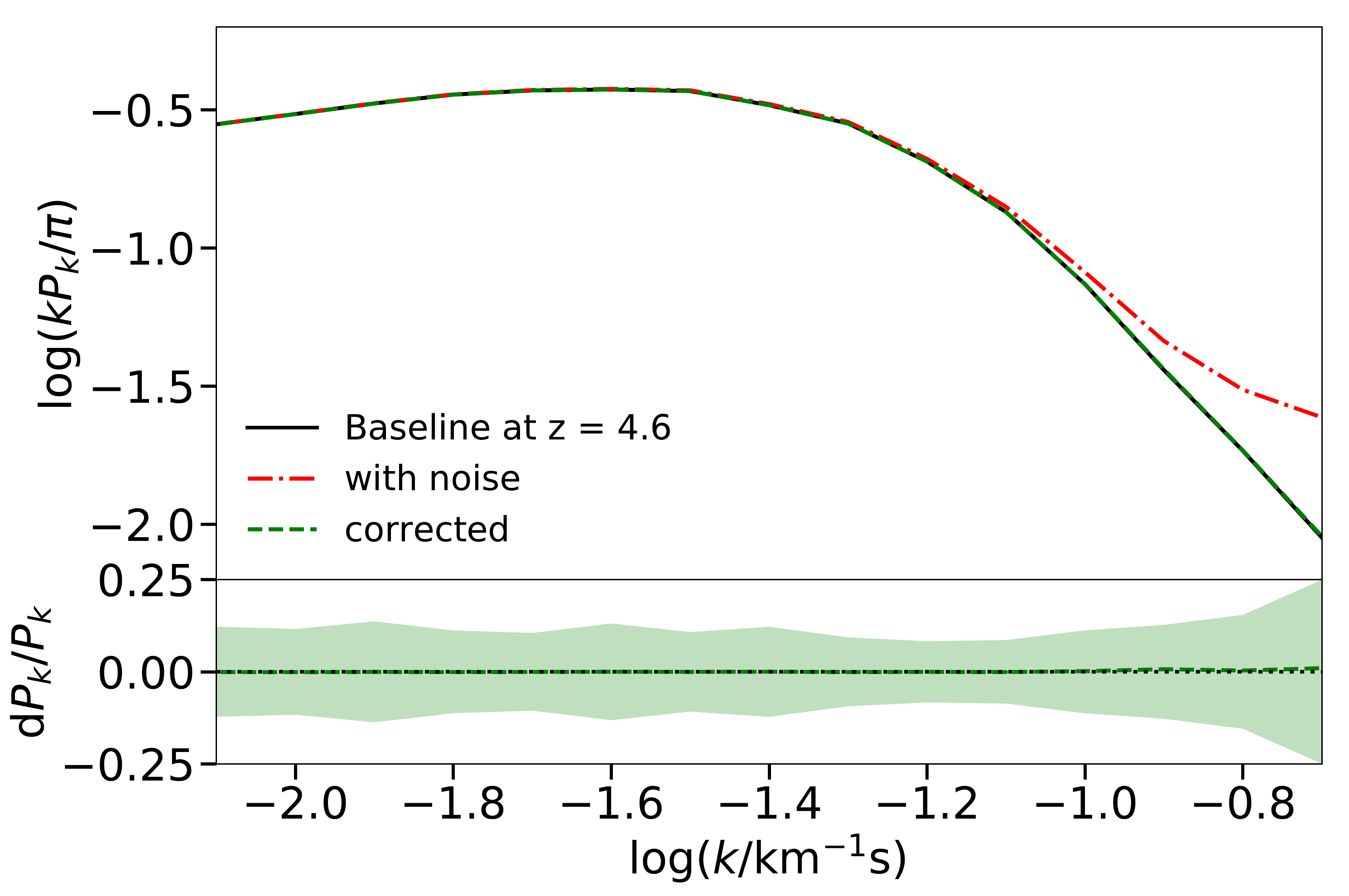} 
\label{fig:z4.6}
} 
\subfigure
{ 
\includegraphics[width=3.4in]{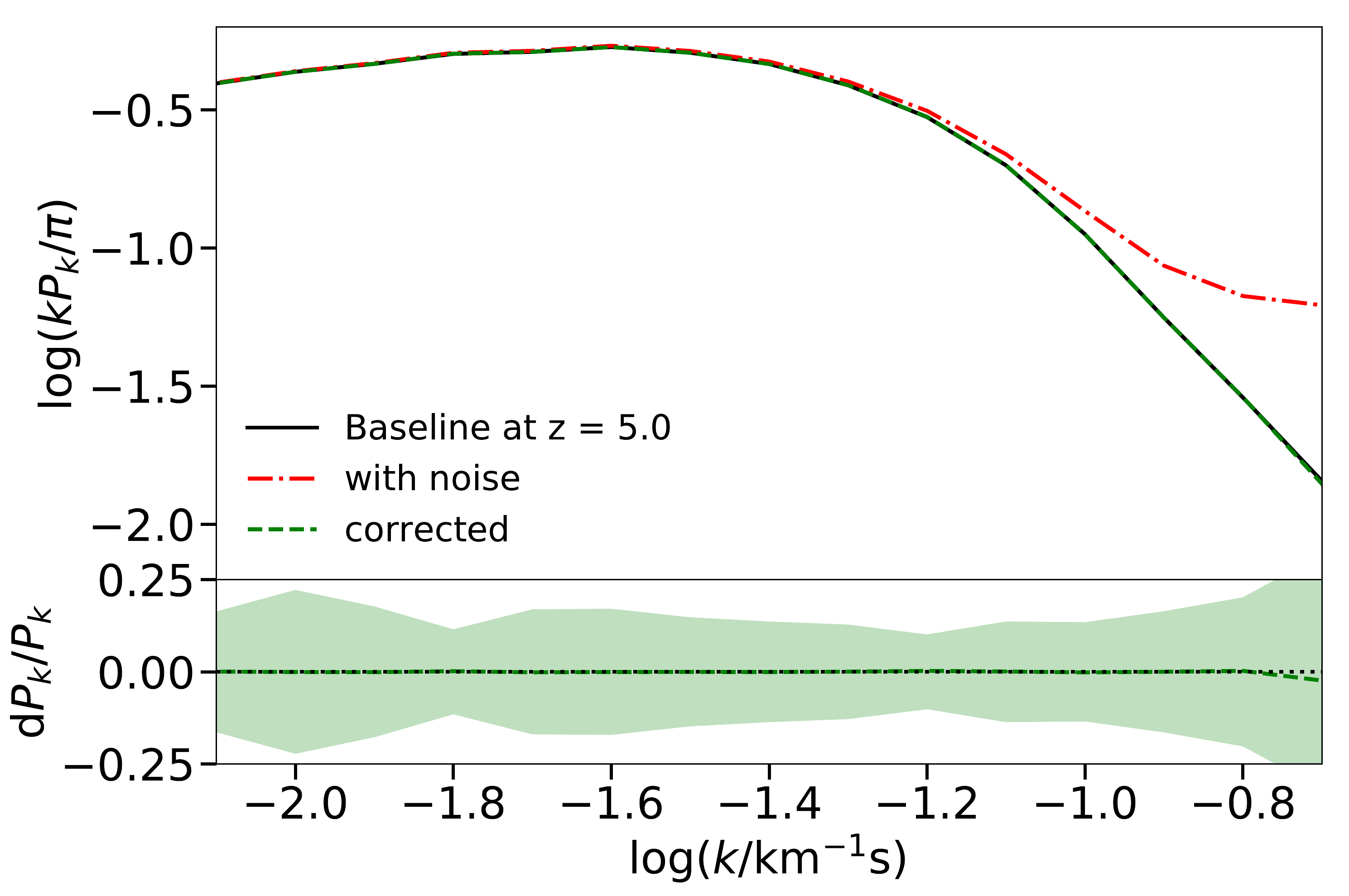} 
\label{fig:z5.0}
} 
 \caption{\small Noise power subtraction test. At each of the three redshift bins we show the power spectrum model for the 20 $h^{-1}$cMpc simulation box with (red dot-dashed line) and without (baseline; black solid line) the addition of realistic noise (see text for details). Applying the noise power subtraction method described in Section \ref{sec:Noise} to the noisy model we obtain the corrected power spectrum (green dashed line), which is compared with the baseline in the bottom panel of each plot. The noiseless power is recovered with high precision, with $ \lesssim 2\%$ changes from the baseline  power at all scales. For comparison, the statistical $68\%$ errors for the observational results are also shown (green shaded regions). }
  \label{fig:NoiseSub}
\end{figure}

In this Appendix we test the noise power subtraction method described in Section \ref{sec:Noise} using synthetic datasets generated from simulations.
For each synthesized line of sight we use the error and the flux arrays of one of our observed spectra to fit a linear correlation between the signal and the noise level. Using these correlations we then add noise to the simulated samples in a flux--dependent way. 
We construct synthetic samples of lines of sight following the procedure described in Section \ref{sec:syntheticLOS} using our 20 $h^{-1}$cMpc simulation box and compute the power spectrum with and without adding noise. In Figure \ref{fig:NoiseSub} we show the noisy (red dot-dashed line) and noiseless (baseline; black solid line) power spectra obtained for each redshift bins. 

We finally applied the noise power subtraction method to the noisy model and compare the corrected power spectrum (green dashed line in Figure \ref{fig:NoiseSub}) to the noiseless baseline. The noiseless power is recovered with errors of $\lesssim 2\%$. This suggests that this step of the data analysis is not introducing relevant systematic effects in the final results.

\subsection{Instrumental resolution}
\label{sec:ResCorr}
In this Appendix we review the effect of uncertainties in the instrumental resolution when smoothing the synthetic spectra to match the resolution of the observed data.
 We calibrate the synthetic Ly$\alpha$ forest spectra using different instrumental resolutions, taking our nominal values of ${\rm FWHM} = 6~{\rm km~s^{-1}}$ as a baseline. 

\begin{figure} 
\centering
\subfigure
{ 
\includegraphics[width=3.4in]{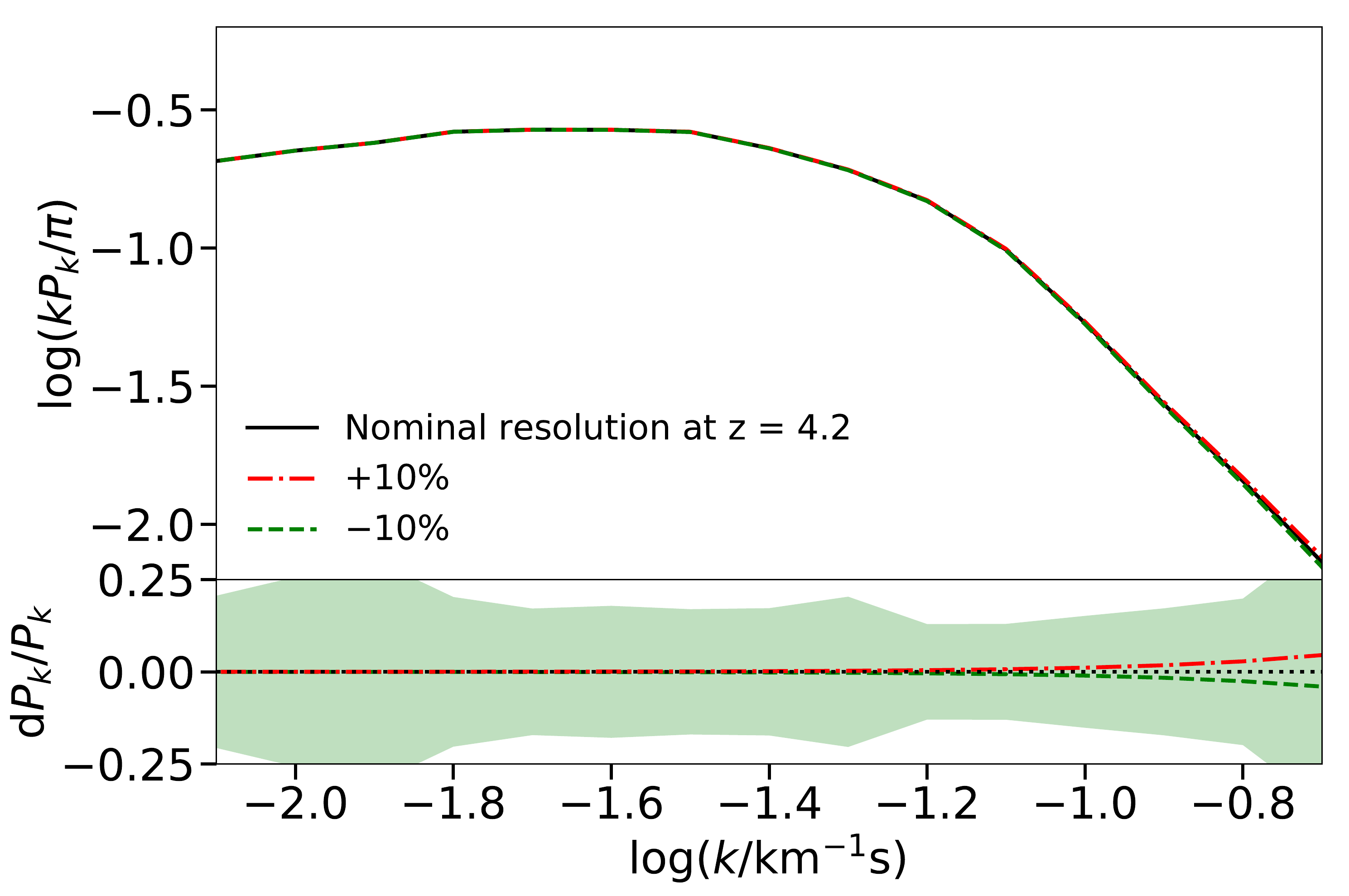}
 
}
\subfigure
{ 
\includegraphics[width=3.4in]{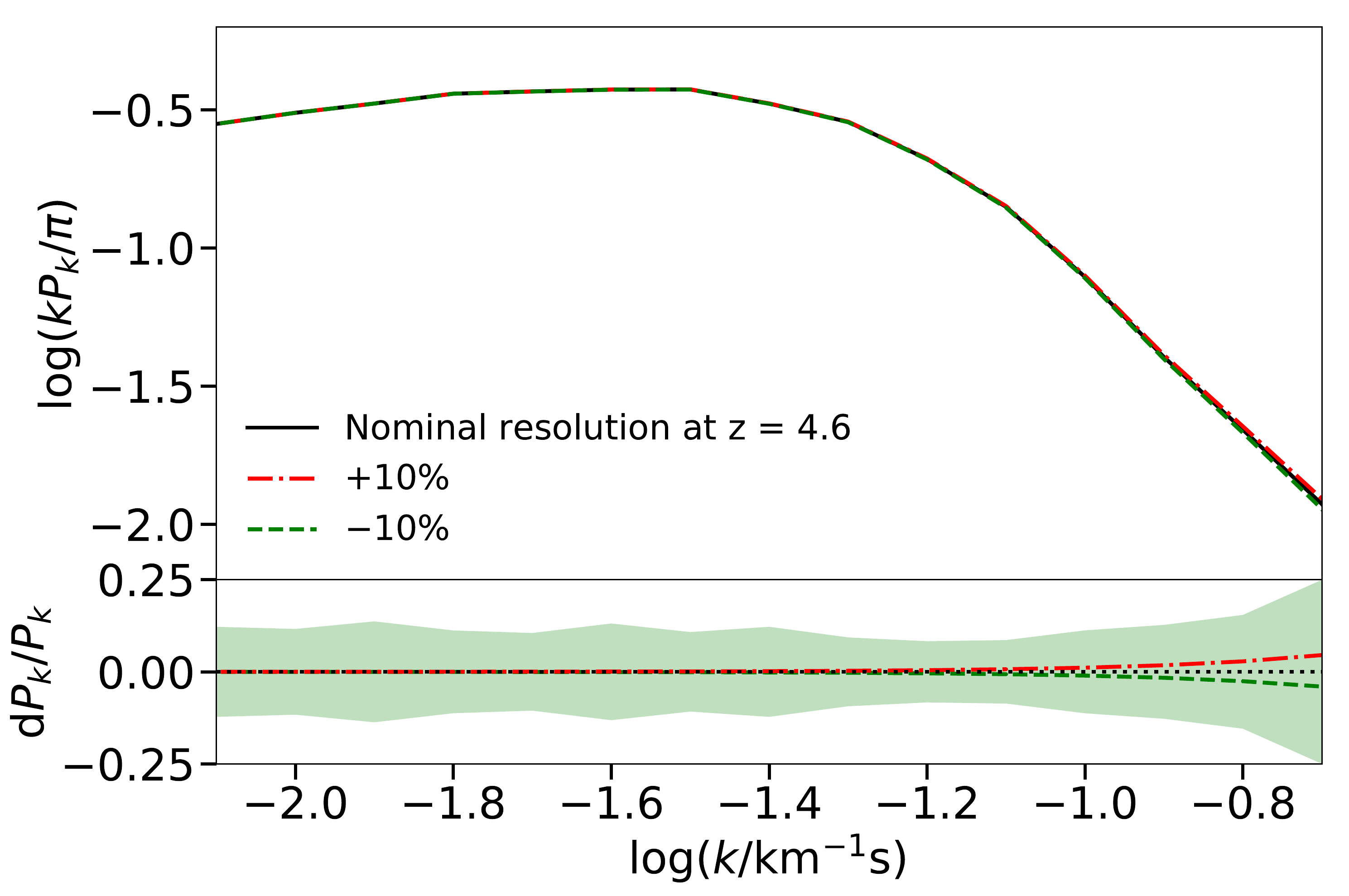} 

} 
\subfigure
{ 
\includegraphics[width=3.4in]{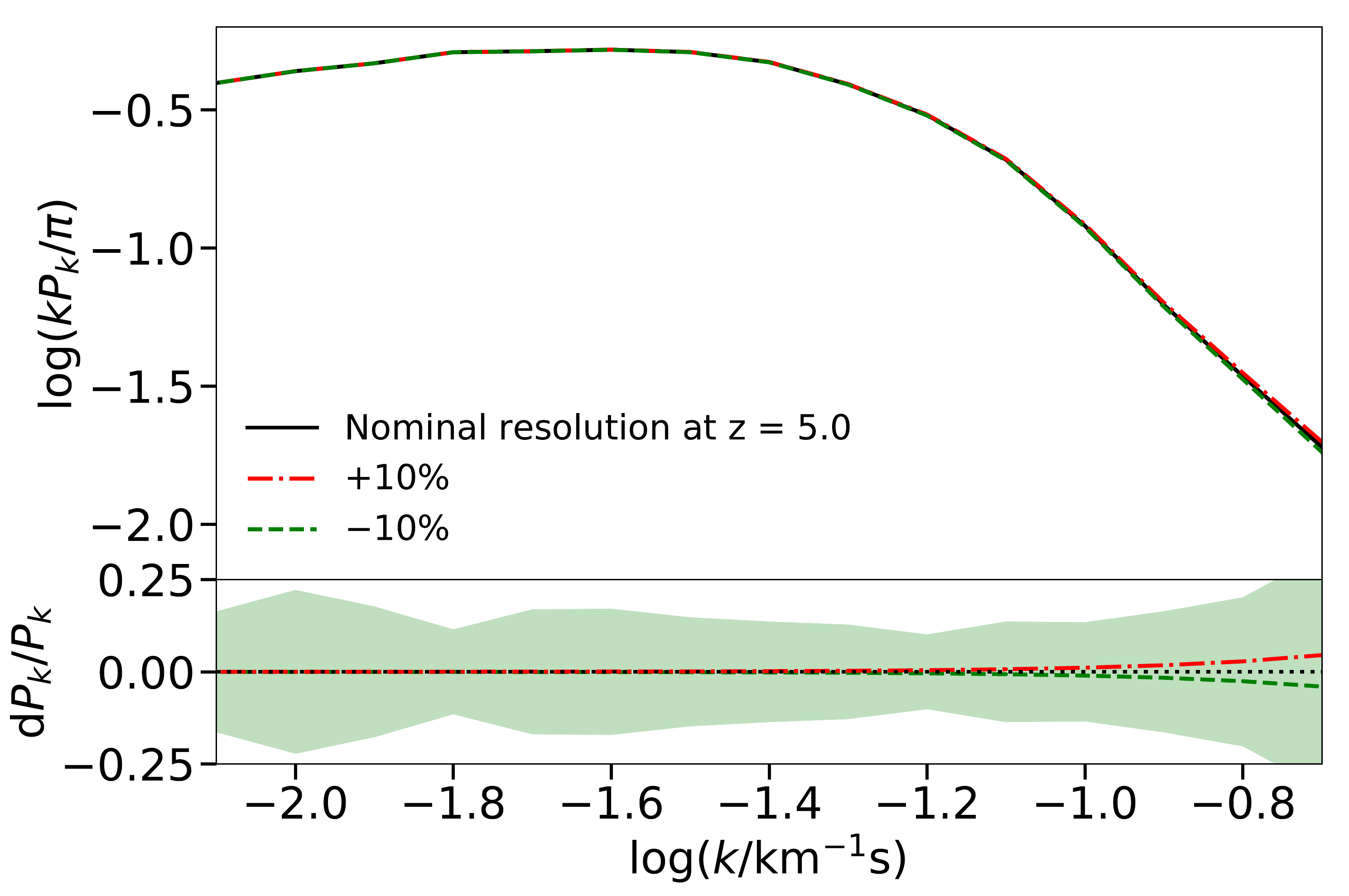} 

} 
 \caption{\small Effect of instrumental resolution on the Ly$\alpha$ flux power spectrum. For each of our redshift bins we show the power spectrum obtained from the 20 $h^{-1}$cMpc simulation box after smoothing the flux spectra by different instrumental resolutions.
The baseline model (black solid line) has been calibrated assuming the correct nominal slit resolution, while a $10\%$ lower (red-dot-dashed line) and higher (green dashed line) resolution has been used for the smoothing of the other two models. Variations in the power spectrum at different scales are reported in the bottom panels as fractions of the baseline power and are $\lesssim$5$\%$ at all scales. For comparison, the statistical $68\%$ errors for the observational results are also shown (green shaded regions).}
 \label{fig:ResCorrection}
\end{figure}

Figure \ref{fig:ResCorrection} shows the variations in the power spectrum expected for a change of $+10\%$ (red-dot-dashed line) and $-10\%$ (green dashed line) in the observed spectral resolution. We note that the power changes by $\lesssim$5$\%$ at all scales (see bottom panel of each plot).  
\subsection{Masking correction function}
\label{sec:mfunction}
In this Appendix we analyze the systematic uncertainties arising from the choice of a particular simulation in computing the masking correction, $C_{m}(k)$, described in Section \ref{sec:MF}. For the fiducial masking correction we adopted the simulation S20$_{-}$1z15 of Table \ref{table:sim}. To demonstrate that this particular choice of model (with $T_{0}\sim7500$ K and $\gamma\sim 1.5$ at the redshifts of interest) does not relevantly affect the final power spectrum measurements, we compare our fiducial results with the measurements obtained when computing the masking correction from post--processed runs with different values of $T_{0}$ and $\gamma$. In particular we tested two extreme cases: a colder model, $C^{test1}_{m}(k)$, with $T_{0}\sim 4000$ K and $\gamma\sim1.5$, and an isothermal model, $C^{test2}_{m}(k)$, with $T_{0}\sim7500$ K and $\gamma \sim1.0$.

\begin{figure} 
\centering
\subfigure
{ 
\includegraphics[width=3.4in]{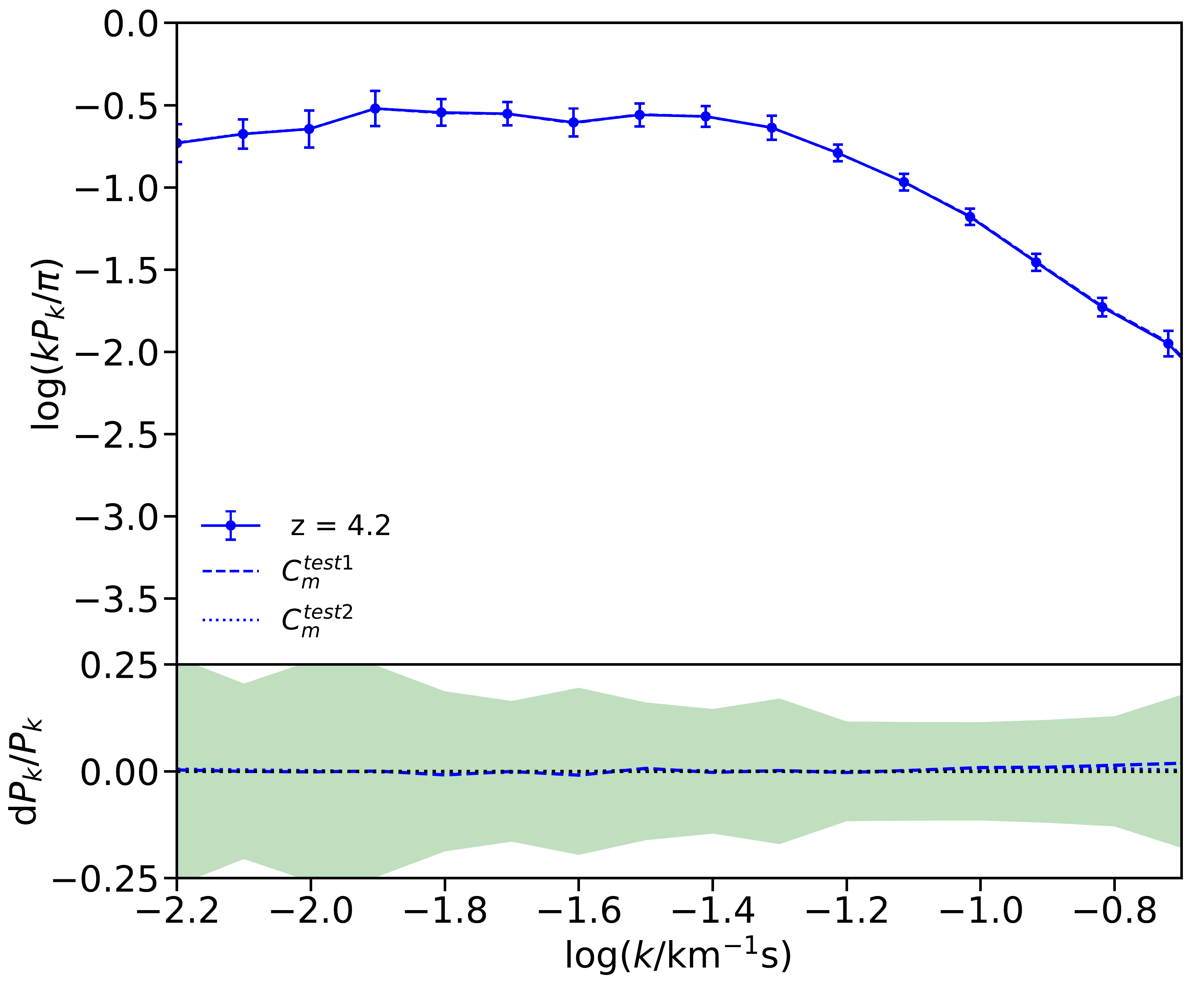}
 
}
\subfigure
{ 
\includegraphics[width=3.4in]{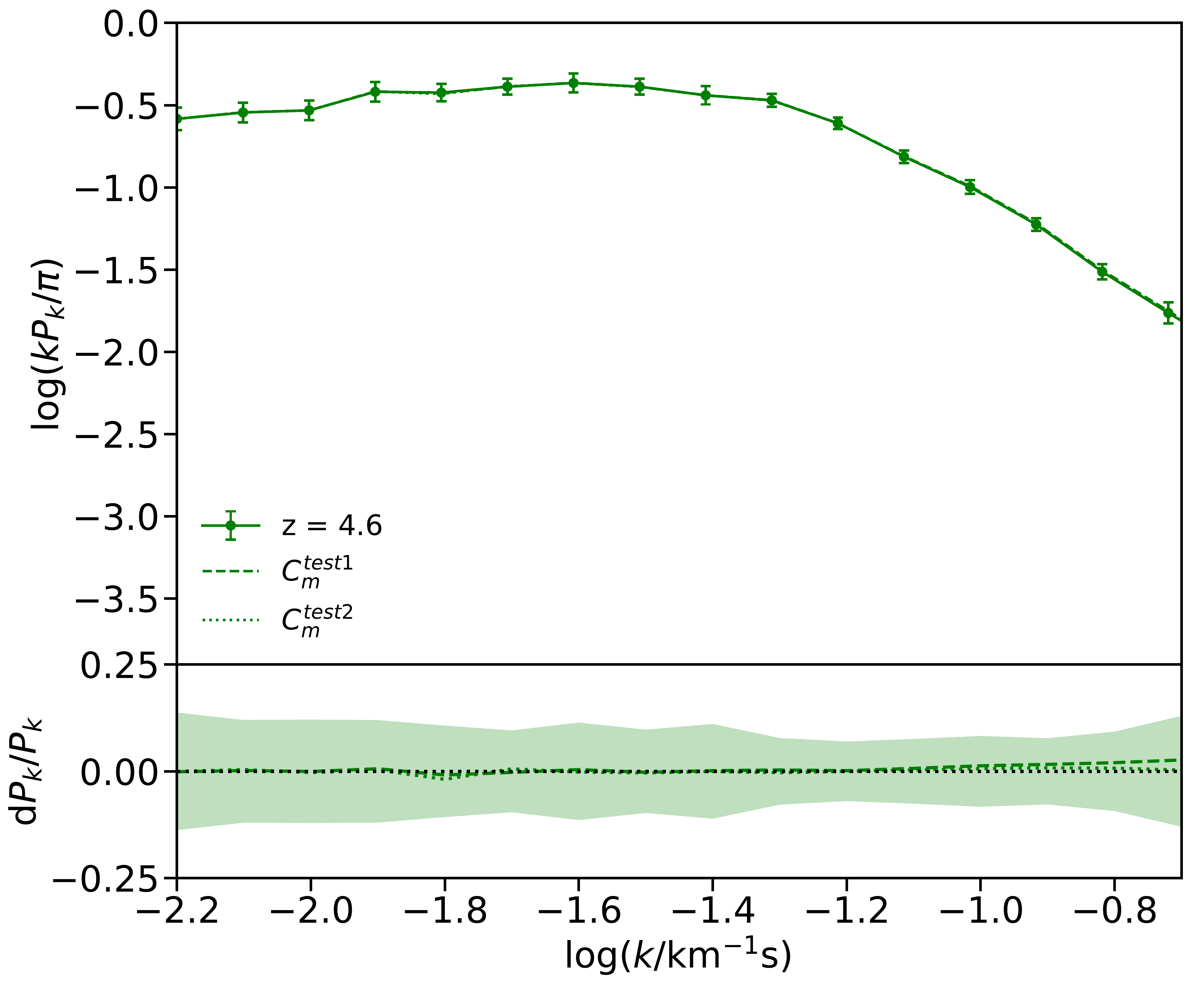} 

} 
\subfigure
{ 
\includegraphics[width=3.4in]{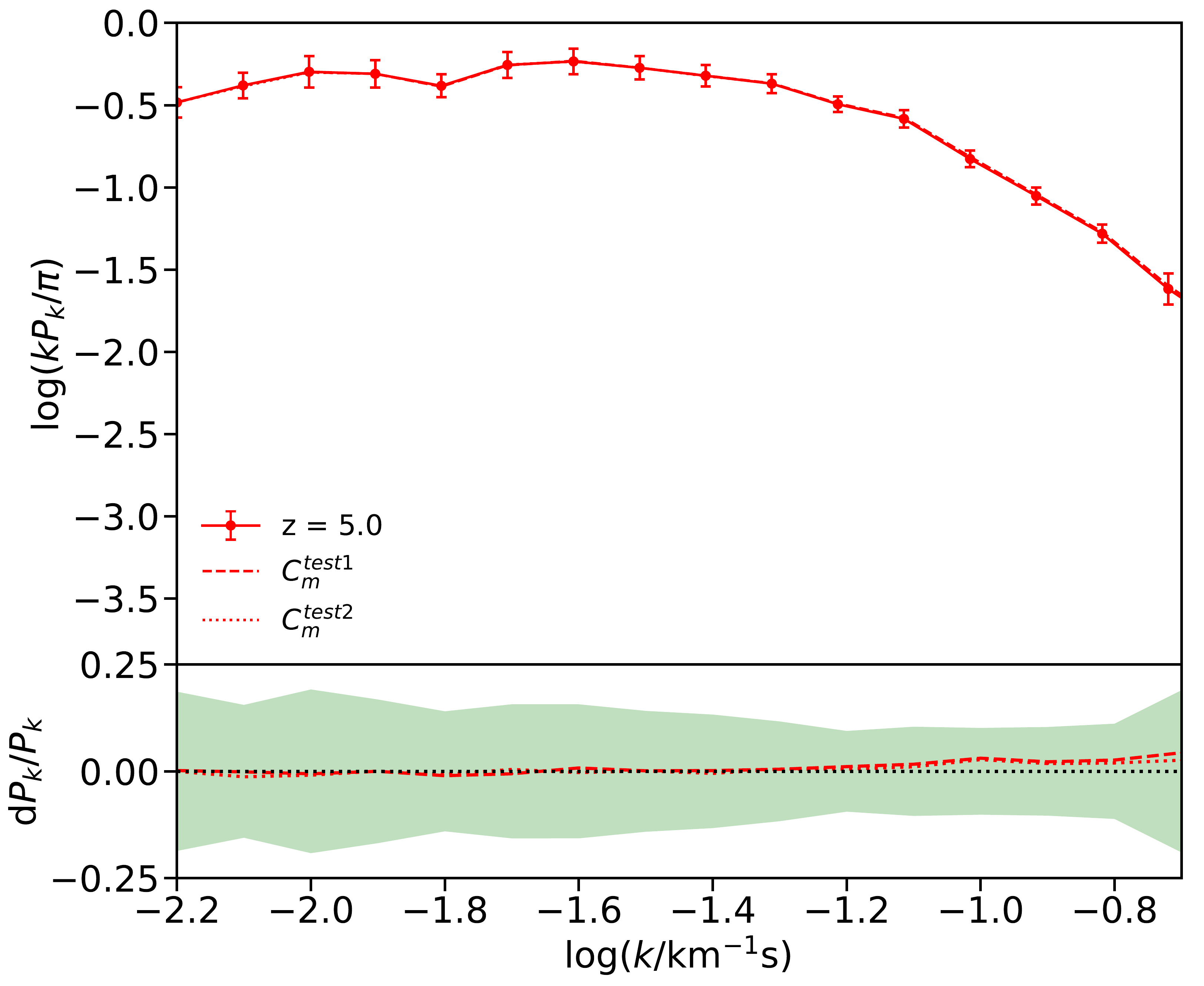} 

} 
 \caption{\small Effect of the choice of thermal model for calculating the masking correction function. For each redshift, the power spectrum corrected with the fiducial $C_{m}(k)$ function (solid colored lines and data points) is compared with the results obtained using masking correction function $C^{test1}_{m}(k)$ computed from a post--processed model with $T_{0}=4000$ K and $\gamma=1.5$ (dashed colored line) and $C^{test2}_{m}(k)$ from a model with $\gamma =1.0$ and a temperature $T_{0}=7500$ K (dotted colored line). Variations in the power spectrum as a fraction of the fiducial power are reported in the bottom section of each panel and compared with the statistical $68\%$ uncertainties (green shaded regions). Uncertainties in the masking correction functions produce negligible variations ($\lesssim4\%$) at the small scales of the final power spectrum measurement that are well within the corresponding statistical errors.}
 \label{fig:MaskCorrTest}
\end{figure}

Figure \ref{fig:MaskCorrTest} shows the power spectrum obtained using different masking correction functions for the redshift bins considered in our analysis. In all the cases, computing the masking correction using models with different thermal parameters affects the small scales only mildly, with variations in the power of $\lesssim$4$\%$ for scales $\log(k$/km$^{-1}$s$)\gtrsim -1.1$. We therefore do not expect that uncertainties in the masking correction function will relevantly affect our final constraints. 

\subsection{Metals correction}
\begin{figure} 
\centering
\subfigure
{ 
\includegraphics[width=3.4in]{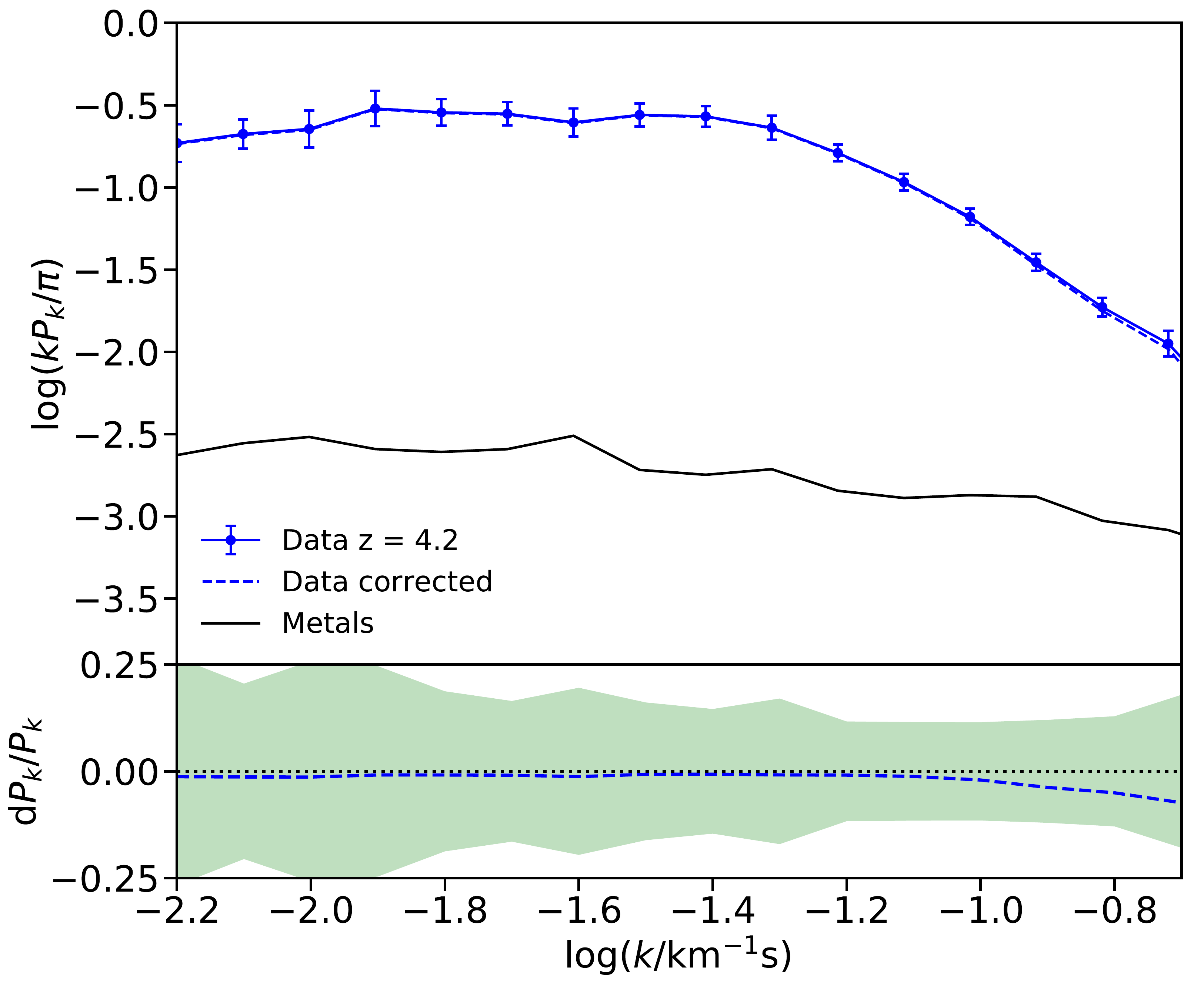}
 
}
\subfigure
{ 
\includegraphics[width=3.4in]{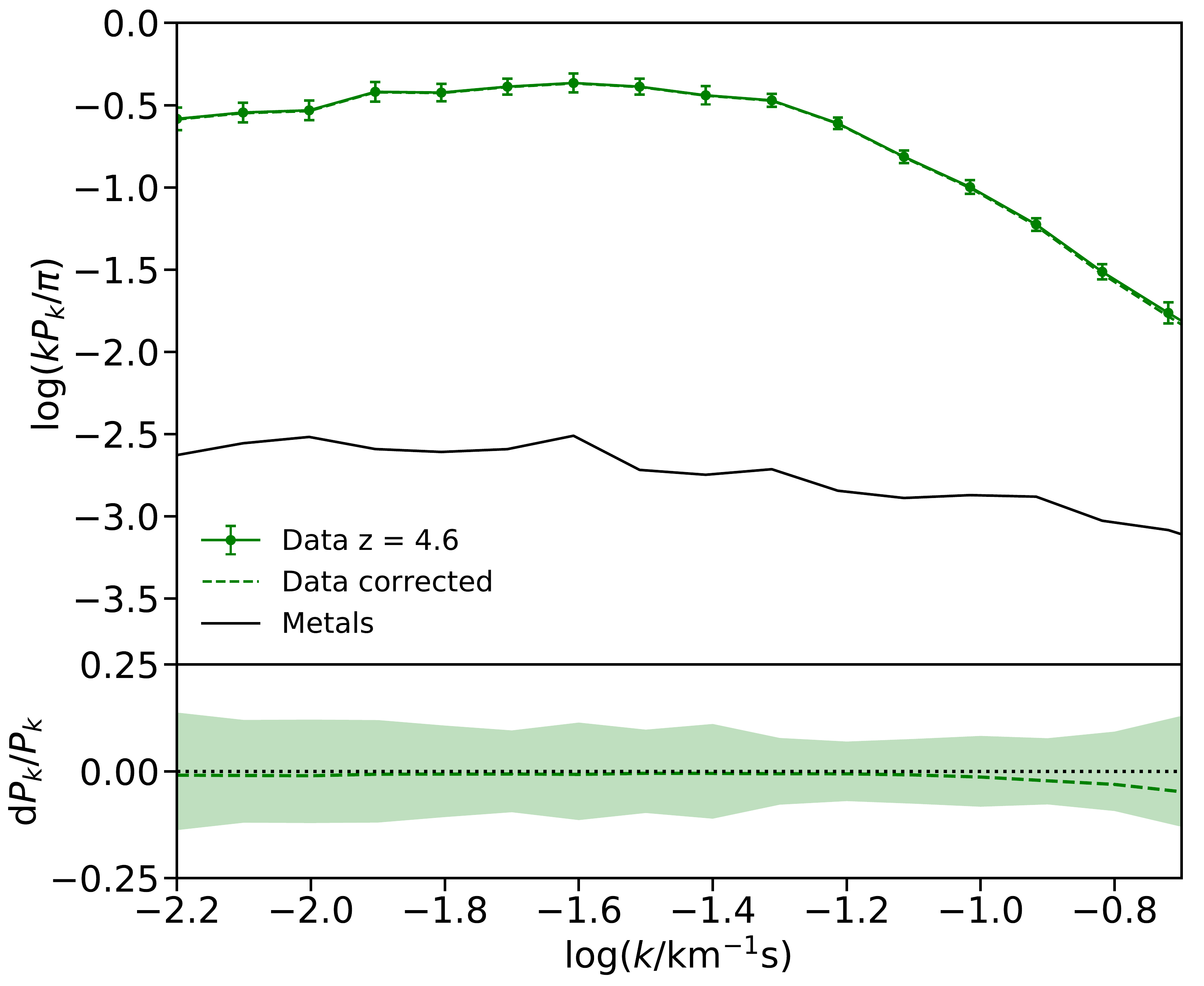} 

} 
\subfigure
{ 
\includegraphics[width=3.4in]{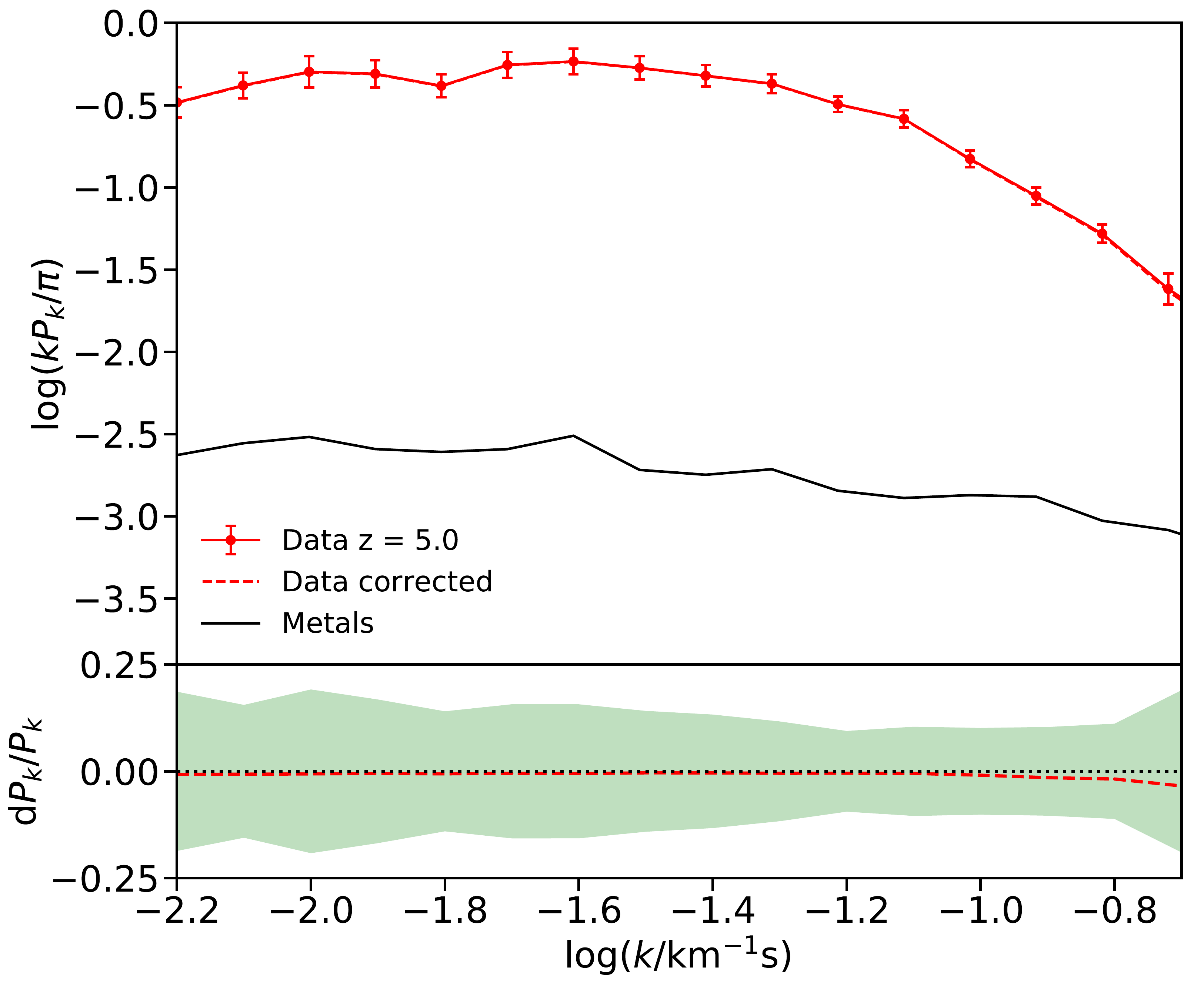} 

} 
 \caption{\small Effect of the metal correction on the Ly$\alpha$ power spectrum measurements for the three redshift bins considered in this work. In each panel the average metal power computed in Section \ref{sec:Metals} (black solid line) is subtracted from the total power (colored solid lines and data points) to obtain the corrected measurement (colored dashed lines). The effect of metals at small scales is more relevant in the lowest redshift bin because the amount of Ly$\alpha$ absorption is lower. Nevertheless, the changes in power due to the metal contribution, reported in the bottom panels as fraction of the total power, is always $<$10$\%$ and well within the statistical errors (green shaded regions). }
 \label{fig:MetCorr}
\end{figure}

The Ly$\alpha$ forest region is affected by narrow metal line contaminants that may increase the flux power at the small scales. To correct for the metal contribution we subtract the metal power evaluated in Section \ref{sec:Metals} from the final power spectrum measurements. 
The effect of the metal subtraction on the power spectrum measurements is presented in Figure \ref{fig:MetCorr} for our three redshift bins.
As expected, only the small scales ($\log(k/$km$^{-1}$s$)\gtrsim-1.0$) are significantly affected. The variations are relatively small ($<$10$\%$) and well within the statitical errors of our final measurements. 
\label{sec:met}
\section{Covariance matrix uncertainties}
\label{sec:cutoffMa}
In this Appendix we test how strongly the choice of simulation model for the covariance matrix regularization affects the final constraints on $T_{0}$ and $u_{0}$.
In principle, off--diagonal coefficients of the covariance matrix will mildly depend on the shape of the power spectrum and therefore on the thermal parameters characterizing the models.

Figure \ref{fig:CovR} shows the results for $T_{0}$ and $u_{0}$ obtained when using a covariance matrix derived from the fiducial model S40$_{-}$1z15 (Covariance matrix$_{-}$1) and from the S40$_{-}$1z9 model (Covariance matrix$_{-}$2). Both the nominal results and errors estimates shows only modest ($\lesssim$3\%) changes.

\begin{figure} 
\label{fig:CovR}
\centering
\subfigure
{ 
\includegraphics[width=3.0in]{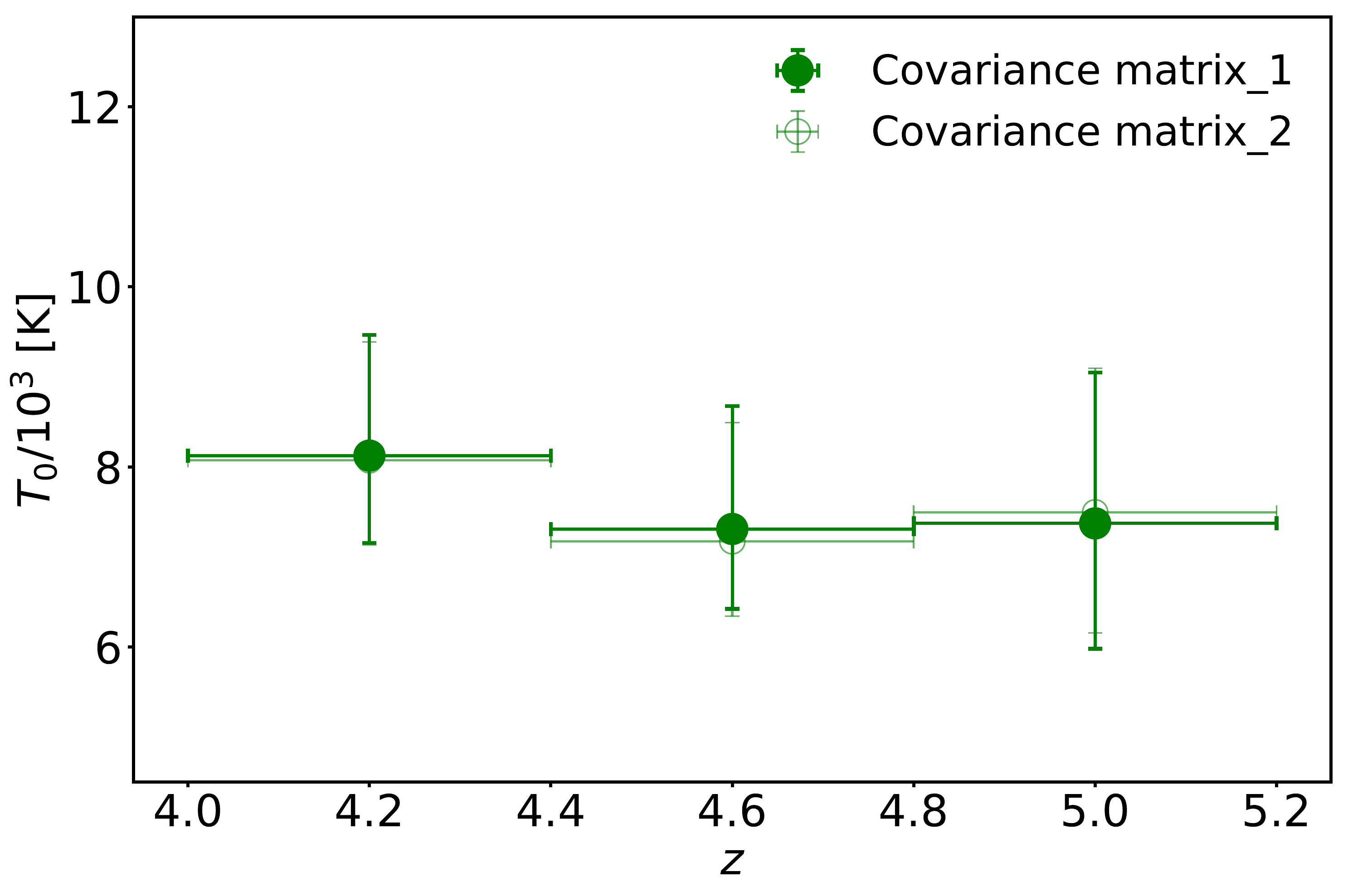}
 
}
\subfigure
{ 
\includegraphics[width=3.0in]{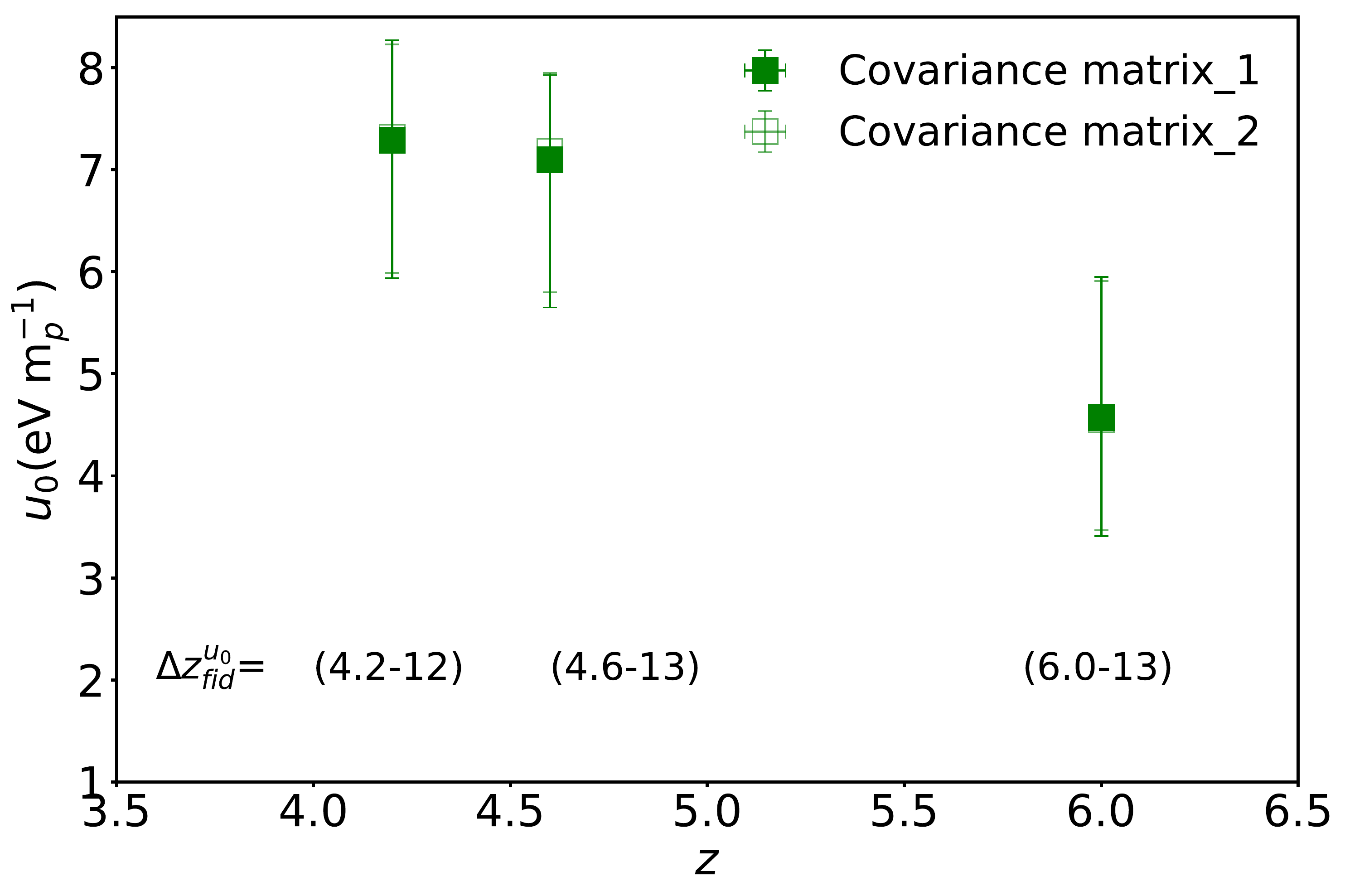} 

} 

 \caption{\small Effect of varying the simulation used for the covariance matrix regularization. The measurements of $T_{0}$ (left panel) and $u_{0}$ (right panel) are reported for two different choices of simulations for the covariance matrix regularization: model S40$_{-}$1z15 (filled markers) and model S40$_{-}$1z9 (empty markers).} 
 \label{fig:CovR}
\end{figure}

\section{Comparison with Viel et al. (2013)}
\label{sec:VielComp}
In this Appendix we compare our power spectrum measurements to those from \cite{Viel13}.
While \cite{Viel13} included somewhat lower resolution spectra from Magellan/MIKE, for consistency we limit our comparison to their measurements obtained with the HIRES spectrograph.

We consider two main differences between our new estimates and the older \cite{Viel13} power spectra: flux contrast estimators and cosmic variance.
 First, \cite{Viel13} normalized the spectra using a spline continuum estimate and then computed the mean flux in large sections of data. To test whether this affected the results we recomputed the \cite{Viel13} power spectra by applying the procedure described in Section \ref{sec:DataAnalysis} to the Viel et al. data, using the same sections of spectra. Note that given the somewhat larger scales probed by Viel et al. we do not expect any relevant effect due to noise or resolution.
We verified that we were able to reproduce the previous results with good precision and no significant bias was introduced by the different flux contrast estimators.

We next consider whether sample size may be playing a role. In each of our redshift bins the number of independent lines of sight (LOS) contributing to our measurement is always more than double the number in \cite{Viel13}. In particular, for the $z=4.2$ bin we used 12 LOS versus 4 LOS in \cite{Viel13}, for the $z=5.0$ bin we used 12 versus 5, while at $z=4.6$, where the largest differences between the power spectra are seen, we used 15 quasars versus 5 in the previous work.

\begin{figure} 
\centering
\includegraphics[width=0.5\columnwidth]{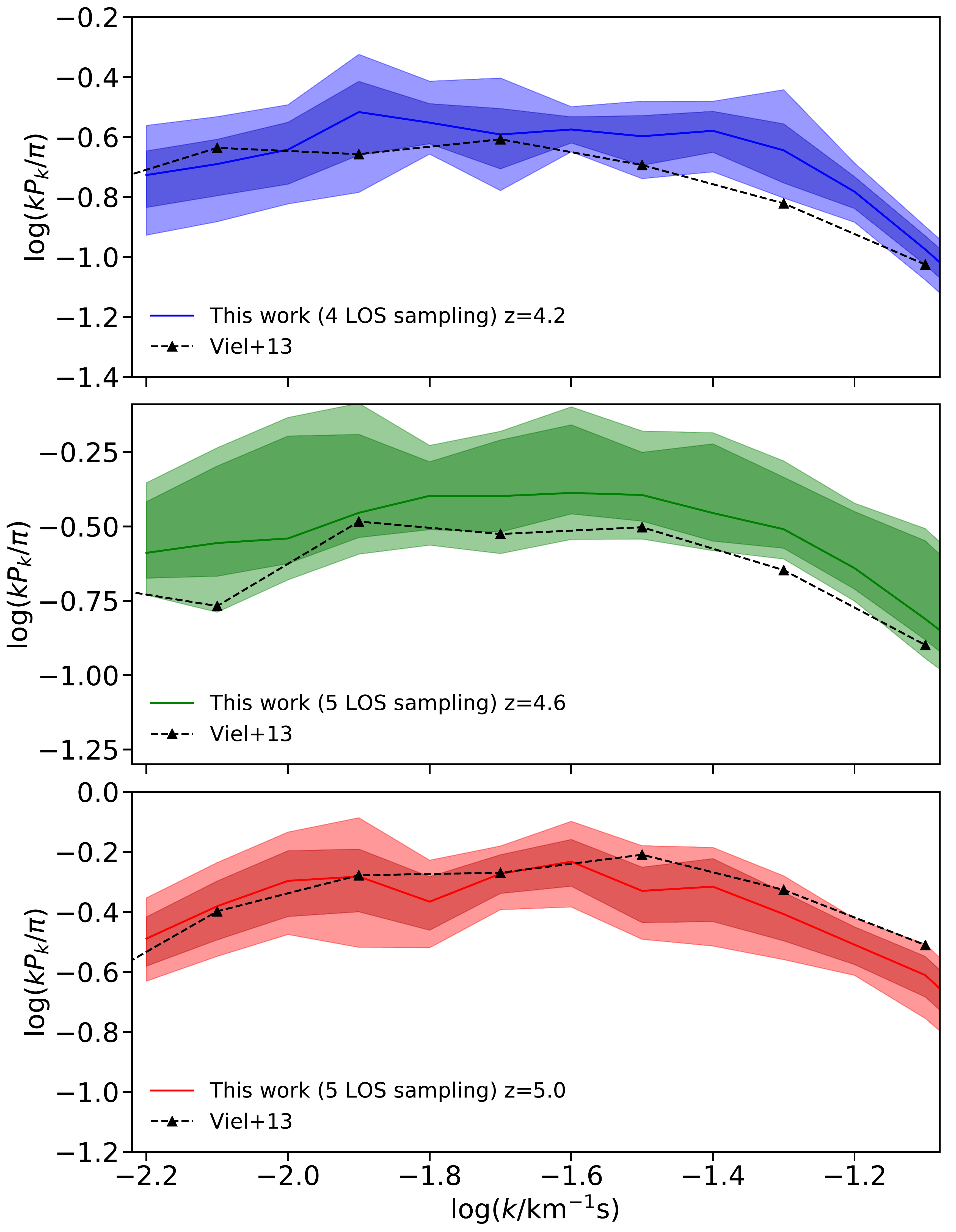} 
\caption{\small Comparison with previous HIRES results. For each redshift bin, the power spectrum measurements obtained by \cite{Viel13} (black dashed lines) are compared with the $68\%$ (darker shaded regions) and $95\%$ (lighter shaded regions) contours of the power spectrum realizations obtained with a Monte Carlo sampling of our lines of sight, when using the sample size of Viel et al.(see text for details). At all redshift there is a broad consistency with the \cite{Viel13} values.}
\label{fig:VielComp}
\end{figure}

To determine whether cosmic variance can explain the discrepancy between our results and those of \cite{Viel13} we computed the power spectrum from subsamples of our data.
 Figure \ref{fig:VielComp} shows the comparison between the \cite{Viel13} power spectra (black dashed lines) and the $68\%$ (darker shaded regions) and $95\%$ (lighter shaded regions) contours of the distribution of power spectrum realizations obtained from a Monte Carlo sampling of our lines of sight. For each redshift bin we randomly select from our sample the same number of LOS contributing to the \cite{Viel13} measurement and use them to compute the power spectrum. Repeating the process for many ($\sim$200) realizations we verify that the \cite{Viel13} results fall within this distribution.
 At $z=4.2$ and $z =5.0$ most of the \cite{Viel13} values fall within the $68\%$ contours. For the $z=4.6$ bin the agreement is slightly worse but still largely within the $95\%$ region.
 We note that the errors in different $k$ bins are correlated, as discussed in Section \ref{sec:CovMatrix}. We further note that our sample size is still relatively modest, and our Monte Carlo technique is likely to underestimate the cosmic variance at the 95\% level.

\section{Comparison with Ir\u{s}i\u{c} et al. (2017)}
\label{sec:IrsicComp}
In this Appendix we compare our power spectrum measurement at $z=4.2$ to the one from \cite{Irsic17}. These authors use spectra from the XQ-100 Legacy Survey
\citep{Lopez16}, collected using the lower resolution VLT/X-Shooter spectrograph (R$\sim11$ km s$^{-1}$ corresponding to a FWHM= 26  km s$^{-1}$ for the VIS arm). We therefore limited our comparison to the power spectrum scales where the resolution correction for the \cite{Irsic17} measurements is $\lesssim$20$\%$, corresponding to $\log(k/$km$^{-1}$s$) < -1.4$.

\begin{figure} 
\centering
\includegraphics[width=0.6\columnwidth]{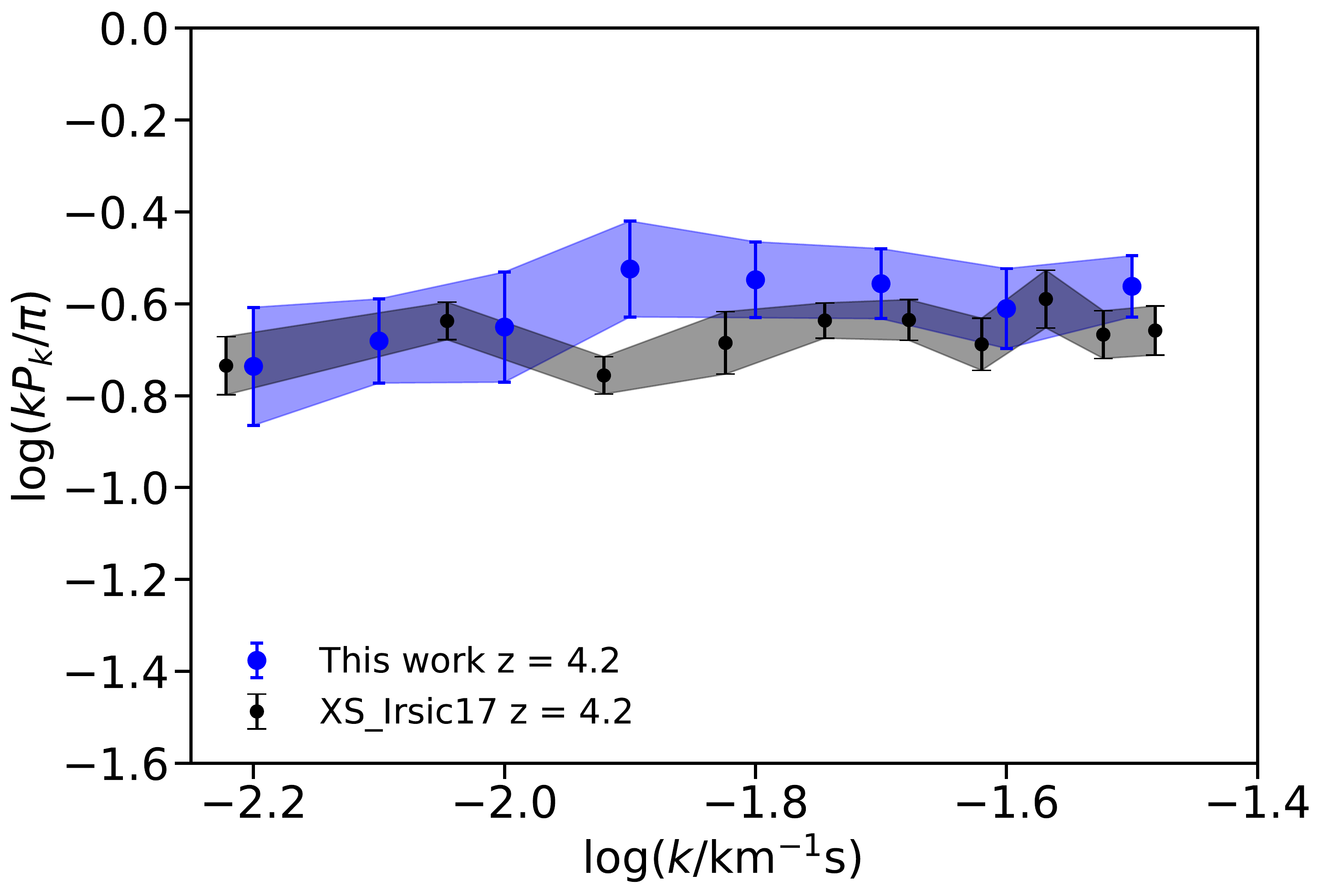} 
\caption{\small Comparison with previous X-Shooter results at $z=4.2$, limited to the scales where the resolution correction for the X-Shooter data is $\lesssim$20$\%$. The power spectrum obtained by \cite{Irsic17} (black points) is compared with our results for the lowest redshift bin (blue points).  
 The two measurements are generally consistent within the $68\%$ uncertainties (colored shaded regions).}
\label{fig:IrsicComp}
\end{figure}

Figure \ref{fig:IrsicComp} presents the comparison between the \cite{Irsic17} power spectrum (black points) and our measurements (blue points) at $z=4.2$.
For both the datasets we also include the final $68\%$ uncertainties (colored shaded regions).
 Even if the X-Shooter power spectrum shows tendentially lower values than ours the two measurements are consistent within the $68\%$ uncertainties at all scales but one point (at the scale  $\log(k/$km$^{-1}$s$)=-1.92$). 
 A possible explanation for the offset in the power at scales $\log(k/$km$^{-1}$s$)\gtrsim -1.9$  may be differences in the spectra samples and in the redshift coverage of the z=4.2 bin: while we include spectra falling within a broad $\Delta z$=0.4 redshift bin, \cite{Irsic17} adopts a narrower $\Delta z$=0.2 bin.   
 Nevertheless, as for the \cite{Viel13} power spectrum, we verified that, using the same X-Shooter spectra, we were able to reproduce the  \cite{Irsic17} results at the considered scales without introducing any significant bias due to possible differences in the analysis.  
\section{Comparison to previous temperature measurements at high--z}
\label{sec:GlobalComp}
\begin{figure} 
\centering
\includegraphics[width=0.6\columnwidth]{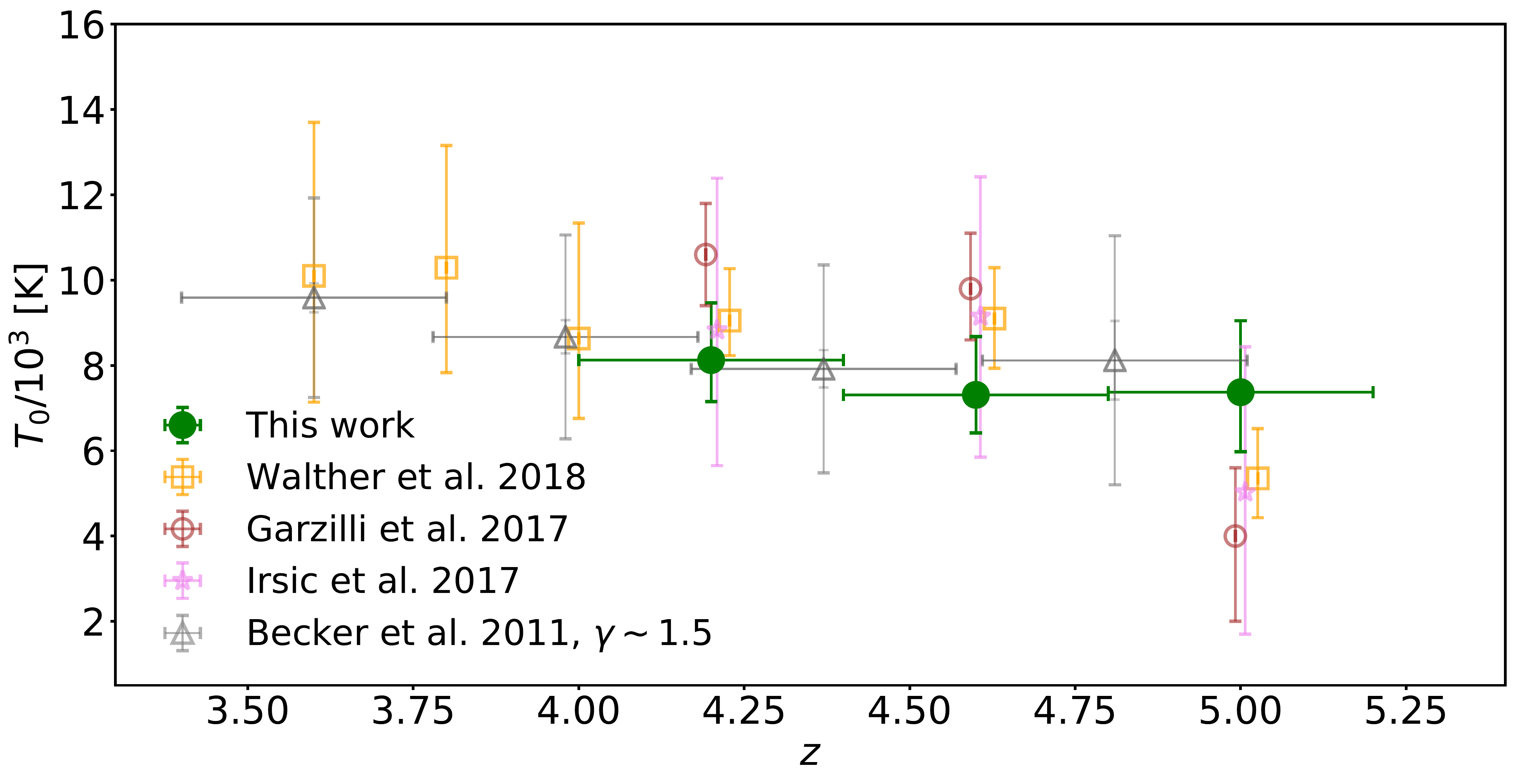} 
\caption{\small Comparison with previous temperature measurements over the redshift range covered by our analysis. Our results (green points) are compared with the  temperature values  obtained from the curvature analysis of  \citealt{Becker11} (gray triangles) and with the thermal estimates from the power spectrum statistic of \cite{Irsic17b} (pink stars), \cite{Garzilli17} (brown circles) and \cite{Walther18} (orange squares). Note that the temperature values from Ir\u{s}i\u{c} et al., Garzilli et al. and the $z\gtrsim4.2$ results from Walther et al. are all obtained from the flux power spectrum measurements of \cite{Viel13} although calibrated with different sets of simulations, therefore, they are not fully independent.
Vertical error bars are $68\%$ statistical uncertainties for all the data. For Becker et al. the nominal errors have been increased to include the Jeans smoothing uncertainty estimated by those authors.}
\label{fig:GlobalComparison}
\end{figure}

In Figure \ref{fig:GlobalComparison}  we compare our IGM temperature measurements to previous constraints from the literature over the redshift range covered by our analysis.
We note that the temperature values from \cite{Irsic17b} (pink stars), \cite{Garzilli17} (brown circles) and the $z\gtrsim4.2$ results from \cite{Walther18} (orange squares) are all obtained from the flux power spectrum measurements of \cite{Viel13} although calibrated with different sets of simulations, therefore, they are not fully independent. 
Among these constraints, the larger error bars reported for \cite{Irsic17b} reflect a 1.5 correction factor applied by these authors to the nominal errors associated to \cite{Viel13} measurements.

While generally within the $68\%$ uncertainties of our new results, previous estimates from the flux power spectrum statistic suggest a significant decrease in temperature at $z\gtrsim4.6$. Our thermal constraints, however, obtained from a larger and higher quality sample of high-resolution spectra, do not show any strong evolution in the temperature at these redshifts (see also Appendix \ref{sec:VielComp} for a comparison between our recent flux power spectrum measurements and the ones from \citealt{Viel13}).
\section{Numerical convergence}
\label{sec:convergence}
In this Appendix we examine the convergence of the Ly$\alpha$ flux power spectrum in the simulations used in this work.
We used multiple simulations with the thermal history model 1$_{-}$z15 in Table \ref{table:sim}. All of the synthetic Ly$\alpha$ forest lines of sight were produced using the procedure described in Section \ref{sec:LOSsin}. The flux power spectrum was then computed as in Section \ref{sec:syntheticLOS}.

\begin{figure*} 
\centering 
\subfigure
{ 
\includegraphics[width=3.35in]{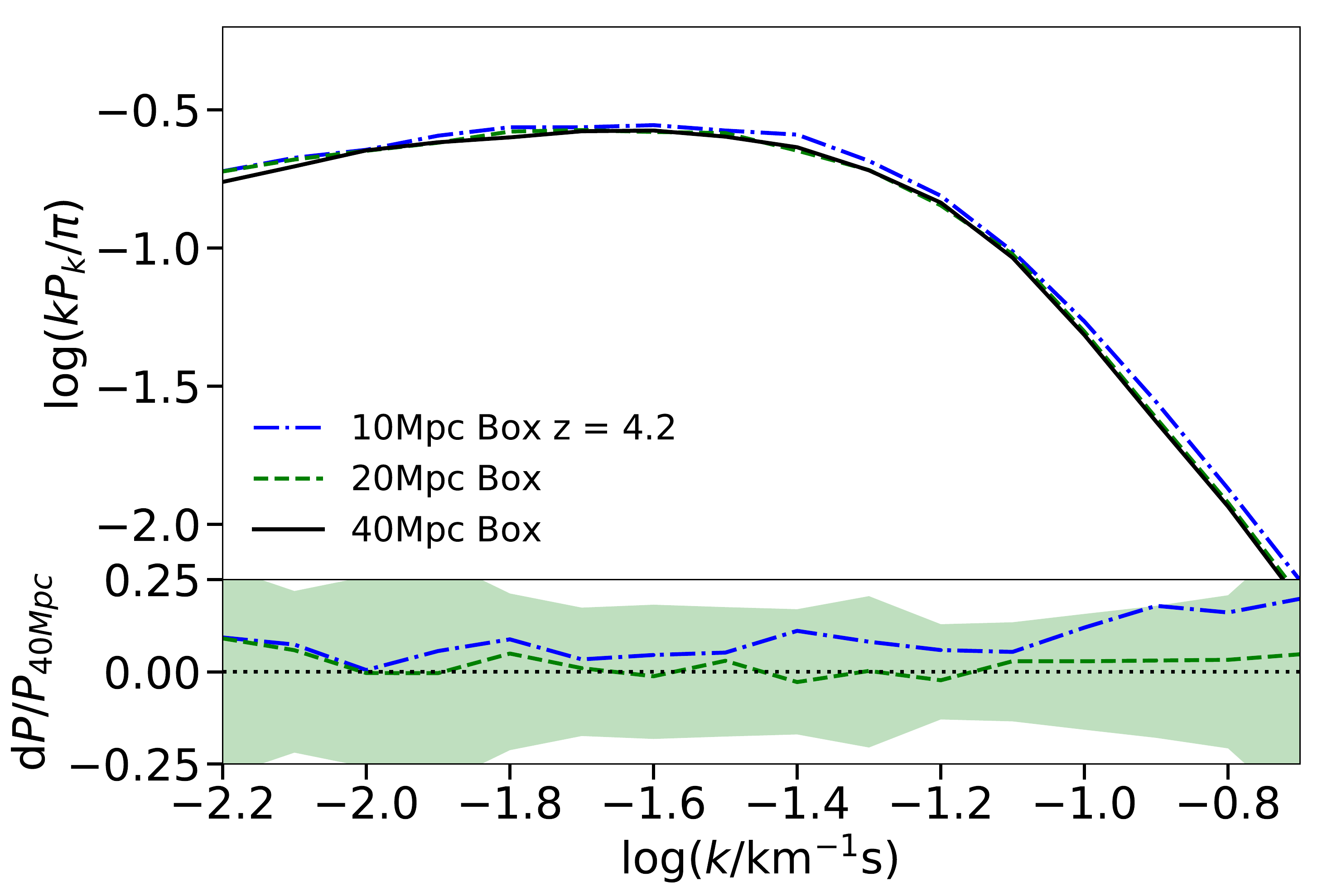}
 \label{fig:T0}
}
\subfigure
{ 
\includegraphics[width=3.35in]{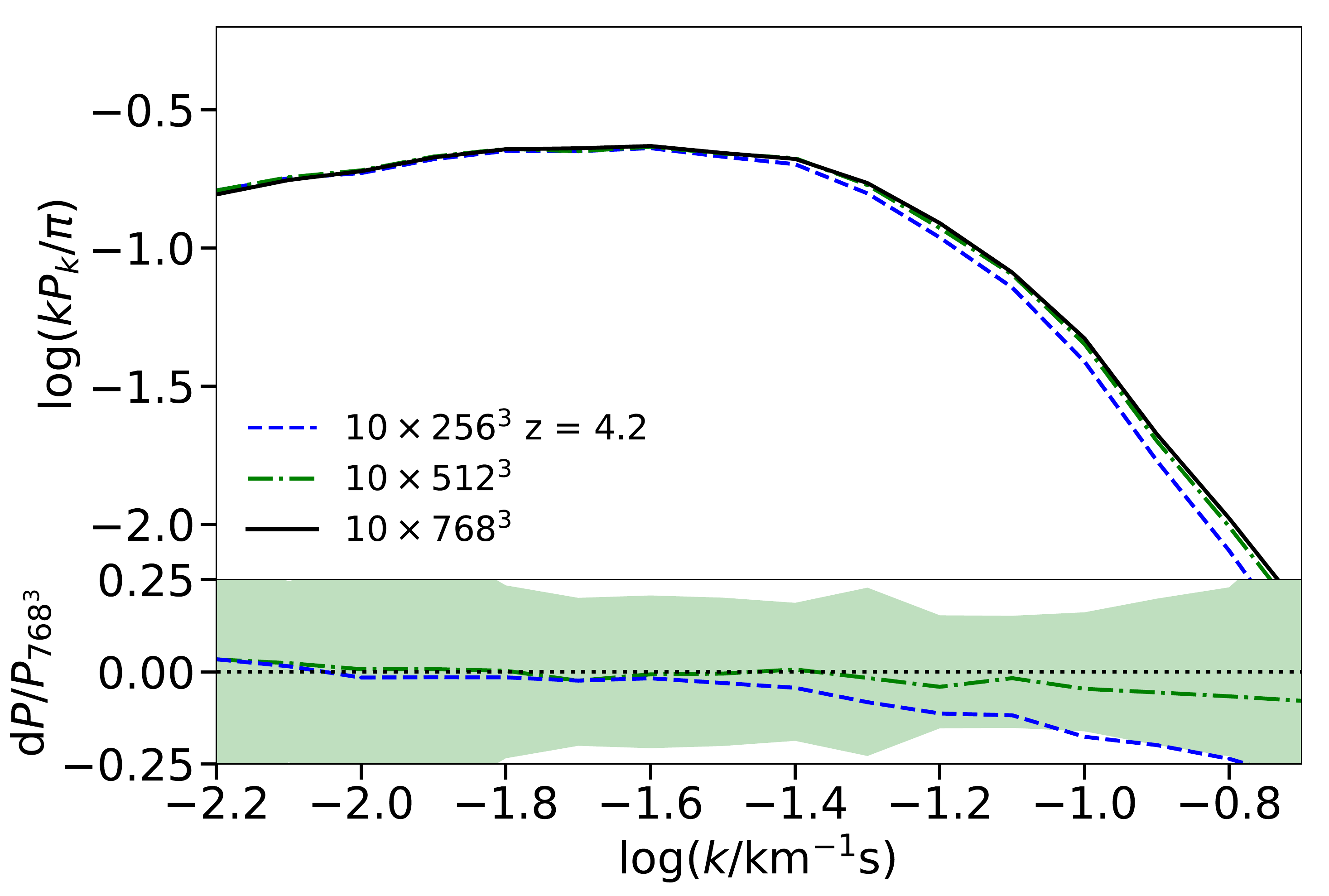} 
\label{fig:u0}
} 
\subfigure
{ 
\includegraphics[width=3.35in]{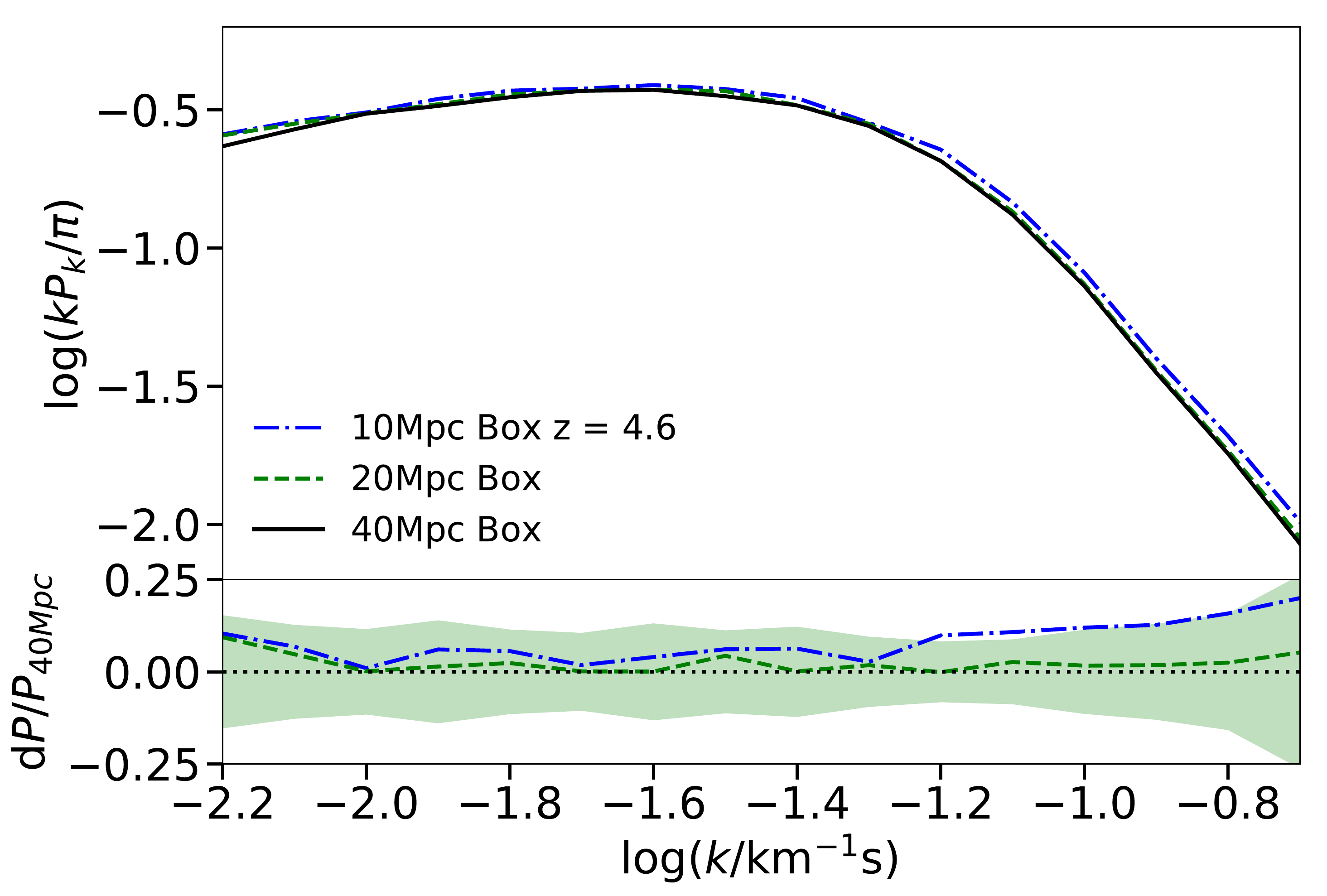} 
\label{fig:u0}
} 
\subfigure
{ 
\includegraphics[width=3.35in]{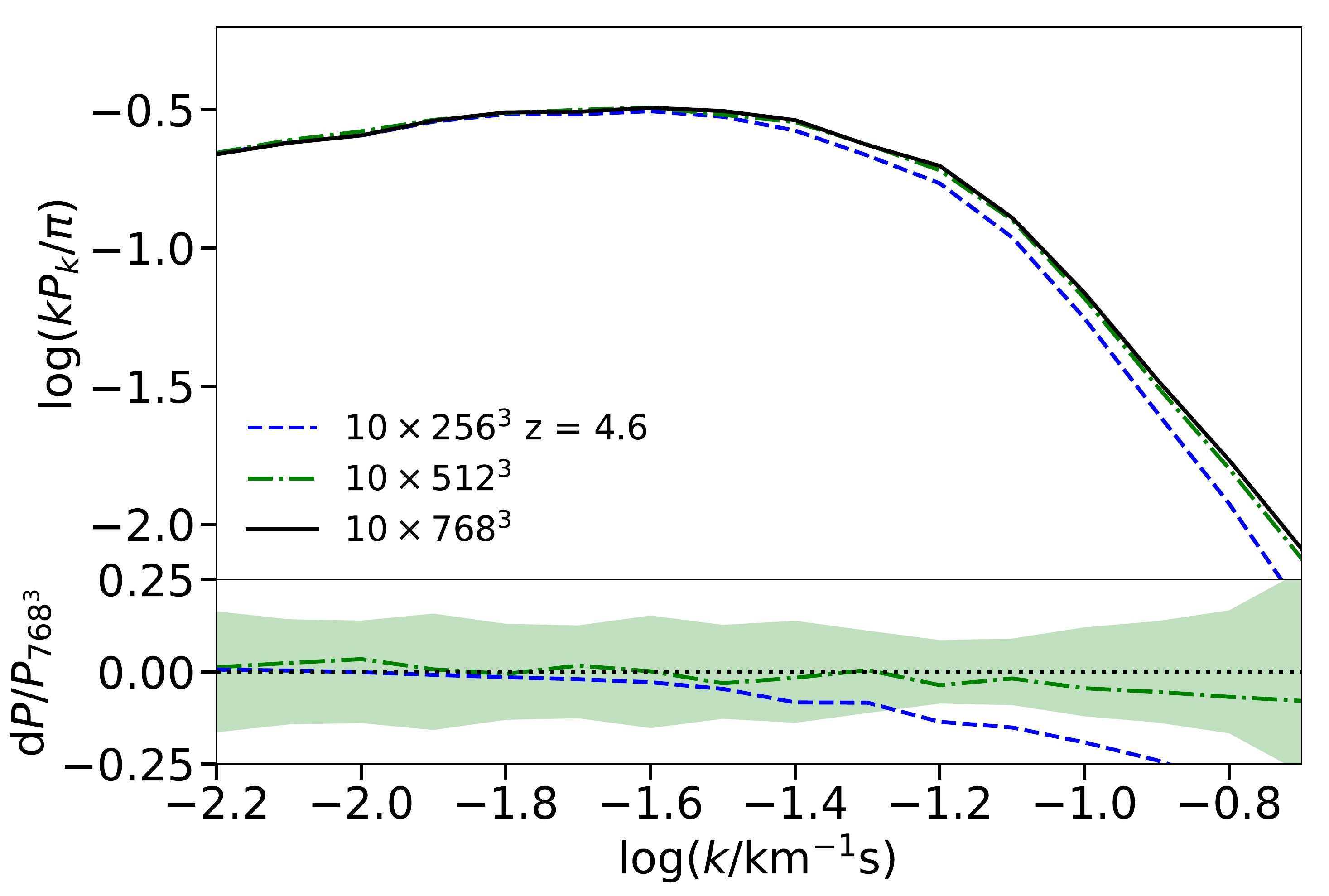} 
\label{fig:u0}
} 
\subfigure
{ 
\includegraphics[width=3.35in]{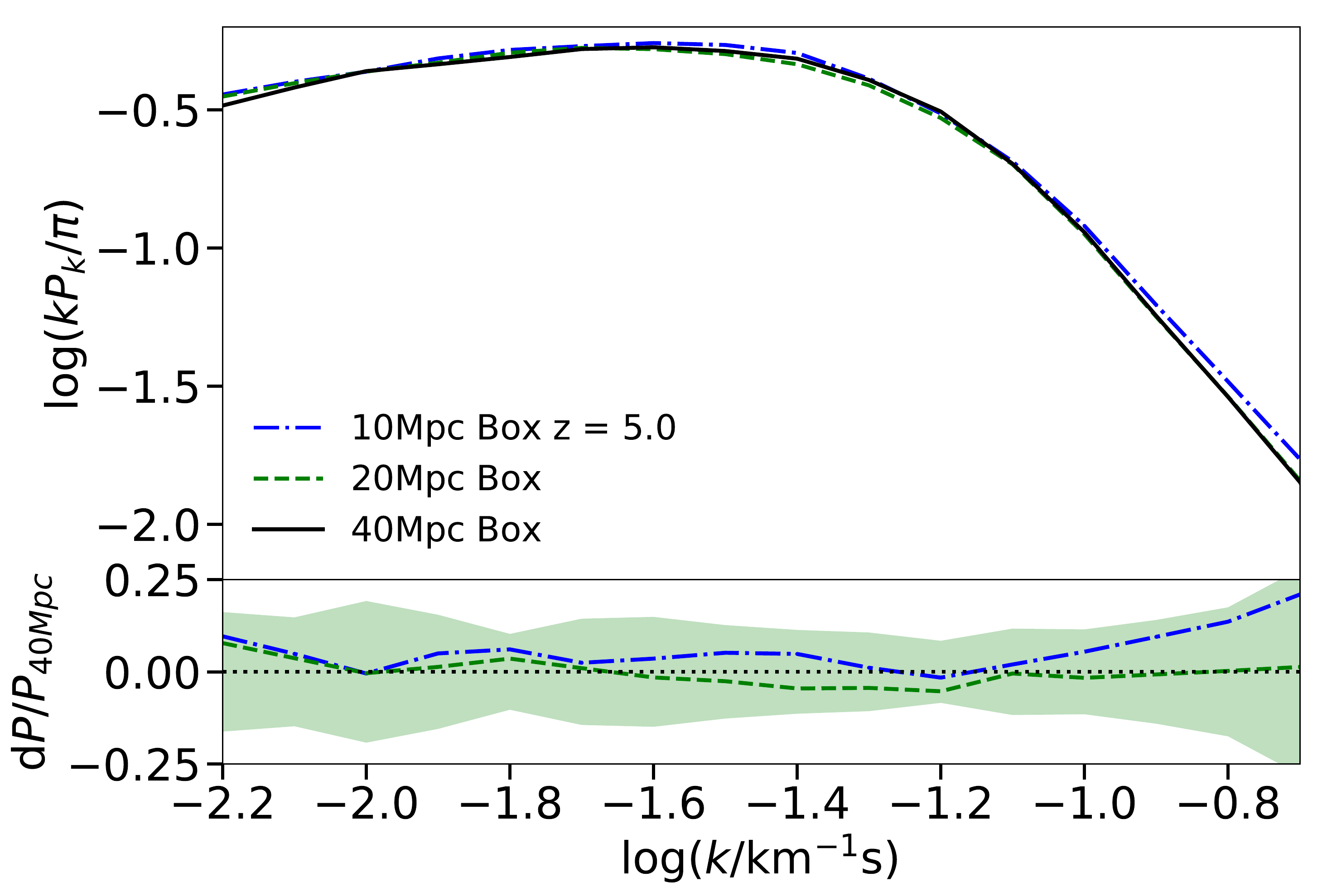} 
\label{fig:u0}
} 
\subfigure
{ 
\includegraphics[width=3.35in]{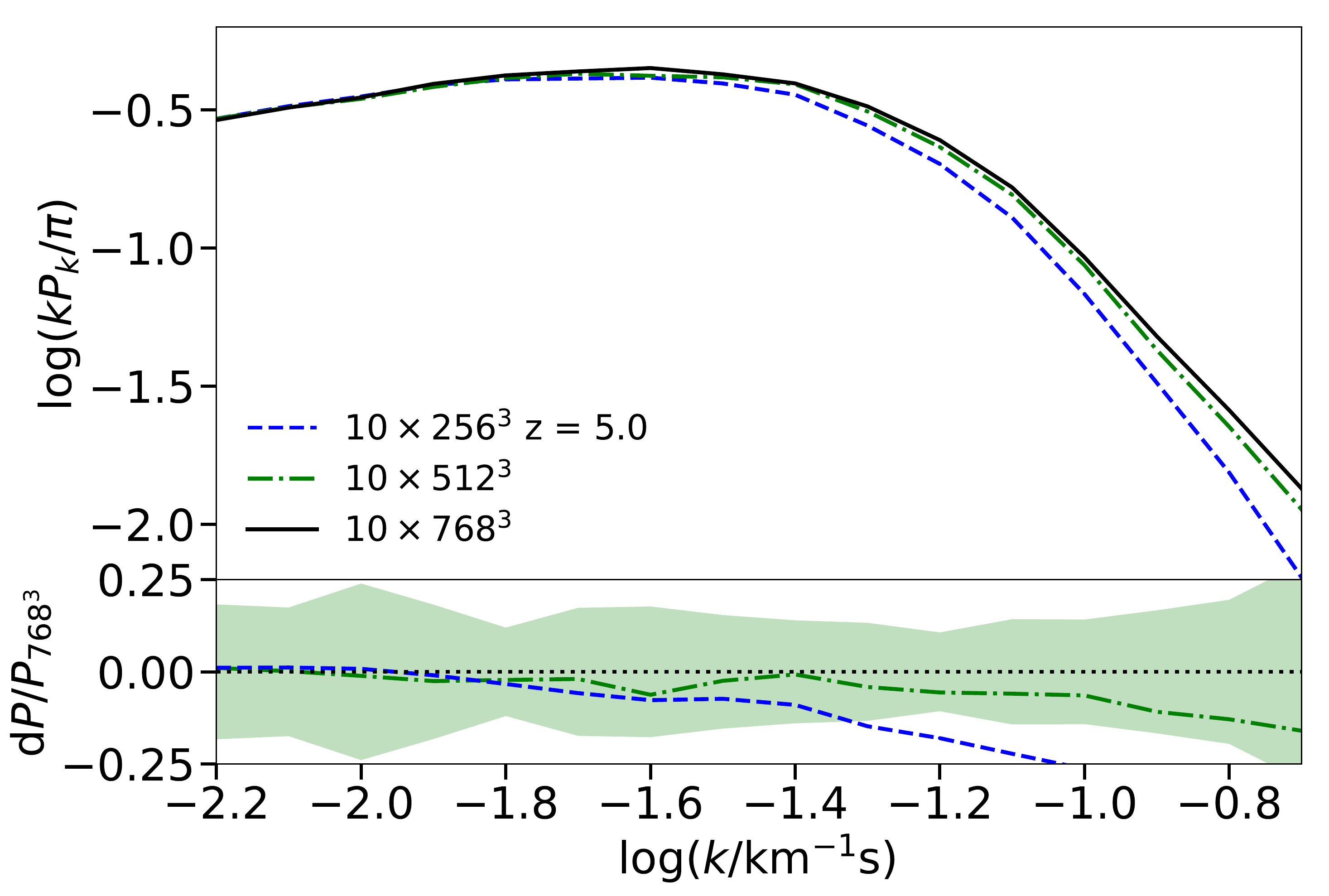} 
\label{fig:u0}
} 
 \caption{\small Convergence of the flux power spectrum with box size and mass resolution for the redshifts relevant in this work. Power spectra for this test were computed from noise--free mock lines of sight using the procedure described in Section \ref{sec:syntheticLOS}. The left column shows the convergence with box size at a fixed mass resolution ($M_{gas}=9.97\times10^{4}$ $h^{-1}M_{\odot}$), while the right column displays the convergence with respect to the highest mass resolution model (S10$_{-}$1z15$_{-}$768 in Table \ref{table:sim}) for a fixed box size ($L=10$ $h^{-1}$cMpc). Variations in the power spectrum as a fraction of the reference power are plotted in the bottom section of each panel and compared with the statistical 68\% uncertainties (green shaded regions). Both resolution and box size corrections have been applied to our fitting models.}
 \label{fig:ConvergenceTest}
\end{figure*}

The tests are shown in Figure \ref{fig:ConvergenceTest}, where the convergence with box size for a fixed mass resolution ($M_{gas}=9.97\times10^{4}$ $h^{-1}M_{\odot}$) is reported in the left column and the convergence with mass resolution for a fixed box size ($L=10$ $h^{-1}$cMpc) is displayed on the right.
The results show that a small correction for both box size and mass resolution needs to be applied to the power spectra derived from our nominal 10 $h^{-1}$cMpc simulations. When increasing the box size, the power decrease up to $\sim$15$\%$, particularly at very small scales ($\log(k/$km$^{-1}$s$) > -1$).
 In contrast, when increasing the mass resolution the power towards small scales ($\log(k/$km$^{-1}$s$)\gtrsim-1.4$) increase progressively, reaching a correction of $\sim$15$\%$ at $z=5.0$ for our nominal mass resolution. 
 Because the corrections are in opposite directions the final scale factor for box and mass resolution convergence is $\lesssim$5$\%$ at all scales.
 However, we note that in principle the mass resolution convergence may depend on the underlying IGM density structure and, consequently, on the choice of a particular thermal history model.  
 We therefore verified that the  entity of this possible systematic effect was negligible when compared to the 68\%  statistical uncertainties characterizing the observational data and that the final thermal constraints were not affected by the numerical corrections.
 
 We further verified, using lower mass resolution simulations that, increasing the box size up to $L=160$ $h^{-1}$cMpc for a fixed mass resolution, does not introduce additional power at the scales considered in this work.
 
\section{Effective optical depth evolution}
\label{sec:opticalD}
\begin{figure} 
\centering
\includegraphics[width=0.6\columnwidth]{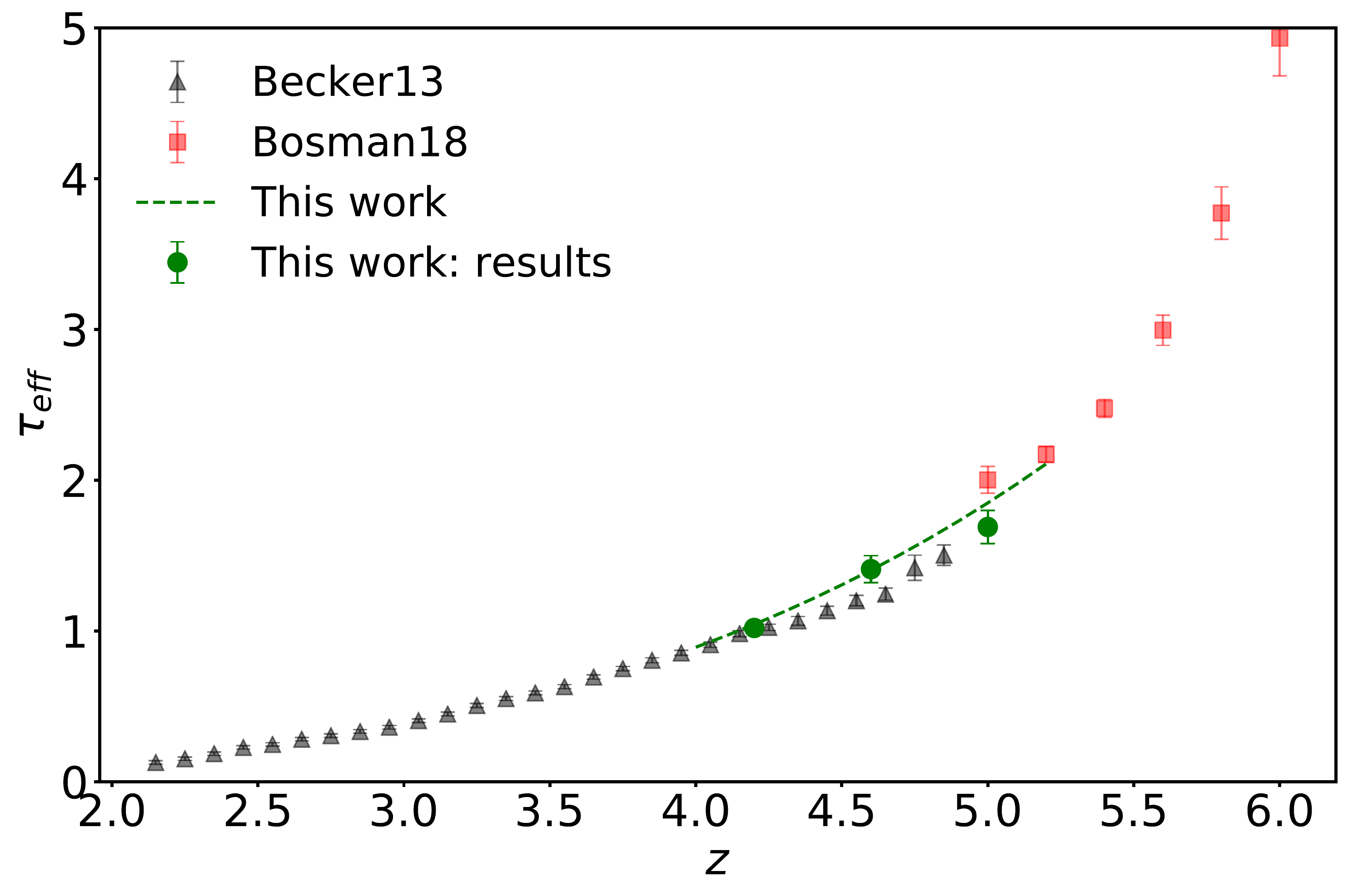} 
\caption{\small Evolution of effective optical depth with redshift. Measurements from \cite{Becker13a} (grey triangles) and \cite{Bosman18} (red squares) are shown along with the fiducial evolution of Eq.\ref{eq:tau} (green dashed line). Constraints from our MCMC analysis are also reported for comparison (green data points). Error bars represent $68\%$ uncertainties for all data points.  }
\label{fig:TauComp}
\end{figure}

As explained in Section \ref{sec:LOSsin}, when constructing mock samples we account for the mild redshift evolution of the mean flux along the line of sight by initially rescaling the effective Ly$\alpha$ optical depth using Eq.\ref{eq:tau}.
In this Appendix we show how the choice of this relation for the $\tau_{\rm eff}$ evolution, while somewhat arbitrary, represents a reasonable transition between the measurements of \cite{Becker13a} and the newer results of \cite{Bosman18}. 
Figure \ref{fig:TauComp} presents the comparison among the different $\tau_{\rm eff}$ evolutions. The results from the analysis of \cite{Becker13a} (gray triangles) at $z<4$ are smoothly connected to the most recent measurements of \cite{Bosman18} (red squares) at $z>5$ by the fiducial fit adopted in this work (green dashed line).
For comparison, we also show the constraints on the optical depth obtained from our MCMC chains and reported in Table \ref{table:Tprob} (green points). We recover values broadly consistent with Eq.\ref{eq:tau} for the three redshift bins.
\section{Interpolation uncertainties}
\label{sec:Interpolation}
In this Appendix we describe the test performed to verify the interpolation scheme implemented in the MCMC analysis.
For this we remove one model from the interpolation grid (in the example below we excluded the model S$10_{-}0.55z9$ of Table \ref{table:sim}) and test how well the thermal parameters for this model are recovered when it is used to generate artificial data.

\begin{figure*} 
\centering 
\includegraphics[width=4.6in]{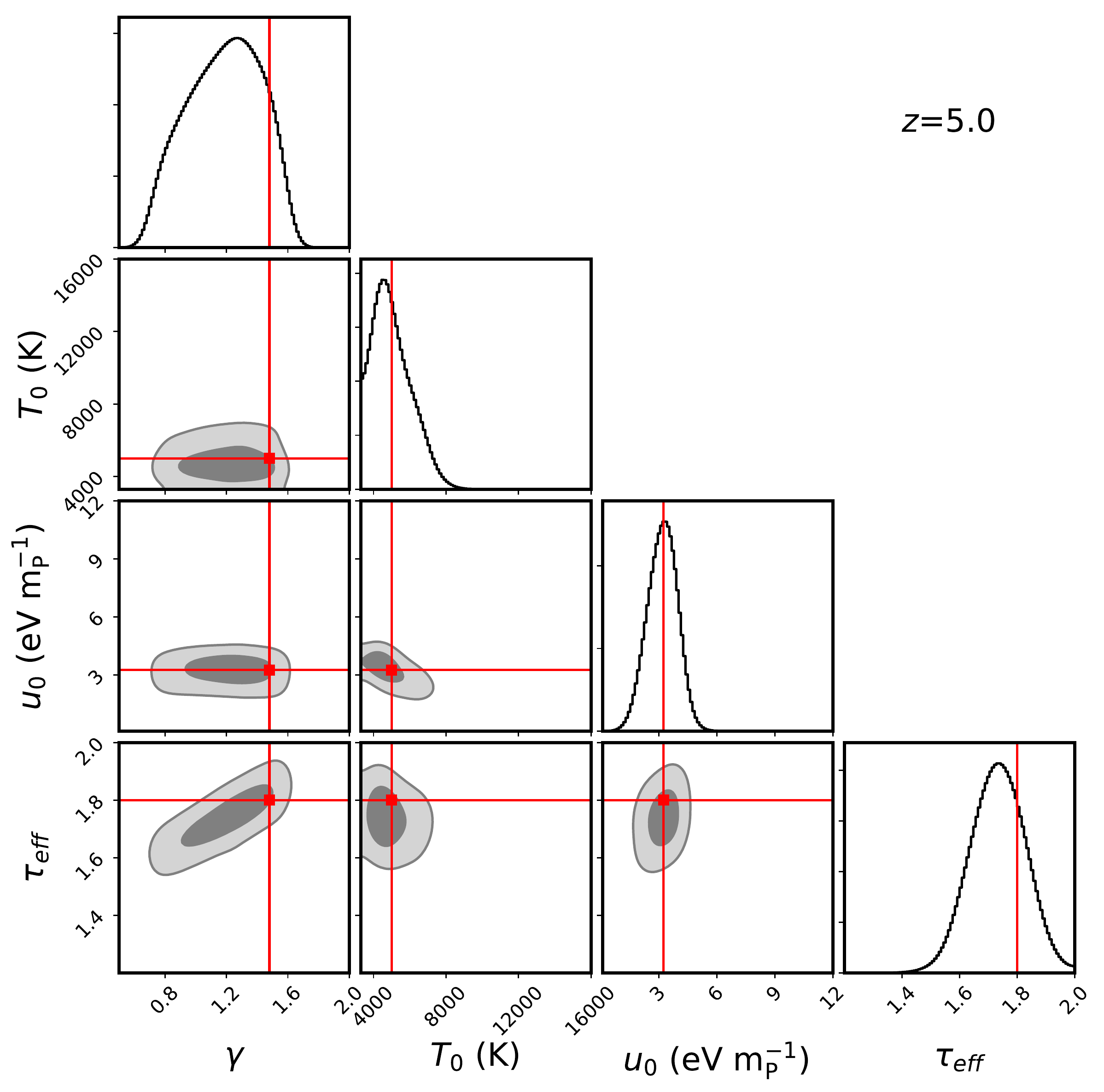} 
\caption{\small Probability distributions for the parameters $T_{0}$, $u_{0}$, $\gamma$ and $\tau_{\rm eff}$ at $z=5$, recovered when fitting the synthetic power spectrum generated from the model S$10_{-}0.55z9$. For this test the model has been removed from the set of comparison simulations. Contours plots show the $68\%$ and $95\%$ two--dimensional probability distributions, while the black histograms display the one--dimensional marginalized posterior distributions for each parameter. The input model parameters are presented for comparison (red squares).}
 \label{fig:IntTest}
\end{figure*}

Figure \ref{fig:IntTest} displays the correct values (red squares) overlaid on the parameters constraints obtained from the MCMC analysis at $z=5.0$. The thermal parameters $T_{0}$ and $u_{0}$ are recovered accurately by the analysis, with discrepancies $\lesssim$5$\%$. The correct values of $\gamma$ and $\tau_{\rm eff}$ fall within the 1$\sigma$ probability distribution, although the peaks of their posterior distributions are somewhat biased towards lower values. As expected, the power spectrum at these redshifts is not sensitive enough to break the degeneracy between $\gamma$ and $\tau_{\rm eff}$. The poor constraints on $\gamma$, however, do not affect our constraints on $T_{0}$ and $u_{0}$.

We have also tested how well our interpolation scheme was able to recover the thermal parameters of a completely independent model. We fit the power spectrum extracted from a simulation using the UV background model of \cite{Puchwein18} assuming non--equilibrium ionization, and verified that the values of $T_{0}$ and $u_{0}$ (see Figure \ref{fig:reiModels}) were recovered within the 68$\%$ uncertainties given by our MCMC method. This gives some reassurance that, as intended, our results do not significantly depend on the specific thermal histories adopted for the modeling in this work.

\section{Thermal histories overview}
\label{sec:TotT}
\begin{figure*} 
\centering
\includegraphics[width=1.0\columnwidth]{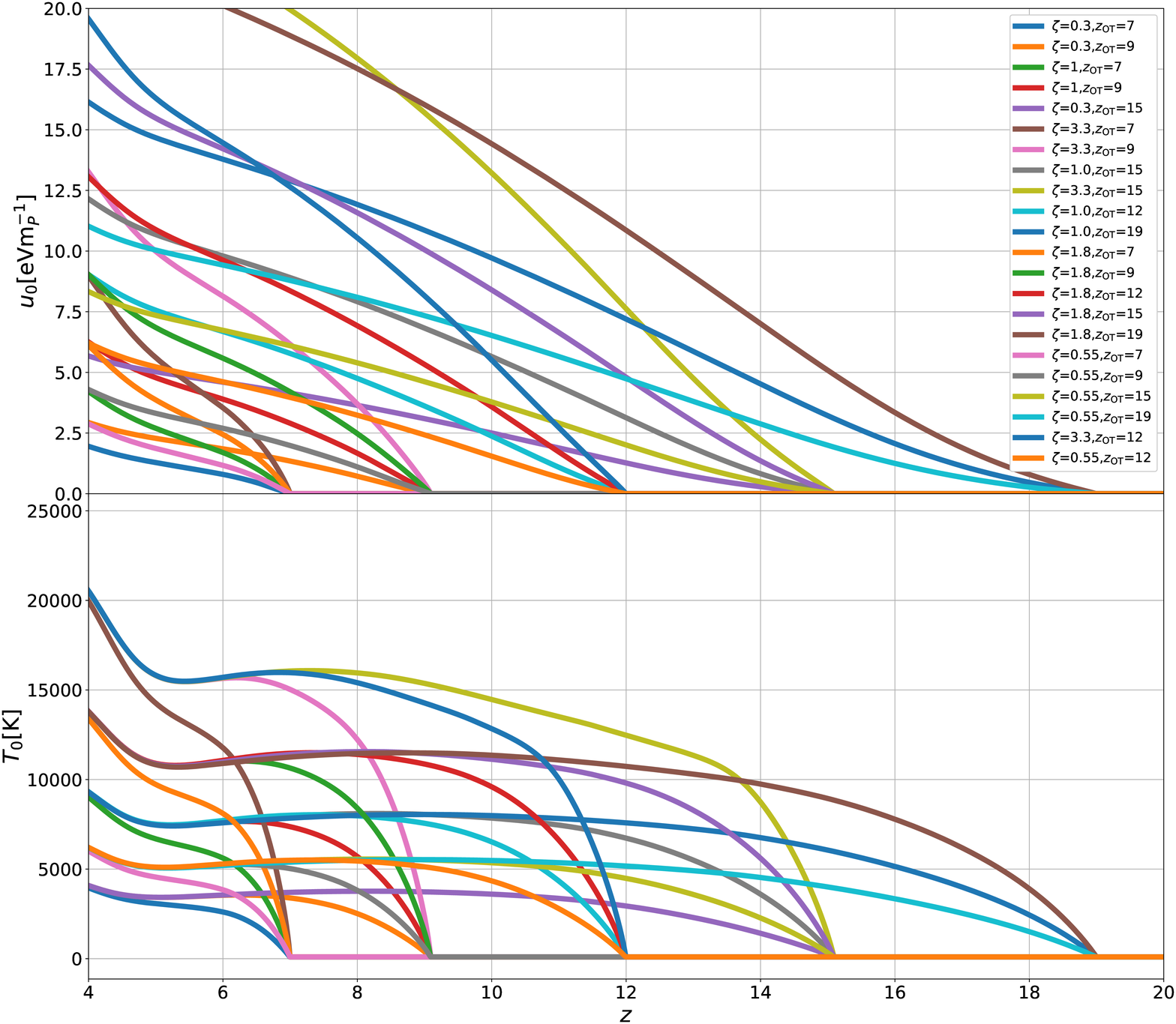} 
\caption{\small Evolution of the parameters governing the thermal state of the IGM in all of the simulations used for this analysis. Top panel: evolution as a function of redshift of the cumulative energy per unit mass, $u_{0}$, for models with a wide range of optically thin redshifts ($z_{\rm OT}$) and photo-heating rates ($\propto\zeta$). Bottom panel: corresponding evolution of the gas temperature at the mean density, $T_{0}$.}
\label{fig:TotThermalH}
\end{figure*}

For illustrative purposes we show in Figure \ref{fig:TotThermalH} the evolution of the thermal parameters $u_{0}$ and $T_{0}$ for all the simulations listed in Table \ref{table:sim}. While these models are not meant to represent realistic reionization scenarios, they provide a wide range of thermal histories and can be used to explore the thermal state of the IGM in a relatively model--independent way. 
\section{Integrated heating vs real space flux cutoff scale }
\label{sec:lambda}
\begin{figure*} 
\centering
\subfigure
{ 
\includegraphics[width=3.4in]{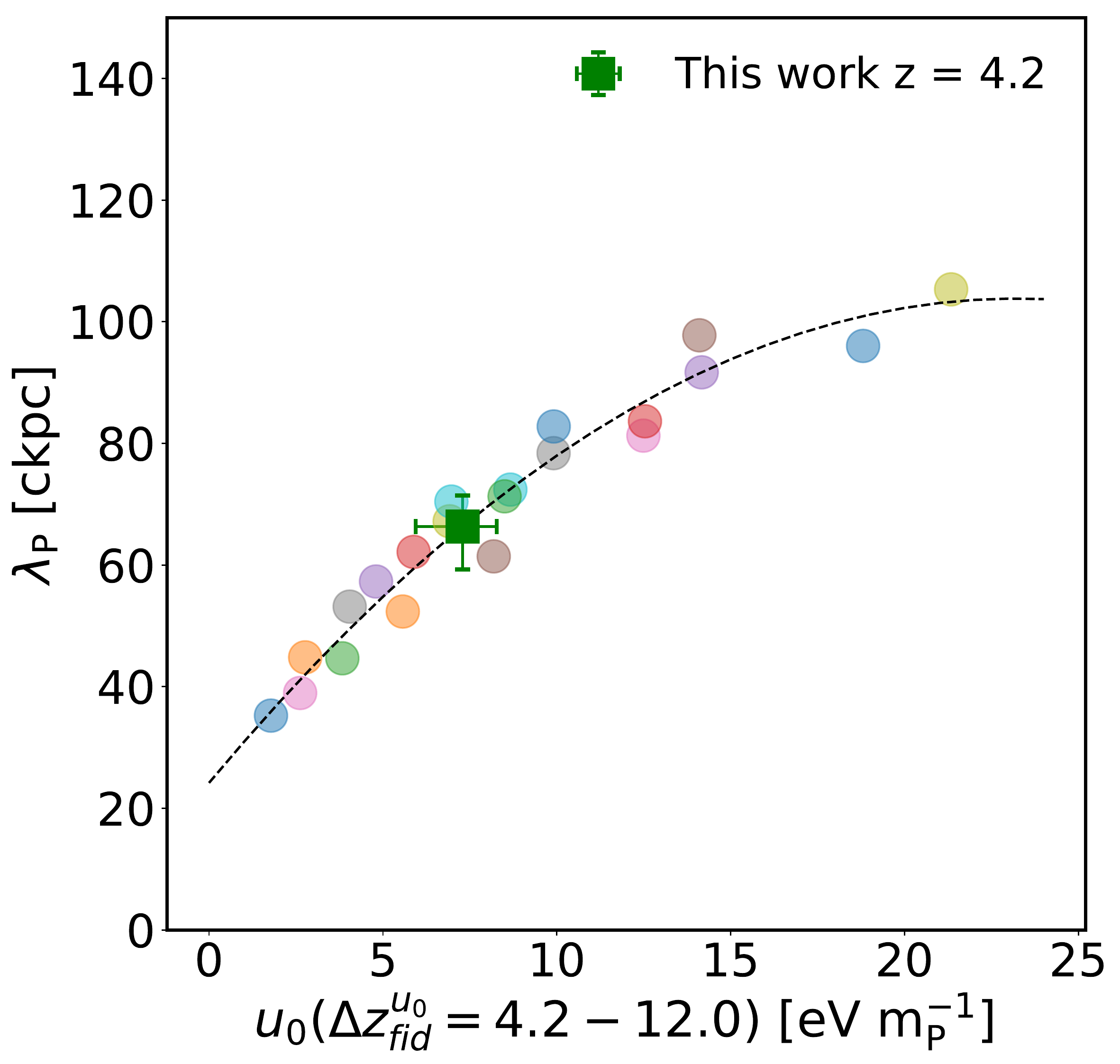}
 
}
\subfigure
{ 
\includegraphics[width=3.4in]{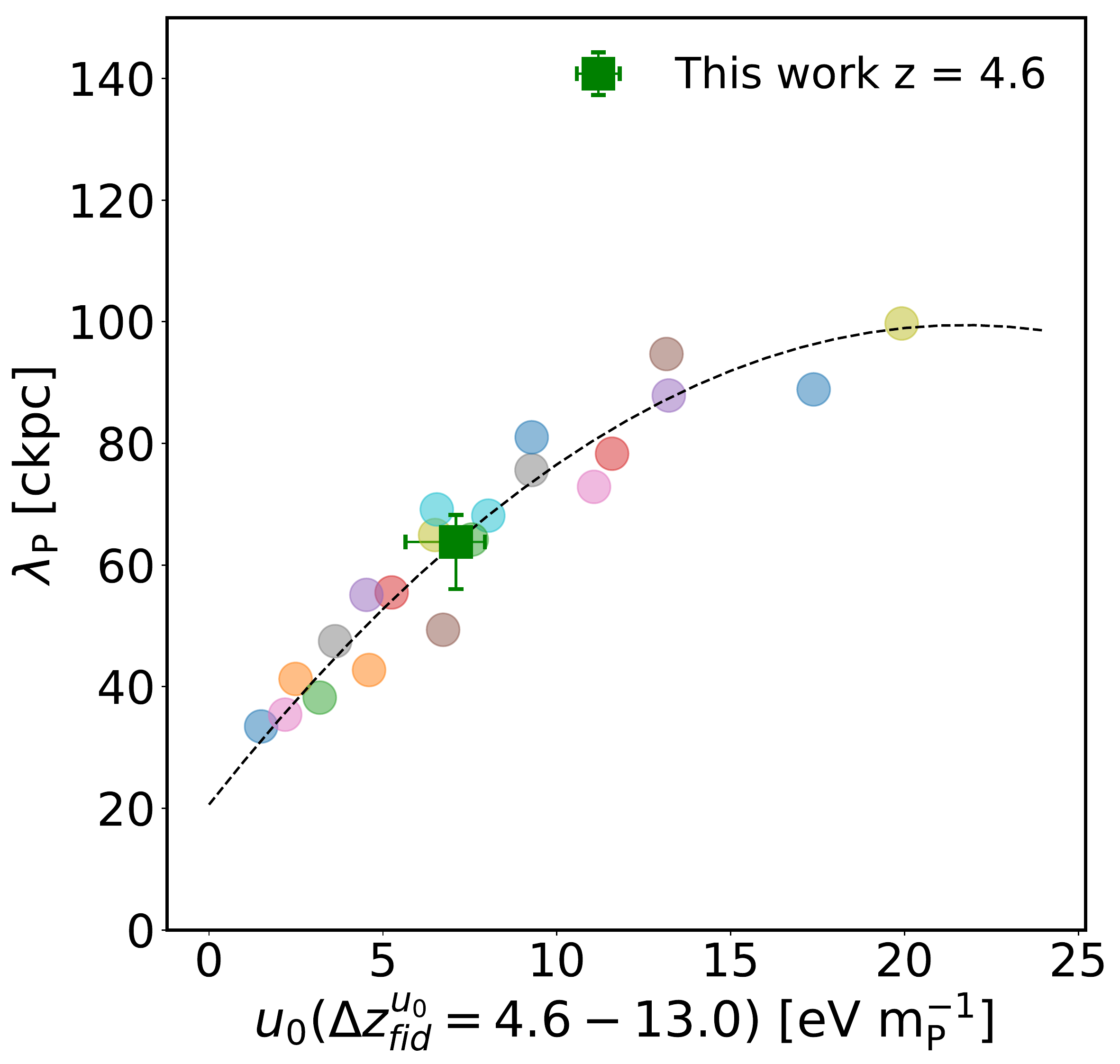} 

} 
\subfigure
{ 
\includegraphics[width=3.4in]{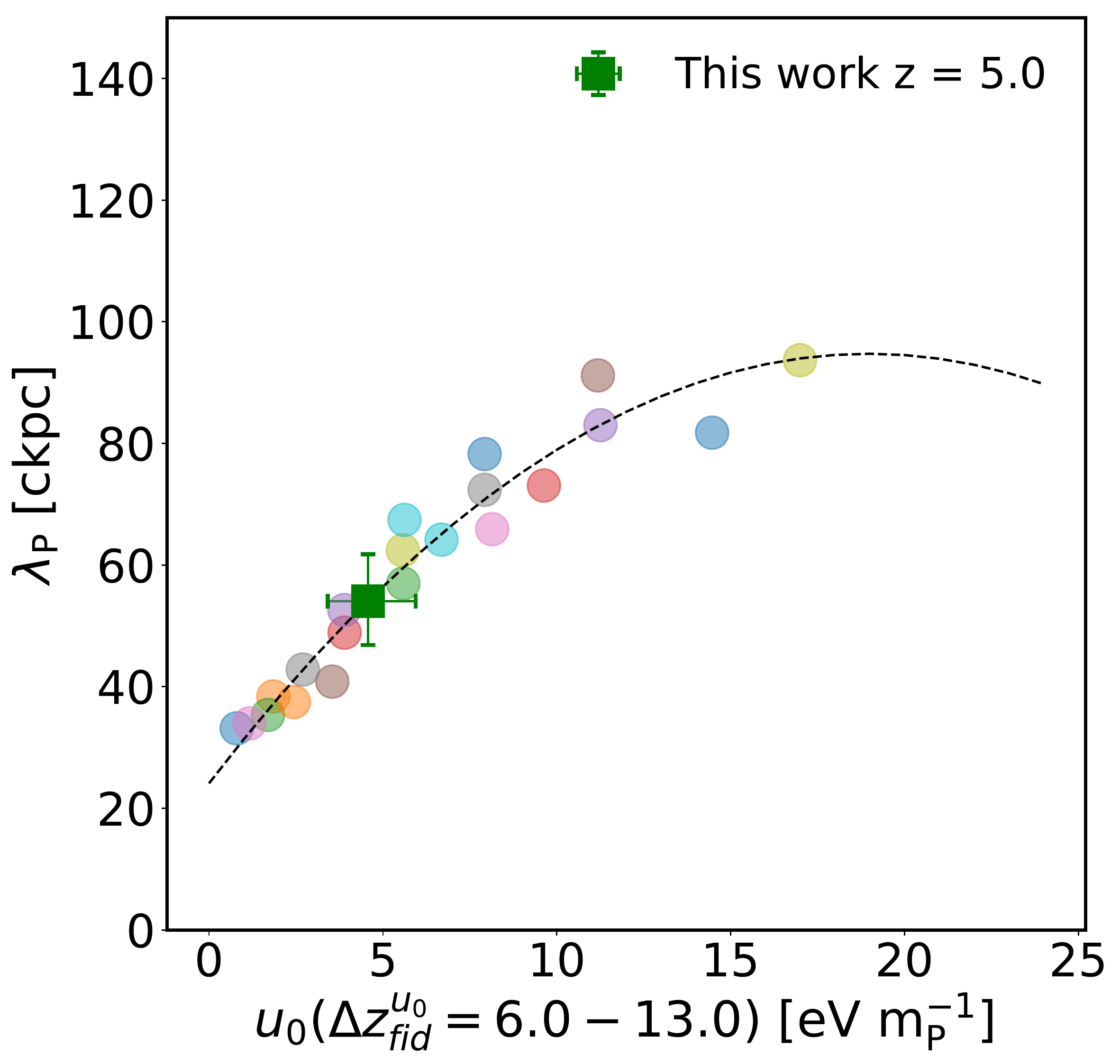} 

} 
 \caption{\small Relationship between the integrated heating per unit mass $u_{0}$ and the real space flux power cutoff scale $\lambda_{\rm P}$ of \cite{Kulkarni15}. Colored points correspond to the different simulations of Table \ref{table:sim}. While some scatter about the fit (black dashed line) is always present, there is a significant positive correlation between the two variables. For reference, along the fit at each redshift we plot our value of $u_{0}$ with the corresponding value of $\lambda_{\rm P}$ (green squares with $68\%$ errors).}
 \label{fig:lp}
\end{figure*}

In this Appendix we show the relationship between $u_{0}$ and the characteristic real space flux power cutoff scale, $\lambda_{\rm P}$, as defined by \cite{Kulkarni15}. 
At each redshift we compute $\lambda_{\rm P}$ for all the models of Table \ref{table:sim} following the method described in Kulkarni et al. We then fit a relationship between the corresponding $u_{0}$ computed over the fiducial redshift range. 
Figure \ref{fig:lp} shows the best fitting relationship between $u_{0}$ and $\lambda_{\rm P}$ for the redshifts considered in this work. While a certain level of scatter about the fit is present at all redshifts, there is clearly a positive correlation between the two variables. Using the current constraints on $u_{0}$ we can then attempt to obtain an rough estimate of the $\lambda_{\rm P}$ (green square with error bars).

We note these estimates for $\lambda_{\rm P}$ are smaller than the recent
constraints at $2<z<4$ from quasar pairs presented by \cite{Rorai17},
suggesting that the pressure smoothing scale increases toward lower
redshift as the IGM is photo--heated further (cf. the instantaneous
temperature measurements presented by \citealt{Becker11})

\section{Ly$\alpha$ flux power spectrum measurements}
\label{sec:Values}
In Tables 6 through 8 we report the power spectrum measurements obtained in this work for the three redshift bins centered at $z=4.2$, 4.6 and 5.0.
In each Table the values of the power spectrum obtained with (column 3) and without (column 2) instrumental resolution and pixel size correction (R.C.) are reported for each scale (column 1).
The corresponding 68$\%$ uncertainties are shown in column 4. 
The covariance matrices for the power spectrum measurements may be found in the on-line version of this article.

\begin{table} 
\centering \begin{tabular}{c c c c} 
\hline
$\log(k/$km$^{-1}$s$)$\ & $P_{k}$[km s$^{-1}]$\ & $P_{k}[$km s$^{-1}$](R.C.) \ & $\sigma$\\
\hline 
-2.2 & 91.4065 & 91.4324 & 27.0528\\
-2.1 & 82.4448 & 82.4819 & 17.3864\\
-2.0 & 70.2289 & 70.2789 & 19.4023\\
-1.9 & 74.6290 & 74.7128  & 17.9354\\
-1.8 & 56.1625 & 56.2625  & 10.6905\\
-1.7 & 43.7497 & 43.8733 & 7.6836\\
-1.6 & 30.6775 & 30.8155 & 6.1441\\
-1.5 & 27.2371  & 27.4304 & 4.1755\\
-1.4 & 21.1838  & 21.4225 & 2.9169\\
-1.3 & 14.3394 & 14.5968 & 2.3634\\
-1.2 & 7.9927 & 8.2213 & 0.9394\\
-1.1 & 4.2090 & 4.4020 & 0.4863\\
-1.0 & 2.0377 & 2.1891 &  0.2347\\
-0.9 & 0.8415 & 0.9444 & 0.1039\\
-0.8 & 0.3525 & 0.4241 & 0.0493\\
-0.7 & 0.1638 & 0.2208 & 0.0281\\

\hline 
\end{tabular}		
\label{table:4.2bin}
\caption{\small Power spectrum measurement for the redshift bin centered at $z=4.2$. Values in the third column have been corrected for instrumental resolution.
The reported values have been obtained from the analysis of 51 sections of 20 $h^{-1}$cMpc of Ly$\alpha$ forest with $z\epsilon [4.0,4.4)$, extracted from a total of 12 quasar lines of sight. The mean redshift for this bin is $\bar{z}=4.24$.}
\end{table}

\begin{table} 
\centering \begin{tabular}{c c c c} 
\hline
$\log(k/$km$^{-1}$s$)$\ & $P_{k}$[km s$^{-1}]$\ & $P_{k}[$km s$^{-1}$](R.C.) \ & $\sigma$\\
\hline 
-2.2 & 128.8440 & 128.8804 & 15.5333\\
-2.1 & 111.7963 &111.8463 & 13.5236\\
-2.0 & 91.4603 & 91.5253 & 10.6477\\
-1.9 & 94.5054 & 94.6114 & 11.2976\\
-1.8 & 74.2880 & 74.4201 & 7.6046\\
-1.7 & 64.1286 & 64.3093 & 6.5800\\
-1.6 & 53.5776 & 53.8172 & 5.3659\\
-1.5 & 40.5199 & 40.8068 & 3.6843\\
-1.4 & 28.5061 &  28.8268 & 3.0231\\
-1.3 & 21.0848 & 21.4623 & 1.6939\\
-1.2 & 12.1394 & 12.4857 & 0.8407\\
-1.1 & 6.0252  & 6.3006 & 0.4470\\
-1.0 & 3.1159& 3.3457 & 0.2371\\
-0.9 & 1.4523 & 1.6273 & 0.1078\\
-0.8 &0.5897 & 0.7071 & 0.0584\\
-0.7  & 0.2588 & 0.3465 &0.0342\\

\hline 
\end{tabular}		
\label{table:4.6bin}
\caption{\small Power spectrum measurement for the redshift bin centered at $z=4.6$. Values in the third column have been corrected for instrumental resolution.
The reported values have been obtained from the analysis of 114 sections of 20 $h^{-1}$cMpc of Ly$\alpha$ forest with $z \epsilon [4.4,4.8)$, extracted from a total of 15 quasar lines of sight. The mean redshift for this bin is $\bar{z}=4.58$.}
\end{table}

\begin{table} 
\centering \begin{tabular}{c c c c} 
\hline
$\log(k/$km$^{-1}$s$)$\ & $P_{k}[$km s$^{-1}]$\ & $P_{k}$[km s$^{-1}$](R.C.) \ & $\sigma$\\
\hline 
-2.2 & 162.4708 &  162.5166 & 30.6076\\
-2.1 &  163.8056 & 163.8787 & 28.1692\\
-2.0 & 157.6143 & 157.7257 & 25.4039\\
-1.9 & 121.8037 &121.9401 & 23.5215\\
-1.8 & 81.6827 &  81.8278 & 11.8521\\
-1.7 & 87.0669 & 87.3118 & 12.4211\\
-1.6 & 72.5758 & 72.8998 & 10.3949\\
-1.5 & 52.7895 & 53.1629 & 7.4238\\
-1.4 & 37.5451 &  37.9670 & 4.5890\\
-1.3 & 26.6637 & 27.1404 & 2.6700\\
-1.2 & 15.8800 & 16.3323 & 1.4250\\
-1.1 & 10.2757& 10.7436 & 1.0259\\
-1.0 & 4.6382 & 4.9787 & 0.5013\\
-0.9 & 2.1834 & 2.4443 & 0.2166\\
-0.8 & 1.0184 & 1.2185 & 0.1182\\
-0.7 & 0.3671 & 0.4897 & 0.0613\\

\hline 
\end{tabular}		
\label{table:5.0bin}
\caption{\small Power spectrum measurement for the redshift bin centered at $z=5.0$. Values in the third column have been corrected for instrumental resolution.
The reported values have been obtained from the analysis of 44 sections of 20 $h^{-1}$cMpc of Ly$\alpha$ forest with $z\epsilon [4.8,5.2)$, extracted from a total of 12 quasar lines of sight. The mean redshift for this bin is $\bar{z}=4.95$.  }
\end{table}
\end{appendices}

\end{document}